\documentclass[bibyear]{aa} %[referee] [bibyear]
\usepackage[varg]{txfonts}
\usepackage{graphicx}
\usepackage{soul}
\usepackage{natbib}
\usepackage[usenames,dvipsnames]{xcolor}
\usepackage{comment}
\usepackage{multirow}

\usepackage[colorlinks=true, linkcolor=blue, citecolor=blue, filecolor=blue, urlcolor=blue]{hyperref}

\newcommand{\tr}{Tr~14}
\newcommand{\Av}{$A_{\rm{V}}$}
\newcommand{\veil}{$r_{750}$}
\newcommand{\teff}{$T_{\mathrm{eff}}$}
\newcommand{\Chi}{$\chi^2_{\rm{red}}$}

\definecolor{red-violet}{rgb}{0.78, 0.08, 0.52}
\definecolor{mayablue}{rgb}{0.45, 0.76, 0.98}

\linenumbers

\begin{document}

\title{The population of young low-mass stars in Trumpler 14\thanks{Based on observations collected at the European Southern Observatory under ESO programme 097.C-0137.}}
%\subtitle{subtitle}
%\titlerunning{<short title>}
\authorrunning{Itrich et al.}

\author{Dominika Itrich\inst{1,2} \and
    Leonardo Testi\inst{3} \and
    Giacomo Beccari\inst{1} \and
    Carlo F. Manara\inst{1} \and
    Megan Reiter\inst{4} \and
    Thomas Preibisch\inst{2} \and
    Anna~F. McLeod\inst{5,6} \and
    Giovanni Rosotti\inst{7} \and
    Ralf Klessen\inst{8,9} \and
    Sergio Molinari\inst{10} \and
    Patrick Hennebelle\inst{11}
    }

\institute{
European Southern Observatory, Karl-Schwarzschild-Strasse 2, 85748 Garching bei München, Germany\\ \email{ditrich@eso.org} 
\and
Universit\"{a}ts-Sternwarte München, Ludwig-Maximilians-Universit\"{a}t, Scheinerstr. 1, 81679 München, Germany
\and
Dipartimento di Fisica e Astronomia "Augusto Righi" Viale Berti Pichat 6/2, Bologna, Italy
\and
Department of Physics and Astronomy, Rice University, 6100 Main St - MS 108, Houston, TX 77005, USA
\and
Centre for Extragalactic Astronomy, Department of Physics, Durham University, South Road, Durham DH1 3LE, UK 
\and
Institute for Computational Cosmology, Department of Physics, University of Durham, South Road, Durham DH1 3LE, UK
\and
Dipartimento di Fisica, Università degli Studi di Milano, via Celoria 16, 20133 Milano, Italy
\and
Universität Heidelberg, Zentrum für Astronomie, Institut für Theoretische Astrophysik, Albert-Ueberle-Straße 2, 69120 Heidelberg, Germany
\and
Universität Heidelberg, Interdisziplinäres Zentrum für Wissenschaftliches Rechnen, Im Neuenheimer Feld 205, 69120 Heidelberg, Germany
\and
INAF-Istituto di Astrofisica e Planetologia Spaziali, Via del Fosso del Cavaliere 100, I-00133, Rome, Italy
\and
Université Paris-Saclay, Université Paris Cité, CEA, CNRS, AIM, 91191, Gif-sur-Yvette, France
}

%\institute{Multidisciplinar de Astrof\'{\i}sica, IST, Avenida Rovisco Pais, 1049 Lisbon, Portugal\email{...}\label{inst1}

\date{Received / Accepted} %1 January 1995

%%%%%%% abstract %%%%%%%
\abstract{Massive star-forming regions are thought to be the most common birth environments in the Galaxy and the only birth places of very massive stars. Their presence in the stellar cluster alters the conditions within the cluster impacting at the same time the evolution of other cluster members. In principle, copious amounts of ultraviolet radiation produced by massive stars can remove material from outer parts of the protoplanetary disks around low- and intermediate-mass stars in the process of external photoevaporation, effectively reducing the planet-formation capabilities of those disks. 
Here, we present deep VLT/MUSE observations of low-mass stars in Trumpler 14, one of the most massive, young, and compact clusters in the Carina Nebula Complex. We provide spectral and stellar properties of 717 sources and {based on the distribution of stellar ages derive the cluster age of} 
$\sim$1~Myr. The majority of the stars in our sample have masses $\leqslant$1~$M_\odot$, what makes our spectroscopic catalogue the most deep to date in term of masses, and proves that detailed investigations of low-mass stars are possible in the massive but distant regions. 
Spectroscopic studies of low-mass members of the whole Carina Nebula Complex are missing. Our work provides an important step forward towards filling this gap and set the stage for follow-up investigation of accretion properties in Trumpler~14.
}

%-- stars: Hertzsprung-Russell and C-M diagrams
\maketitle 

%\section
%\sub(sub)section
%\paragraph

%%%%%%%%%%%%%%%%%%%%%%%%%%%%%%%
%%%%%%%%%%%% text %%%%%%%%%%%%%
%%%%%%%%%%%%%%%%%%%%%%%%%%%%%%%

%%%%%%% introduction %%%%%%%
\section{Introduction}
\label{sec:intro}
\indent

Star formation takes place in both, low-mass and massive complexes of molecular clouds. The latter is considered to be a more common star-forming environment in the Galaxy \citep[e.g.,][]{miller1978,lada2003, winter2018}. Those giant regions form hot and massive OB stars, which can significantly affect the formation and evolution of less massive cluster members. Copious amounts of ionising far- (FUV) and extreme-ultraviolet (EUV) photons, as well as an enormous volume of ejected mass via outflows or strong stellar winds, can create a so-called {\it negative} feedback. By injecting large amounts of energy into the surrounding medium, massive stars ionise and disperse the natal molecular cloud producing expanding H{\sc ii} regions \citep[e.g.,][]{freyer2003,krumholz2011, winter2018} and remove the matter from circumstellar disks that could otherwise be used to form a planetary system \citep[e.g.,][]{adams2004,anderson2013,facchini2016,eisner2018,winter2022}. 
On the other hand, expanding shock and ionisation fronts can also compress molecular clouds and in that way trigger the formation of new stars \citep[{\it positive} feedback, e.g.,][]{gritschneder2010, haworth2012}. 
There is a strong evidence that the formation of the Solar System took place in such a large cluster and was heavily influenced by close-by massive stars \citep[e.g.,][]{adams2010,pfalzner2015}. It is therefore of great importance to understand how the presence of massive stars in the cluster influences the intrinsic star formation, particularly of low-mass stars, as well as the initial mass function (IMF), star formation efficiency, or the planet formation capacity. 

Despite the importance of understanding the global picture of star formation, a substantial part of the investigation was so far focused on nearby ($<$300~pc), and therefore low-mass, star-forming regions \citep{manara2022}. While they are very important to construct and test a theory of formation of Sun-like stars, those studies neglect the role of the cluster environment. 
The closest massive region, the Orion Nebula, although 
providing excellent first examples of photoevaporating disks \citep[\lq\lq proplyds\rq\rq,][]{odell1993}, might not be representative of the most extreme environments where most of the stars in the Galaxy are forming \citep{smith2006}.

The greatest problem of studying massive star-forming regions is that they are all relatively far away from us ($>$1~kpc) and usually suffer from high extinction. These two factors significantly hinder the characterisation (and even detection) of individual members of these complexes, especially those less massive and fainter. Since the stellar content of any cluster is dominated by low-mass stars, lack of those objects can essentially impact results of studies of massive star-forming regions, as well as their interpretation. Moreover, low-mass stars are more vulnerable to the harsh environment than the massive ones \citep[see e.g.,][]{whitworth2004,almendros-abad2023}. 
Environmental conditions like high UV radiation impacts also the capability of protoplanetary disks around young stars to form planets \citep[e.g.,][]{throop2005,anderson2013,facchini2016, winter2018, parker2020, winter2020, qiao2023}.  

Another observational problem that often accompanies the studies of massive regions is a bright and variable emission from the surrounding H{\sc ii} region. Assessment of this emission in most cases cannot be done globally but requires the knowledge of the local variation of this emission. This issue is particularly profound when studying emission lines from the young stars, e.g., tracing accretion, winds, or jets. Stellar spectrum is contaminated with the nebular emission what can lead to potentially incorrect conclusions. Due to that reason, fiber-fed spectroscopy is not a good approach 
to study star-forming regions. A significantly more efficient way to obtain local and wavelength-dependent sky emission is to employ integral field spectroscopy (IFU) instruments. Current instrumentation offers several IFU spectrographs with medium to high spectral and spatial resolution \citep[e.g., ERIS, MUSE, KMOS at the Very Large Telescopes, or GMOS, NIFS at the Gemini Telescopes,][]{allington-smith2002, mcGregor2002, bacon2010,sharples2013, davies2023}. They allow obtaining position-dependent spectra of sources of interest as well as surrounding background and with that characterising faint objects in distant regions. 

The Carina Nebula Complex (CNC) is one of the biggest sites of star formation and one of the most massive H{\sc ii} regions in our Galaxy. It is located in the  plane of the Galactic Disk at a distance of 2.35 kpc from the Sun \citep{shull2021,goeppl2022}, which makes it the closest analog of a typical environment in which stars form.
Most if not all the clusters within the CNC are located at similar distances with very small distance dispersion of 1-2\% \citep{smith2006, cantat-gaudin2018, maizApellaniz2020, goeppl2022,berlanas2023}. 
Low interstellar extinction towards the region \citep[e.g.,][]{walborn1995,hur2023} makes it an even more suitable target for observational studies of massive clusters in the optical wavelengths. However, it was noticed that the reddening law towards the CNC is anomalous \citep[$R_{\rm V}$ = 4--5, e.g.,][]{smith2002} combined with the variable intracluster extinction \citep[by $\sim$9~mag][]{tapia2003, rowles2009, preibisch2012b}. Additionally, the CNC is located close to the galactic plane, what causes serious problems with contamination of the field stars (both foreground and background). 

The CNC contains more than 5$\times$10$^4$ stars \citep{povich2019} with a total mass of $\sim37 000~M_\odot$ \citep{preibisch2011a} immersed in a massive H{\sc ii} region.
While part of the CNC population is spread over a wide area characterised by a low stellar density regime, most of the star formation is confined in a number of star clusters, with Trumpler (Tr) 14, 15, and 16 being the most massive ones. 
These clusters host the greatest concentrations of O-type stars, which are expected to highly influence the evolution of their low-mass neighbours. 
There are at least 74 O-type stars in the CNC \citep{smith2006, berlanas2023}, including some of the most massive stars known: prototypical O2 (HD 93129A in \tr) and O3 (in \tr\ and Tr~16) stars, luminous blue variable $\eta$ Carinae (in Tr~16), as well as several Wolf-Rayet stars \citep{walborn1973,walborn2002,smith2006}.

Current observational campaigns of the Carina {Nebula} focused mostly on massive or intermediate-mass stars. Photometric studies included optical, near infrared (NIR), and X-ray observations. 
More than 100 stars in \tr\ and Tr~16 were observed by \cite{feinstein1973} down to G-type stars which allowed the authors to obtain distance values to the clusters close to the most recent ones. 
\cite{degioia-eastwood2001} investigated over 500 stars in \tr\ and Tr~16 with optical photometry detecting stars down to $\sim$1~$M_\odot$. 
\cite{tapia2003} presented the optical and NIR photometry of 4150 stars in the Carina {Nebula} with mass limit of 2~$M_\odot$. 
Multi-wavelength observations (optical + NIR) were also analysed by \cite{beccari2015} who built spectral energy distributions (SEDs) of 356 stars, obtained their stellar parameters (down to $<$0.4~$M_\odot$) and estimated their mass accretion rates. %$\lesssim$
Optical photometry of stars in \tr, Tr~16, and Collinder 232 was analysed by \cite{carraro2004} down to $\sim$1~$M_\odot$. 
\cite{hur2012} showed visual CCD photometry of the two most massive clusters in CNC and investigated their stellar content together with IMF with the limit of 1.5~$M_\odot$. They recently extended this catalogue by deep photometry of 135 000 stars down to 0.2~$M_\odot$ in $I$-band assuming the CNC age of 7~Myr \citep{hur2023}.

NIR photometry of massive and intermediate-mass stars was published by \cite{ascenso2007} together with a study of the mass function. \cite{povich2011} investigated mid-IR excess of $\sim$1400 young stars in Carina based on {\it Spitzer} observations. Extensive, wide-field, and deep NIR photometry of CNC from VISTA and HAWK-I of more than 600 000 sources down to $<$0.1~$M_\odot$ was published by \cite{preibisch2011a,preibisch2011b,preibisch2014}. %\lesssim
Later, \cite{zeidler2016} investigated their NIR excess. These surveys were often cross-matched with the deep X-ray imaging of the {\it Chandra} Carina Complex Project \citep[CCCP,][]{broos2011,townsley2011}, which identified $\sim$14~000 sources and help confirming the youth of low-mass Carina stars. The survey was preceded by study of Tr~16 with {\it Chandra} by \cite{albacete-colombo2008} and XMM-Newton observations of early-type stars \cite{antokhin2008}.

Individual, high-mass members of {the Carina Nebula} were classified first photometrically \citep{walborn1973,massey1993, walborn1995}, then spectroscopically \citep{levato1982, morrell1988, walborn2002,vaidya2015,maizApellaniz2016, preibisch2021,berlanas2023}. The first spectroscopic properties of massive, intermediate-mass, and solar-like stars in \tr\ and Tr~16 were obtained in the {\it Gaia}-ESO survey by \citep{damiani2017}, who used high-resolution (R$\sim$17~000) observations from FLAMES/Giraffe spectrometer at the ESO Very Large Telescope (VLT) to characterise more than 1000 stars and to portray characteristics of those two clusters. 

The photometric and spectroscopic works listed above are not a comprehensive list of all studies of the CNC. However, up to date no spectroscopic surveys targeting stars below 1~$M_\odot$ were conducted in the Carina {Nebula} leaving the most important part of the region uncharacterised. The aim of this work is to fill this gap and provide a spectroscopic catalogue of low-mass stars in one of the main clusters in the Carina {Nebula}, Trumpler 14.

%%%%%% Trumpler 14 %%%%%%

\tr\ is the most compact and youngest among the three main clusters in the CNC. Its structure was recognised to consist of a dense core ($r$ of 0.5\arcmin--0.9\arcmin corresponding to 0.3--0.6~pc at the distance of 2.35~kpc) and an extended halo population of possibly slightly older age \citep[4\arcmin--7.8\arcmin corresponding to 2.7--5.2~pc][]{tapia2003, ascenso2007, kharchenko2013}. 
Its age was estimated to be $\sim$1~Myr \citep{penny1993,vazquez1996,degioia-eastwood2001,carraro2004}; 2~Myr younger than Tr~16 \citep{walborn1995,smith2008}. It contains $\sim$20 O-type \citep{shull2021,berlanas2023} and several tens B-type stars. As a result, its ultraviolet luminosity is $\sim$20 times higher than $\Theta^1$Ori~C in the Orion Nebula \citep{smith2006,smith2008}. 
High UV field, high cluster density and mass, young age, and low reddening towards the cluster make \tr\ a perfect target to investigate the role of environment on star formation.

Here, we present the optical photometry and spectroscopy of young, low-mass stars in \tr. We define the methodology to detect those faint sources, extract their spectra, assess contamination from the sky emission, and conduct the spectral classification. Subsequently, we describe how the stellar properties are obtained. We conclude our work with a more global outlook on the \tr's properties. This study will be followed up by the detailed characterisation of accretion properties of the young stars presented here.

%%%%%%% observations %%%%%%%
\section{Data}
\subsection{Observations and data reduction}
\indent

Observations were carried out in 2016 with the Multi Unit Spectroscopic Explorer (MUSE), a second generation integral field unit (IFU) instrument on the VLT in Paranal, Chile \citep{bacon2010}, under the programme ID 097.C-0137 (PI: A. Mc~Leod). MUSE covers the wavelength range of 4650--9300~\AA~with a spectral resolution $R\sim4000$ (sampling of 1.25~\AA). Observations were performed without adaptive optics in Wide Field Mode (WFM) with spatial sampling of 0.2$\arcsec$ and a total field of view of $1\arcmin\times1\arcmin$. A wide region around \tr\, including the core of the cluster and the surrounding molecular cloud, was covered with 22 pointings with a total integration time per pointing of {39~min}. 
In the ESO archive, there are also available \lq\lq short\rq\rq\ exposures of total integration time of 15~sec, designed for observations of the brightest stars, which we do not use here because we are investigating only the low-mass content of \tr. The goal of the project was not only to capture the cluster members, but also to study the kinematics of the gas in the pillar-like structures north-east and south-west from \tr. The observations were designed to have small spatial overlap between individual pointings, and with that smoothly cover the whole area. Each of the pointings was observed three times with a 90$^\circ$ rotation dither pattern to better remove instrument artefacts. Observational logs are presented in Appendix~\ref{app:weather}.

Observations were reduced using the dedicated ESO pipeline v.~2.8.3 \citep{weilbacher2020} embedded in the {\tt EsoReflex} environment \citep{freudling2013}. The pipeline provides wavelength and flux calibrated IFU cubes. 
The standard stars used for calibration are listed in Appendix~\ref{app:weather} together with the observing conditions. Calibrated exposures were combined into the 3D data cubes, one per field, that were used for the analysis presented in this paper.

\begin{figure*}
\includegraphics[width=\textwidth,angle=0,trim={4.2cm 0.5cm 4.cm 0.5cm}, clip]{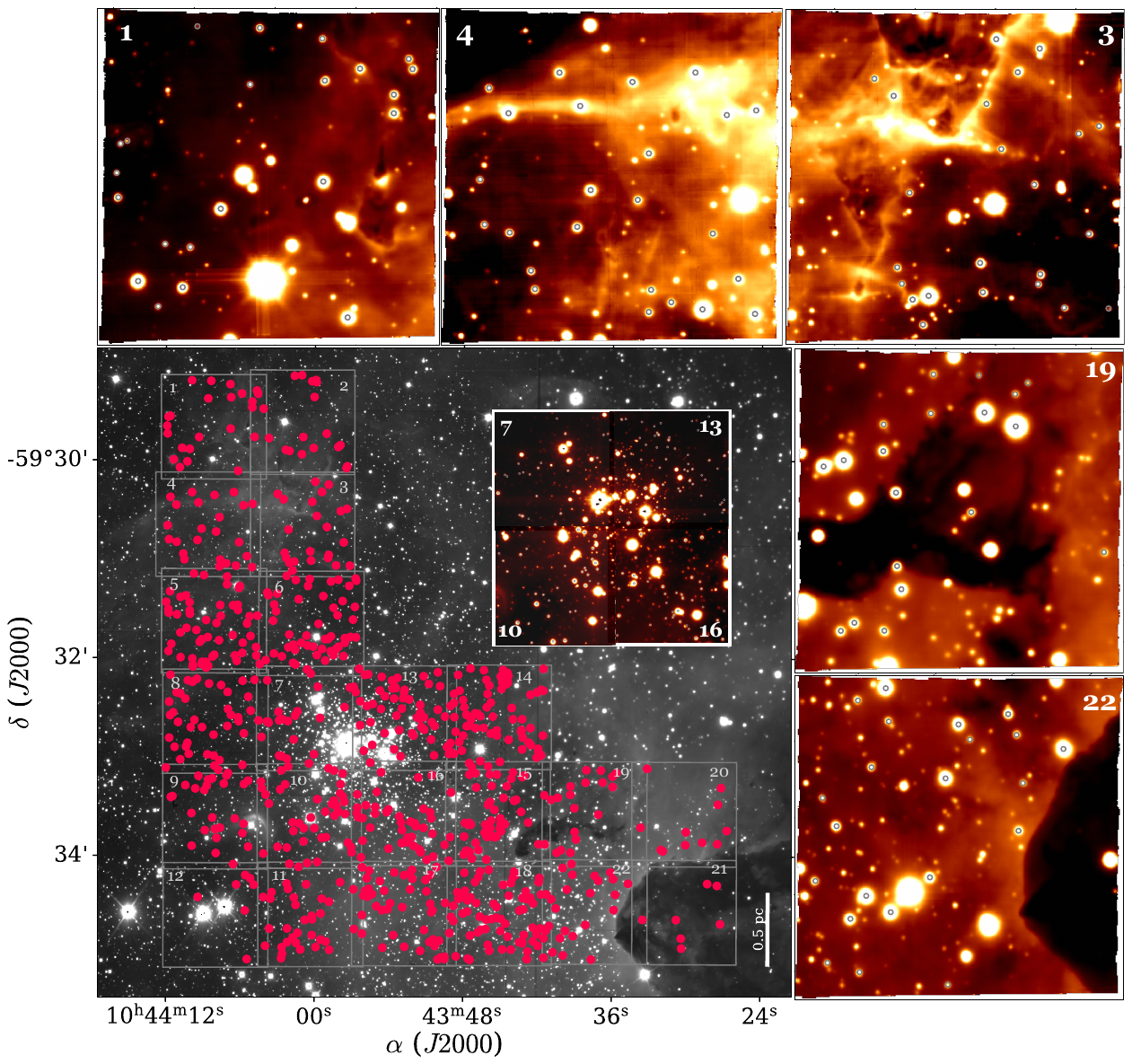}
\caption{Trumpler 14 cluster studied in this work. 
Grey sky image from HAWK-I $H$-band observations \citep{preibisch2011a,preibisch2011b} shows the whole region with the light grey grid of MUSE pointings and stars studied in this work marked as red circles. The five panels above and on the right show selected MUSE $I$-band images. The panel inside consists a mosaic of four MUSE pointing of the \tr\ center. Their numbers are indicated in the panels. Positions of the stars on MUSE images are marked with empty grey circles. The bar on the lower right corner of the HAWK-I image indicates the projected distance of 0.5~pc at the assumed distance of 2.35~kpc towards \tr\ \citep{goeppl2022}.
}
\label{fig:Tr14detect}
\end{figure*}

\subsection{Identification of sources and extraction of spectra}
\label{subsec:id}
\indent

In addition to producing the 3D data cubes, the ESO pipeline allows the user to extract photometric images, among other ones, in the standard Johnson-Cousins bands. We use these images as reference to guide the selection of bona-fide sources for the extraction of spectra from the IFU cubes. 
We use {\tt SExtractor} \citep[Source--Extractor\footnote{\url{https://sextractor.readthedocs.io/}},][]{bertin1996} to perform the source identification on the photometric images. 
{\tt SExtractor} is a free software designed to perform aperture photometry on astronomical images, 
suited also for crowded regions. It estimates and subtracts the background emission assuming its smooth variation. We tested the background estimation varying the size of the mesh cell and found that the size of 16 pixels gives the best performance in terms of recovering large gaseous features on the sky and at the same time not creating artificial ones. The same pixel size is used throughout the aperture photometry and extraction of the spectra. We in fact anticipate here that the latter is done by performing aperture photometry on each individual slice of the MUSE cube at the position of the bona-fide sources identified on the $I$-band images.

We first perform the identification of the sources on the $I$-band images, which we use as the reference. We employed a fixed aperture size of 5~pixels in diameter and the background mesh size of 16 pixels. Hence using the {\tt SExtractor} we create a catalogue which includes for each source the X and Y  position in the MUSE CCD reference frame together with {\it I}-band aperture magnitudes. Based on \lq\lq identification\rq\rq~image, we run the {\tt SExtractor} in double image mode on all photometric images obtaining magnitudes from other bands, namely $R$ and $V$. This is possible as all the images have identical dimensions being all extracted from the same IFU cube. With that we create the initial photometric catalogue of 5428 objects with $I$-band magnitude measurements. 

We used the same approach to extract spectra from the MUSE cubes of each source detected in the $I$-band images. 
Upfront, we slice MUSE datacubes with {\tt MissFITS}\footnote{\url{https://www.astromatic.net/software/missfits/}} \citep{marmo2008} into individual images, one per spectral element. Then, we run {\tt SExtractor}  on each slice and estimate flux for each source within the same aperture as for photometric images. Measured fluxes per spectral element are then combined into a single spectrum for each target. In the same way we extract wavelength-dependent flux uncertainties and sky spectra. Both are later used to evaluate goodness of stellar spectra.

\subsubsection{Coordinates correction}
\label{subsec:coord}
\indent

We transform astrometric coordinates of stars extracted from the MUSE cubes to the coordinate system of the recently released {\it Gaia} DR3 catalogue \citep{gaia-vallenari2022}. We first perform a match between the catalogues with a large separation limit (1-2\arcsec), separately for each MUSE pointing. We estimate median offsets of right ascension and declination for every pointing and adopt them as coordinate corrections. Absolute corrections range between 1.46\arcsec~and 5.75\arcsec~for right ascension, and between 0.08\arcsec~and 2.98\arcsec~for declination. We list corrections and show distributions of offset for each pointing in Appendix \ref{app:coord}. We apply the corrections to coordinates of our stars and list them in Table \ref{tab:cat}.

Consecutively, we match our catalogue with corrected coordinates once again with {\it Gaia} \citep{gaia-prusti2016,gaia-vallenari2022}. We find 1902 counterparts within the separation of 0.5\arcsec. In Appendix \ref{app:coord} we show the distribution of separations between {\it Gaia} and MUSE and argue the selection of the separation limit. Based on this distribution we also find that the accuracy of astrometry of our stars is of $\sim$0.1\arcsec.

\subsubsection{Identification of spurious sources}
\label{subsec:spurious}
\indent

Before analysing the sample we remove from the catalogue spurious sources, which  
may be due to low signal to noise, confusion near the very bright stars, and contamination from the structures in the nebular emission. 

Due to different atmospheric conditions the sensitivity of different pointings is uneven. Additionally, some images contain bright stars whose luminosity dominates at the images making the detection of the faint sources in their vicinity challenging. 
We first use flags issued by {\tt SExtractor} on photometric measurements to exclude potentially incorrect magnitudes. Those flags mark cases when neighbouring source likely bias estimation, when the light from the object has been deblended, when the position of the object is too close to the edge of the image, when one or more pixels were saturated, or when the photometry process was corrupted\footnote{The full description of {\tt SExtractor} flags is available on the website at \url{https://sextractor.readthedocs.io/en/latest/Flagging.html\#flags-def}}. With this approach, we remove 21\% of the sources from the catalogue. 
We additionally only accept stars with the photometric uncertainty in $I$-band of less than 0.1~mag. We remove photometric measurements not fulfilling those criteria in other bands. 
This procedure leaves 3082 sources in our catalogue. 

The Carina Nebula is an H{\sc ii} region, remarkably bright in some atomic lines (H$\alpha$, H$\beta$, O{\sc I}, He{\sc I}, etc.). 
In particular, small, compact gas concentrations can mimic light from the stars having stellar-like point spread function (PSF). 
We performed visual inspection of the {\it I}-band images comparing them to the other broad-band MUSE images, as well as to the HAWK-I {\it H}-band image \citep{preibisch2011a,preibisch2011b}, identifying all possible \lq\lq spurious\rq\rq\ detections that are not present in other images; we flag them accordingly and remove from the catalogue.  

We pay particular attention to the images affected by the presence of nearby saturated stars. Their light is spread on the pixels around their PSF on the CCD detector. This artificially changes the local background  making the measurement of the stellar flux falling on those regions unreliable.
Additionally, we identify elongated spikes near those bright stars, due to the diffraction pattern of the secondary mirror support. Combined, those effects significantly hinder the analysis of fainter stars in the closest neighbourhood of the bright ones. Based on visual inspection, we identify stars where light is not separated spatially on {\it I}-band MUSE images from the bright stars, flagged them as ,,illuminated'' and remove from the final photometry catalogue leaving 2727 stars.

\subsubsection{Detections in overlapping pointings}
\indent

Edges of some neighbouring pointings overlap causing double detection of the same stars in our photometry catalogue. After correcting coordinates, we defined a threshold separation of 0.5$\arcsec$ within which we looked for stars present in two (or more) pointings. We find 55 pairs of doubly observed targets. We exclude from the catalogue those sources that have worse signal-to-noise ratio (snr) of spectrum around 7500~\AA. We show comparisons of spectra between two detection in the Appendix \ref{app:double}.

\subsubsection{Background emission}
\label{subsec:bkg}
\indent

All the images of Carina Nebula, as an H{\sc ii} region, suffer from bright and highly variable background emission. The presence of a strong background with flux variations of the order of one stellar PSF makes the estimation of the local background in the vicinity of the stars very uncertain. Hence, the aperture photometry from which the stellar spectra are built, can be very imprecise and unreliable. 
During the stellar flux extraction (Sec. \ref{subsec:id}), background emission was estimated by {\tt SExtractor} assuming it is varying smoothly across the field. To make sure that the photometry is robust, we estimate the local background variation for each star in our catalogue 
based on the standard deviation (std) of the background estimates for stars within a radius of 20\arcsec. 
We adopt the threshold of 3$\sigma$, where $\sigma$ is the standard deviation, and remove all photometric measurements where the stellar flux is below the threshold. In Appendix \ref{app:bkg} we discuss the applied definition in depth. The final number of stars with well defined and reliable $I$-band magnitudes is 804 (Figure \ref{fig:Tr14detect}).

\subsubsection{Magnitude correction}
\indent

To check the flux calibration we compare the MUSE photometry with the optical photometry from the Wide Field Imager (WFI) at the MPG/ESO 2.2m telescope published by \cite{beccari2015}. The catalogues were matched adopting a 0.5\arcsec maximum separation radius between the stars from the two catalogs. We find 613 common stars in the two catalogs. We perform the comparison of magnitudes for each field and band separately. 
For the $I$-band the corrections vary between 0.16 and 1.28 mag depending on the MUSE pointing with the mean value of 0.55~mag. This is mainly due to the fact that each MUSE field was observed in very different weather conditions. All corrections are provided in Appendix \ref{app:mag}. 
We discard $B$-band magnitudes as highly uncertain for our faint stars. We provide in the Table \ref{tab:cat} the photometry of the sources extracted from the MUSE images with the magnitudes corrected to match the WFI ones. 

As a next step, we investigate the distribution of the magnitudes. We show the observed luminosity function for $I$, $R$, and $V$-bands from MUSE observations in Appendix \ref{app:mag}. The distribution of $I$-band magnitudes peaks at $\sim$18~mag and falls to $\sim$21~mag. If we adopt the cluster age of 1~Myr \citep{smith2008}, the distance to the cluster of 2.35~kpc \citep{goeppl2022}, and extinction of 2.3~mag (see Sec. \ref{sec:stars}), those magnitudes will correspond to the stellar masses of $\sim$0.8 and $\sim$0.14~$M_\odot$ according to the theoretical evolutionary models of \cite{baraffe2015}. Even though we do not correct luminosity function for completeness, those rough mass estimates show depth of our catalogue.

\subsubsection{Cross-match with other photometry catalogs}
\label{subsec:cross-match}
\indent

To complete our catalogue with information from other wavelength ranges, in addition to the {\it Gaia} DR3 and WFI,   
we cross-match the MUSE catalogue with VISTA \citep{preibisch2014}, HAWK-I \citep{preibisch2011a,preibisch2011b}, {\it Spitzer} \citep{povich2011}, and {\it Chandra} \citep{preibisch2011b,townsley2011} observations. For consistency, we define the same maximum separation of 0.5\arcsec~for all the catalogues. In Appendix \ref{app:coord} we explain the use of this separation limit. We find 658, 766, 26, and 309 stars in common between MUSE and VISTA, HAWK-I, {\it Spitzer}, and {\it Chandra}, respectively. We present in Table \ref{tab:cat} the example of the catalogue together with flags indicating the presence of the counterpart in any other catalogue. The full content of the catalogue is available online. 

We emphasise that we adopt a very conservative approach and apply severe photometric quality thresholds and checks to select only bona-fide stars with high quality spectra. In doing so we are aware that a number of real stars which have not passed our photometric quality criteria might have been removed from the final catalogue. In fact, a large number of these stars do have a counterpart in one or more of the catalogues used to complement the MUSE photometry. 
Among discarded sources are 841 {\it Gaia}, 212 WFI, 913 VISTA, 1809 HAWK-I, 2 {\it Spitzer}, and 120 {\it Chandra}  counterparts. 
We are aware that within this limitation our catalogue is not complete in terms of cluster members. We report the list of probable members with uncertain photometry due to the background contamination in Table \ref{tab:highbkgcat} and assess the completeness in the next section.

\subsection{Completeness of the catalog}
\label{subsec:complete}
\indent

\begin{figure}%[h!]
\includegraphics[width=\columnwidth, trim={0.4cm 0.45cm 0.55cm 0.4cm}, clip]{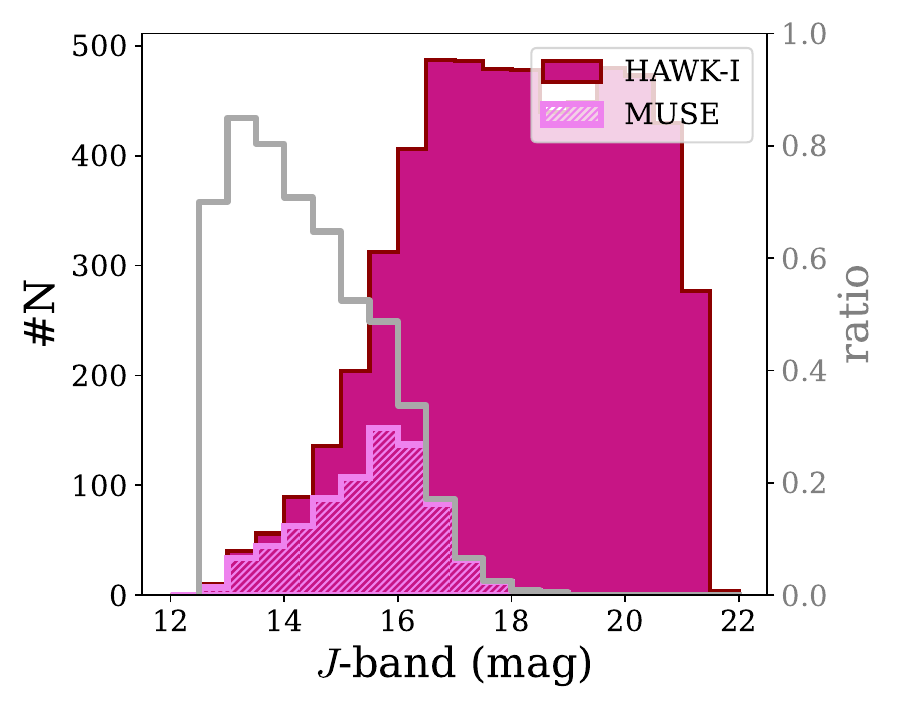} 
\includegraphics[width=\columnwidth, trim={0.4cm 0.45cm 0.55cm 0.4cm}, clip]{figures/Jband-compl-Hawki-full\_bkg\_frg-hatched-no_line.pdf} 
\caption{The distribution of $J$-band magnitudes from HAWK-I {in the MUSE field} \citep{preibisch2011a,preibisch2011b}. The dark berry histogram shows all HAWK-I measurements taken in the same area that was covered by MUSE. The violet distribution presents {point source detections from Sec. 2.2.6,} 
({upper panel}) and 
{in combination} with the one excluded due to the high background variation and foreground stars ({lower panel}). In both panels, the grey line shows completeness of our catalogue defined as a ratio of the number of stars in a given magnitude bin (0.5~mag wide) in our catalogue and in the HAWK-I catalogue. 
}
\label{fig:JcomplHAWKI}
\end{figure}

The goal of this work is to have a high quality spectroscopic sample of low-mass members of \tr. In order to achieve this goal we apply a number of quality cuts to the photometric catalogue (Sec. \ref{subsec:bkg}), which can affect the interpretation of our results.

To better understand the limitations of our work, we estimate 
the completeness by comparison to the photometric catalogue from HAWK-I \citep{preibisch2011a,preibisch2011b}. Figure \ref{fig:JcomplHAWKI} shows distribution of $J$-band magnitudes observed by HAWK-I in the exact same region as covered by our MUSE pointings (dark berry histogram). The stars in common between the two catalogues are shown in violet. {With the grey line we show the ratio between the stars retrieved in our catalogue and the number of stars observed by HAWK-I per magnitude bin of 0.5~mag. The ratio gives us a rough estimate of the completeness of our catalogue.} 
The upper panel shows only those stars for which three sigma level of background variation did not exceed $I$-band flux measured with MUSE.  
Their $J$-band magnitudes range from 12.5 to 19.0~mag, corresponding at the low-mass end to 0.065~$M_\odot$ at 1~Myr \citep{baraffe2015}.  Assuming that the HAWK-I catalogue is complete down to $\sim$21~mag in $J$-band \citep{preibisch2011a}, we can adopt this ratio as the rough estimate of the completeness of our catalogue. Based on this assumption, we reach 50\% level of completeness  
at $\sim$15.5~mag corresponding to 0.8~$M_\odot$ at 1~Myr. The 30\% completeness level is achieved at $\sim$16.5~mag corresponding to {$\sim$0.4~$M_\odot$} 
at 1~Myr. As we indeed see later in Sec. \ref{subsec:MUSEphot} or \ref{subsec:HR}, our deep observations allow us to detect and characterise very low-mass stars in \tr. However, due to our conservative approach to the background emission (Sec. \ref{subsec:bkg}), the final sample is highly incomplete in the low end of mass spectrum.

In the lower panel of Figure \ref{fig:JcomplHAWKI} we include the \lq\lq full MUSE\rq\rq\ sample consisting the one with robust $I$-band MUSE photometry and the one discarded due to the highly variable background emission. Here, flagged or uncertain MUSE photometry sources (Sec. \ref{subsec:spurious}) were not included. We see immediately that with our approach we removed mostly faint, low-mass objects. The \lq\lq full\rq\rq\ catalogue extends to the $\sim$21 $J$-band magnitude \citep[0.018~$M_\odot$ at 1~Myr adopting the evolutionary models of][]{baraffe2015} and reaches 50\% level of completeness at $\sim$18.5~mag, corresponding to 0.085~$M_\odot$ at 1~Myr. The 30\% completeness level is achieved at $\sim$19~mag (0.065~$M_\odot$). 

The significant difference between the two distributions (\lq\lq MUSE\rq\rq\ sample with robust $I$-band photometry and \lq\lq full MUSE\rq\rq\ sample affected by the background emission) shows the possible impact of the adopted procedure on the final results and the estimated global properties of \tr. Since the deep NIR photometric observations can contain significant fraction of contamination from background sources, especially in the faint end (see the discussion in Sec. 3.3 in \citet{preibisch2011a} and in Sec. 2.3 in \citet{preibisch2011b}), we do not correct our analysis for the incompleteness. We assume that our study is complete at the level of 50\% for stars more massive than 0.8~$M_\odot$ and at the level of 30\% for stars more massive than 0.4~$M_\odot$.

%%%%%%% Table 1 %%%%%%
\begin{table*}
\caption{catalogue of low-mass in Trumpler 14 stars analysed with MUSE observations.}
\label{tab:cat} 
\resizebox{\textwidth}{!}{
\begin{tabular}{cccccccccccccccccccccccccc}
\hline
\hline 
ID & coordinates & $I$-band & $R$-band & $V$-band & snr$_{I,\mathrm{bkg}}$ & snr$_{R,\mathrm{bkg}}$ & snr$_{V,\mathrm{bkg}}$ &  possible\_frg\_bkg & gaia\_flag & wfi\_flag & VISTA\_flag & hawki\_flag & spitzer\_flag & chandra\_flag & NIR excess & SpT & $T_{\mathrm{eff}}$ & \Av & \veil & $\log{(L_{\mathrm{bol}})}$ & $M_{\mathrm{*,PARSEC}}$ & $M_{\mathrm{*,B15S00}}$ & Age$_{\mathrm{PARSEC}}$ & Age$_{\mathrm{B15S00}}$ \\ 
 & (h:m:s ~d:m:s) & (mag) & (mag) & (mag) & & & & & & & & & & & &  & (K) & (mag) &  & $L_\odot$ & ($M_\odot$) & ($M_\odot$) & (Myr) & (Myr) \\ 
\hline
F01N009 & 10:44:09.93 -59:29:11.72 & 20.49$\pm$0.12  &   -- &  --  &  3.96   &   --   &   --  &   False   &   None  &  False  &  True   &    True   &  False   &  False  & True & M5.0$^{+1.6}_{-0.7}$ & 3125$_{-201}^{+103}$ & 2.70$_{-1.07}^{+1.22}$ & 0.08$_{-0.07}^{+0.55}$ & -1.05$_{-0.85}^{+0.51}$ & 0.38 & 0.18 & 6.2 & 2.3 \\
F01N010 & 10:44:08.38 -59:29:12.03 & 17.37$\pm$0.09 & 18.11$\pm$0.09 & 19.01$\pm$0.08 & 17.26  &  4.18  &  5.00 &  False  & Good   &  True   &    True    &   True   &  False  &  False  & False & K4.0$^{+0.6}_{-0.6}$ & 4561$_{-110}^{+108}$ & 1.20$_{-0.43}^{+0.41}$ & 0.00$_{-0.00}^{+0.56}$ & -0.32$_{-0.36}^{+0.34}$ & 1.02 & 0.99 & 13.0 & 19.8 \\ 
\hline
\end{tabular}}
\tablefoot{The first column gives IDs of the detected sources, the second one lists coordinates. The third, fourth, and fifth columns give apparent magnitudes in $I$, $R$, and $V$-band, respectively. The sixth, seventh, and eighth columns provide signal-to-noise of the flux with respect to the background variation in a given band, as indicated by the lower script (see Sec. \ref{subsec:bkg} for details). 
The ninth marks possible foreground or background stars (see Sec. \ref{subsec:frg} for definitions). 
The next five columns flag matches with other catalogs: {\it Gaia} \citep{gaia-vallenari2022}, WFI \citep{beccari2015}, VISTA \citep{preibisch2014}, HAWK-I \citep{preibisch2011a,preibisch2011b}, {\it Spitzer} \citep{povich2011}, and {\it Chandra} \citep{preibisch2011a,townsley2011}. The following indicates if the star has an NIR excess as defined by \cite{zeidler2016}. In the consecutive {nine} columns are given stellar parameters: spectral type, effective temperature, visual extinction, constant veiling at 7500~\AA, bolometric luminosity, {and stellar mass and stellar age estimated from PARSEC \citep{bressan2012} and \cite{baraffe2015} / \cite{siess2000} tracks, as indicated by the subscript}. 
A full version of this table will be available at the CDS upon publication. The first few rows are shown as an example.}
\end{table*}

%%%%%%%%%%%%%%%%%%%%%%%%%%%%%%%%
%%%%%%%%%%%% stars %%%%%%%%%%%%%
%%%%%%%%%%%%%%%%%%%%%%%%%%%%%%%%
\section{Stellar population} 
\label{sec:stars}

\subsection{Identification of foreground stars}
\label{subsec:frg}
\indent

To exclude possible contamination from foreground and background stars we use accurate {\it Gaia} astrometry for our sources. We first perform a number of quality checks on the matched objects. We first exclude all stars that have goodness of fit parameter RUWE $>1.4$ \citep{lindegren-ruwe}, astrometric\_gof\_al $>5$ \citep{lindegren2021}, parallax over error lower than 5, and uncertainty of the proper motion above 20\%. We flag them as object with poor {\it Gaia} astrometry, {\tt gaia\_flag}=\lq poor\rq. We find 175 good objects out of the 794 {\it Gaia} counterparts. None of them is flagged as a non-single star or a duplicated object, reassuring us about the good quality of the astrometry. All of them have 5- or 6-parameter solution. We select fore- and background stars based on parallaxes corrected for bias, as described in \cite{lindegren2021}. This correction is possible only for stars with $G$-band magnitude between 5 and 21, and for the effective wavenumber (for 5-parameter solution) or the pseudocolour (for 6-parameter solution) between 1.24 and 1.72~$\mu$m$^{-1}$. Wavenumbers are calculated using calibrated BP/RP spectra, while pseudocolour is an approximate colour of the source based on its astrometric solution utilising the chromaticity of  the instrument.

\begin{figure}
\includegraphics[width=\columnwidth]
{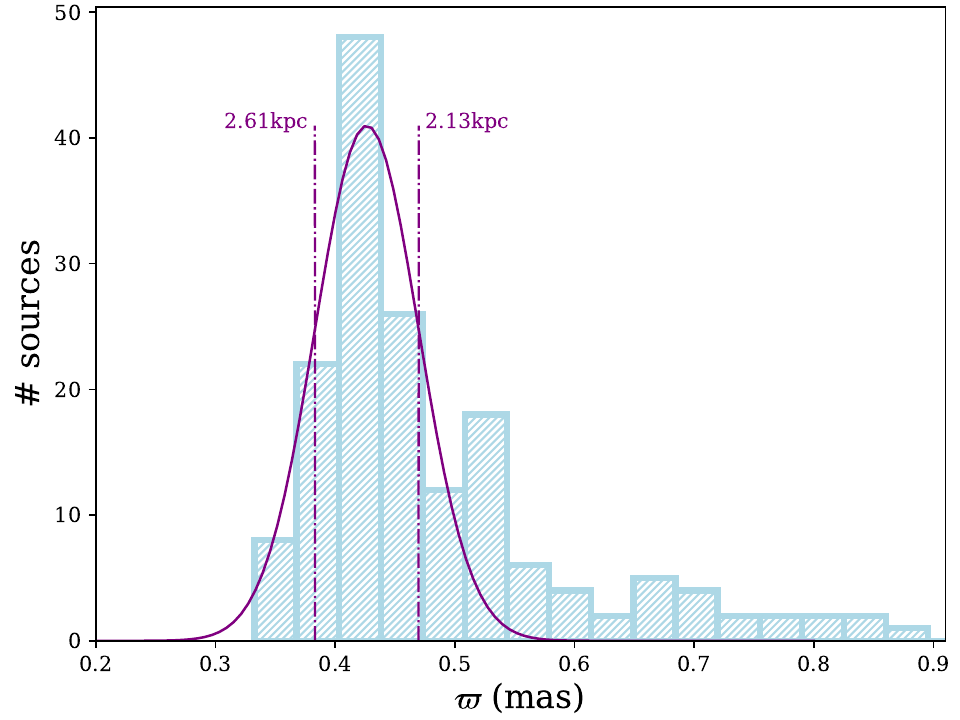}
\caption{Distribution of 
corrected parallaxes (light blue histograms) with fitted normal distribution 
(purple line). Dashed-dotted lines mark 1$\sigma$ width of the fitted distribution and applied ranges for excluding fore- and background stars.}
\label{fig:parallax}
\end{figure}

We illustrate the distribution of corrected parallaxes of stars with good {\it Gaia} astrometry in Figure \ref{fig:parallax}. The Gaussian profile fit to the distribution is centred at $\varpi=0.43$~mas and has a width of $\sigma=0.04$~mas. The centroid parallax corresponds to the distance of 2.35~kpc, in very good agreement to the findings of \cite{goeppl2022}. We follow their procedure to identify fore- and background stars. We define the background stars as those, whose 3$\sigma$ extend of the parallax value (i.e. parallax value plus its error) is smaller than the $\varpi_{\rm{min}}$ ($\varpi+3\sigma_\varpi<\varpi_{\rm{min}}$), while foreground stars as those with 3$\sigma$ extend (i.e. parallax value minus its error) higher than $\varpi_{\rm{max}}$ ($\varpi-3\sigma_\varpi>\varpi_{\rm{max}}$). We adopt as $\varpi_{\rm{min}}$ and $\varpi_{\rm{max}}$ values corresponding to range of parallaxes defined by the width of the Gaussian distribution (0.43$\pm$0.04~mas), further corresponding to the distance range of (2.61~kpc, 2.13~kpc). 
Out of 175 'good' {\it Gaia} counterparts we identify 0 background stars and 24 foreground stars. 
However, if we take into account all possible {\it Gaia} matches with corrected and positive parallaxes and possibly high astrometry uncertainties (784 stars), then for the same parallax ranges we find 35 possible foreground stars and 10 possible background stars. We remove from our catalogue 24 foreground stars with robust astrometry and flag remaining 21 stars as possibly contaminated (fore- or background) stars ({\tt possible\_frg\_bkg}). Summarising, we find 794 {\it Gaia} counterparts, including 175 with good astrometry; 24 of them are foreground stars. At the end our catalogue contains 780 sources.

\subsection{Colour--magnitude diagram}
\label{subsec:MUSEphot}
\indent

\begin{figure*}
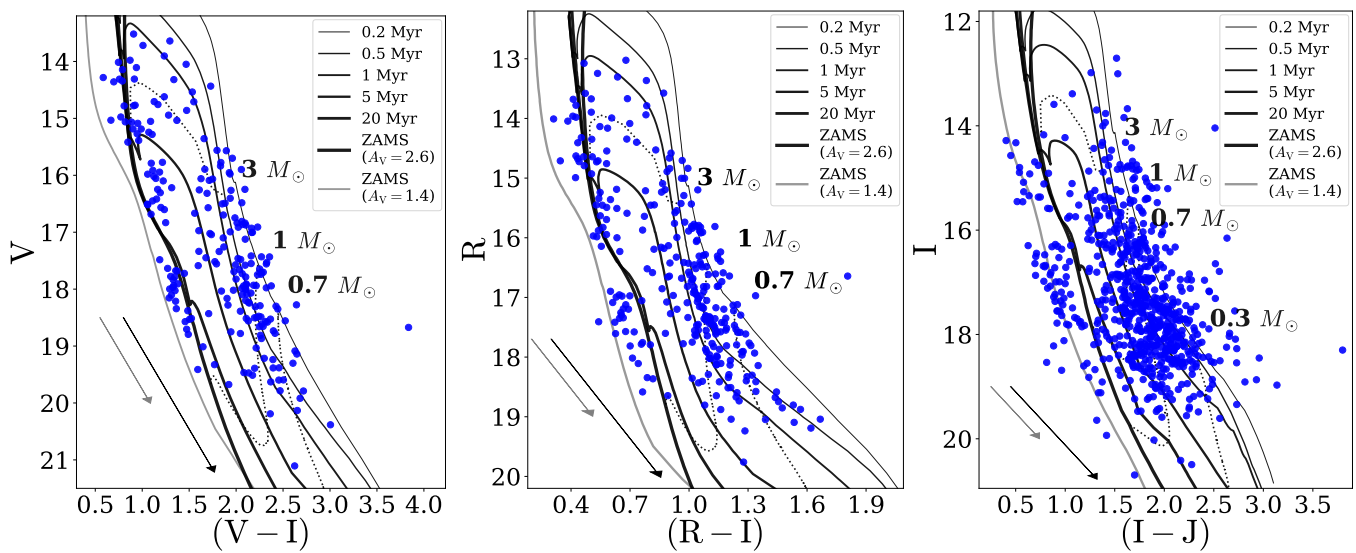

\includegraphics[width=.65\columnwidth, trim={0 0 0cm .cm}, clip]{figures/CMD-V\_I-final-MASTER-vPARSECf2.pdf}
\includegraphics[width=.65\columnwidth, trim={0 0 0cm .cm}, clip]{figures/CMD-R\_I-final-MASTER-vPARSECf2.pdf}
\includegraphics[width=.65\columnwidth, trim={0 0 0cm .cm}, clip]{figures/CMD-I\_J-final-MASTER-vPARSECf2.pdf}
\caption{Colour--magnitude diagrams from MUSE broad-band filters images in $V$ and $I$ ({right panel}), $R$ and $I$ magnitudes ({middle panel}), and $I$ and $J$ magnitudes ({left panel}). $J$-band magnitudes are from VISTA and HAWK-I instruments. Shown are only data points with $I$-band magnitudes above 3$\sigma$. {Solid lines show PARSEC isochrones from 0.2 to 20~Myr and ZAMS, dotted lines show isomasses of 0.3, 0.7, 1, and 3~$M_\odot$, as labeled \citep{bressan2012}.} 
Isochrones were reddened by the average extinction \Av\ = 2.6~mag measured from MUSE spectra (see Sec. \ref{subsec:Av}). Additionally, we plot the ZAMS reddened by \Av=1.4~mag with grey lines.}
\label{fig:CMD}
\end{figure*}

We first present the colour-magnitude diagrams (CMDs) based on corrected MUSE photometry. Figure \ref{fig:CMD} shows two CMDs based on (V-I) and (R-I) colours from MUSE, exclusively, and one based on $I$-band magnitude from MUSE and $J$-band magnitudes from HAWK-I or VISTA. 
We also plot {PARSEC v1.2S\footnote{\url{http://stev.oapd.inaf.it/cmd}} theoretical isochrones with black solid 
lines \citep{bressan2012, chen2014}} 
reddened by \Av=2.6~mag, best matching our observations (see also Sec. \ref{subsec:Av}), using extinction law of \cite{cardelli1989} and $R_{\mathrm{V}}$=4.4 \citep{hur2012}. We apply a distance modulus of 11.86~mag, equivalent of the distance to \tr\ \citep{goeppl2022}. We show tracks for 0.3, 0.7, 1, {and 3}~$M_\odot$ stars with dotted lines. Our observational CMDs already demonstrate that, despite very conservative quality control, our MUSE data allows us to sample stars in \tr\ 
with robust $I$-band photometry 
down to $\sim0.3~M_\odot$. We note, however, that due to the decrease of snr at shorter wavelengths, number of robust $R$ and $V$-band magnitudes is smaller than in $I$-band, and at the same time the number of very low-mass stars detected in those bands reduced.

\cite{ascenso2007} used high resolution near-IR data to study the core of Trumpler 14. Based on their photometry they found a global visual extinction towards \tr\ of \Av=2.6$\pm$0.3~mag and a sparse foreground population with \Av\ of 1.4~mag. Additionally to the isochrones reddened by the visual extinction matching our observations, we show also in Fig.~\ref{fig:CMD} the location of the Zero Age Main Sequence (ZAMS, grey solid line) reddened by the visual extinction of a sparse population (1.4~mag). The authors suggested that this population of older stars comes from the nearby young clusters. In our CMDs we see indication of two separate populations, one concentrated around 1~Myr isochrone, and the other following ZAMS. We note similar, two-population CMDs of more massive stars in the work of \cite{carraro2004} (see their Fig. 5). We will examine this feature with spectral classification in the following sections. 

\subsection{Spectral classification}
\label{subsec:spt}
\indent

The goal of this paper is to characterise the low-mass members of Trumpler 14 and provide a catalogue of their stellar parameters. As shown in Fig.~\ref{fig:CMD}, our dataset samples a wide range of stellar masses and colours, and thus spectral types. 
Hence, we split the procedure of spectral classification into two cases and give the detailed description in the forthcoming sections. We note that our procedure is comparable to the one adopted by \cite{fang2021} in the study of the Trapezium cluster.

We base the spectral classification on Class III templates observed with VLT/X-Shooter and published by \citet{manara2013, manara2017}. The list of templates is provided in Appendix \ref{app:templ}. Those sources where previously studied in the literature and their spectral types are well known. We note here that they all have a negligible extinction \Av$<$0.3~mag. 
We will refer to the Class III templates as \lq\lq templates\rq\rq~later in the text. We degraded the templates spectra (with natal resolution of $R\sim7500$--18200) to the MUSE resolution convolving them with a Gaussian kernel and then re-sampled on the common for the both instruments spectral range ($\sim$5500--9350~\AA). The comparison between the spectra is done in the aforementioned range after normalising to the flux at 7500~\AA, $f_{750}$.

%%%%%%%%
\subsubsection{M-type stars}
\label{subsubsec:Mstars}
\indent 

The spectra of M-type stars have a characteristic shape in the optical range due to the presence of TiO and VO absorption bands. The depth of those features changes with the spectral sub-type and increases with later stellar types. 

In this work, only spectra that have sufficient signal-to-noise (snr$>$10) are used and classified. From the whole sample of spectra we pre-select those, that might be of M-type based on spectral indices from \cite{riddick2007}, \cite{jeffries2007} and \cite{oliveira2003} (TiO feature at 7140~\AA) and \cite{herczeg2014} (TiO 7140, 7700, 8465~\AA). We require that at least half of the indices suggest an M-type spectrum. We remove from the spectra prominent emission and absorption lines to prevent confusion in the fitting to the templates. Spectral classification is performed together with the estimation of the visual extinction, \Av, and veiling at 7500~\AA, \veil. We veil and redden templates using the Cardelli extinction law \citep{cardelli1989} and a pre-defined grids of \Av\ and \veil\ values. The extinction is sampled with a step of 0.1~mag between 0.0 and 7.0~mag, while the veiling is assumed to change between 0.0 and 1.9 with a step of 0.02. The average extinction towards \tr~was found to be 2.6$\pm$0.3~mag \citep{ascenso2007}, thus we do not expect huge variation of \Av\ for cluster members. The adopted sampling of the extinction and veiling is smaller than the typical uncertainty of these parameters assessed later. 

We minimise the value of a reduced $\chi^2$-like metric, defined as 
\begin{equation}
\chi^2_{\rm{red}} = \frac{1}{N}\sum_i \frac{(O-T)_i^2}{err_i^2}
\end{equation}
to find the best combination of spectral type, \Av, and \veil. The $O$ is the observed spectrum, $T$ is the fitted template, $err$ defines the extracted uncertainty of the observed spectrum per spectral bin $i$, and $N$ is the number of degrees of freedom (number of all spectral bins subtracted by three free parameters). Figure \ref{fig:spec-Mstar} shows an example of the result from the fitting procedure, whereas in the Appendix~\ref{app:spt} we show the corresponding $\chi^2_{\rm{red}}$ maps. We notice that the worst fits usually have marginal values of \Av\ and \veil. In total, we classify 269 
M-type stars.

To asses the uncertainty of the estimated parameters, we run the fitting again keeping each time one of the parameters fixed at the best value. We draw the 1--$\sigma$ curves on the $\chi^2_{\rm{red}}$ maps between each two parameters. Examples are shown in Fig. \ref{fig:chi2-Mstar}. The maximum and minimum values within 1$\sigma$ from the best fit of two parameters are indicated by the extreme points of the 1--$\sigma$ curve. With this procedure we get two pairs of uncertainties for each parameter. We combine them taking the minimum. 
The lower and upper uncertainties are reported together with the best-fit values in the Table \ref{tab:cat}. The uncertainties of the spectral types obtained in this way are on average 2-3 sub-classes, also confirmed by the visual goodness of the fit to the spectra. We note that due to the uneven sampling of spectral types, the $\chi^2_{\rm{red}}$ maps do not represent the true $\chi^2_{\rm{red}}$. 
We also emphasise that our method of error assessment introduces a bias towards the values close to the borders of the adopted ranges and causes underestimation of the uncertainty on that side. Therefore, uncertainties for parameter values close to the border should be treated with caution.

\begin{figure}
\includegraphics[width=\columnwidth]{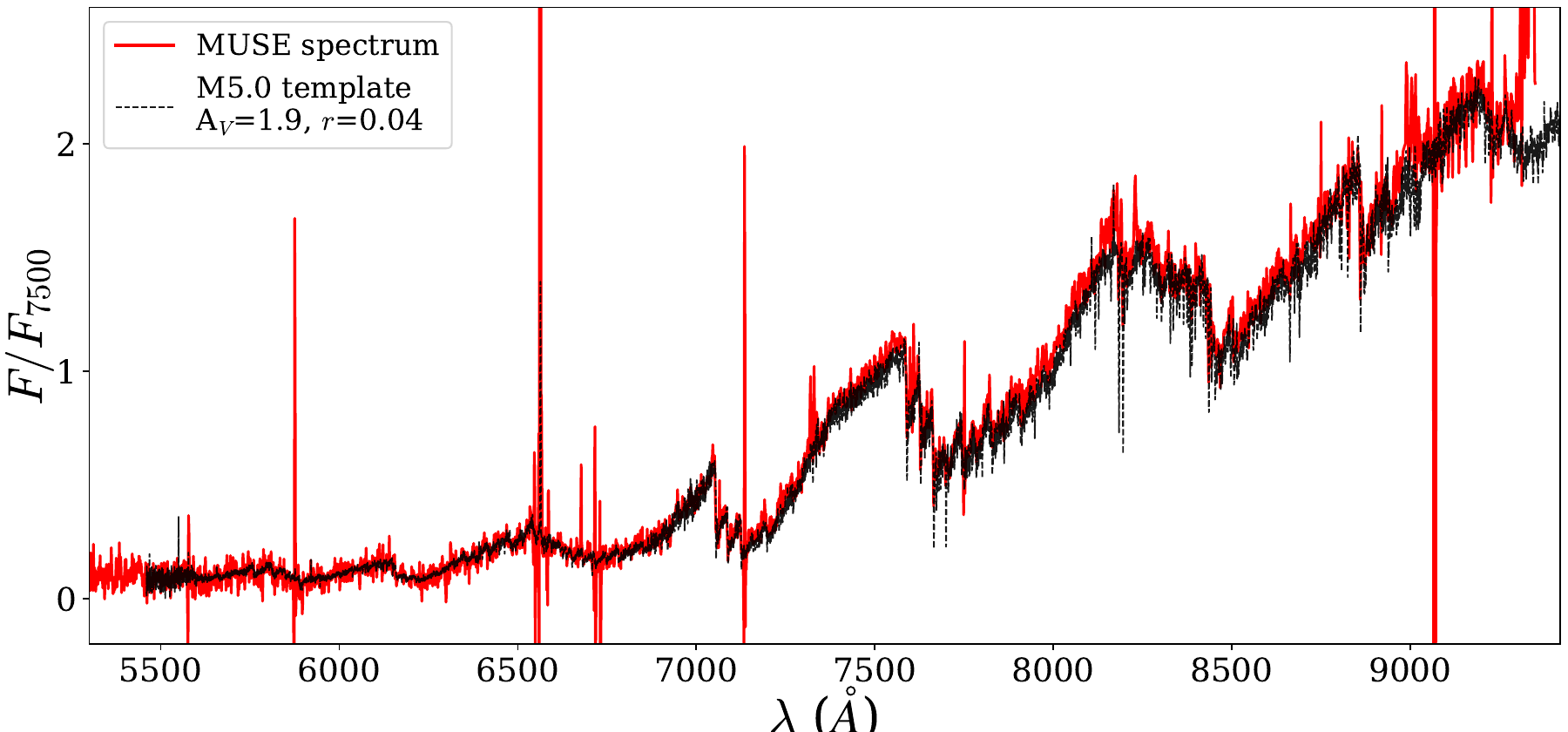}
\caption{Example of a MUSE spectrum (red solid line) of an M-type star and the matching spectral template (black dashed line). Both spectra are normalised at 7500~\AA.}
\label{fig:spec-Mstar}
\end{figure}

%%%%%%%%
\subsubsection{K and late G-type stars}
\label{subsubsec:Kstars}
\indent

The prominent TiO and VO bands in the spectra of M-type stars fade away in mid-K-type stars, while the overall shape of the spectrum flattens. We identify hotter stars in our sample based on the equivalent widths (EWs) of selected absorption lines; we list them in Table \ref{tab:ind}. We first calibrate the change of the EWs as a function of spectral type using the Class III templates (Sec. \ref{subsec:spt}) assuming a linear correlation. For each line we adopt a single value of the uncertainty of our calibration based on the fit's uncertainty. Additionally, for late K-type stars we use the spectral index TiO~(7140~\AA) identified by \cite{oliveira2003} and \cite{jeffries2007} and add it to a pool of previous estimates. 
The final spectral type is assigned as a weighted mean of types from single EWs and indices. Similarly, the uncertainty of the spectral type is a weighted mean of the uncertainties assigned to all of the indices. A single index error is the root of the sum of the squared uncertainties on individual EW measurements and EW calibrations. The resulting values are listed in Table \ref{tab:cat}. 

\begin{figure}
\includegraphics[width=\columnwidth, trim={0 0 0cm 0cm}, clip]{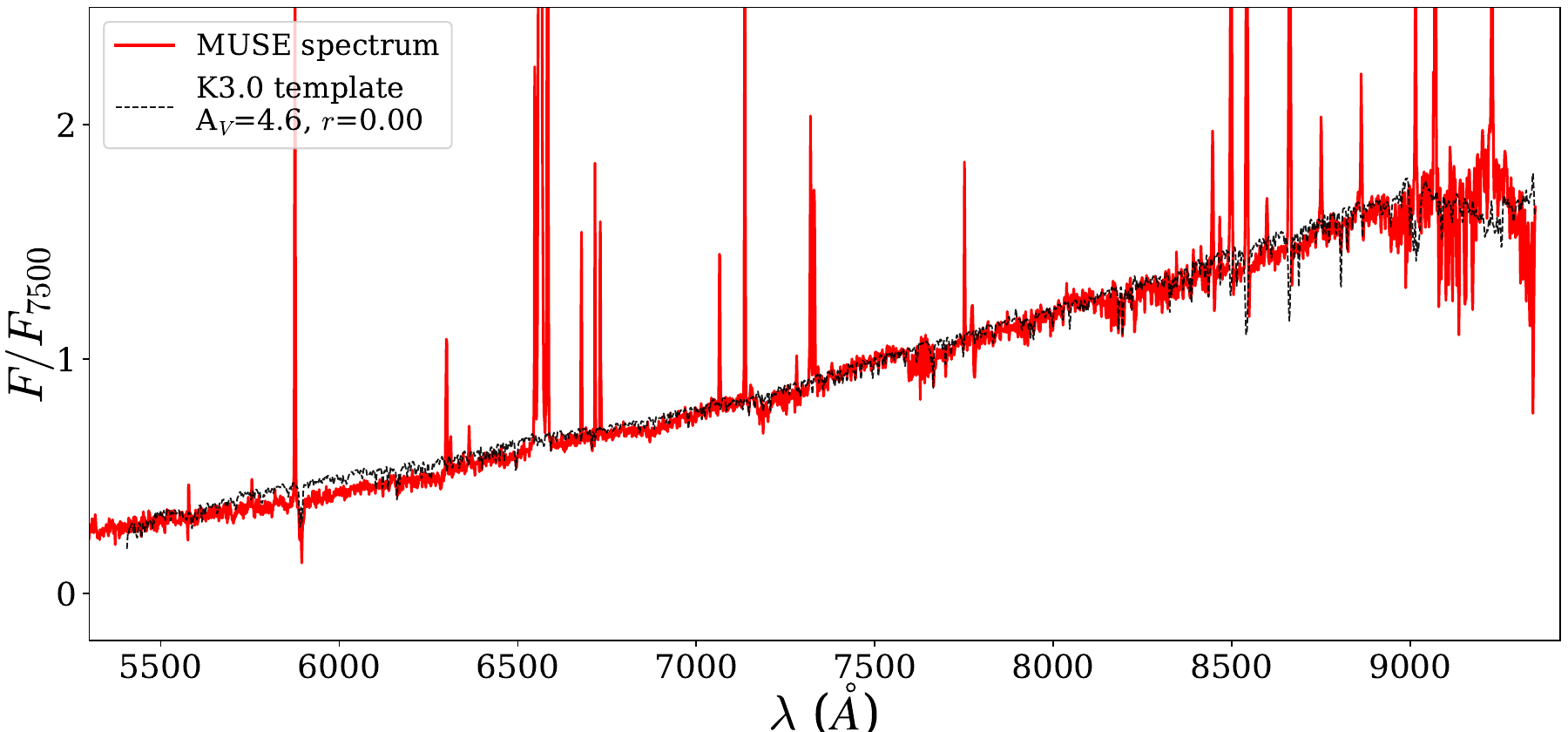}
\caption{Example of the MUSE spectrum (red solid line) of the K-type star and matching spectral template (black dashed line). Both spectra are normalised at 7500~\AA.}
\label{fig:spec-Kstar}
\end{figure}

Once the spectral type is assigned, we perform an estimation of the extinction and veiling following the same approach used to classify the M-type stars (see Sec. \ref{subsubsec:Mstars}). We fit to observed spectra the templates closest to the estimated spectral types varying \Av\ and \veil. The best values are those, for which the value of pseudo-\Chi, $\chi^2_{\rm{red,ps}} =$ \Chi/min$(\chi^2_{\rm{red}})$, is minimum. The uncertainties of \Av\ and \veil\ are estimated based on $\chi^2_{\rm{red,ps}}$ maps, similar to the procedure described in Sec. \ref{subsubsec:Mstars}. Example of a MUSE spectrum of a K-type star with the matched template is shown in Fig.~\ref{fig:spec-Kstar}. We find 14 early M-, 339 K-, and 95 late G-type stars.

\medskip 

We note that the non-homogeneous sampling of the templates can cause an error on the estimates of stellar parameters that is hard to estimate properly (see the examples of $\chi^2_{\rm{red}}$ maps in Appendix \ref{app:spt}). That applies not only to the spectral types, but also for extinction and veiling. However, for the lowest mass stars in our sample, the low snr dominates over any other source of uncertainty. For this reason, we do not interpolate between spectral types of templates to create a homogeneous grid. K- and G-type stars have on average smaller uncertainties of the spectral classification than M-type stars, since for these stellar types the estimate is based on absorption lines and is independent of the density of the grid sampling. A possible source of large error is the assumption of a linear correlation between the spectral types and EWs. Those relations are usually quadratic \citep[e.g., Ca{\sc i}][]{herczeg2014} or higher-order polynomial \citep[e.g.,][]{oliveira2003,riddick2007}. 
Overall, our estimations of spectral type are accurate within 2-3 sub-classes.

%%%%%%%
\subsection{Extinction corrected colour-magnitude diagrams}
\label{subsec:Av}
\indent

As described in Sec. \ref{subsec:MUSEphot}, the MUSE data presented here allow us to sample the stellar population in \tr\ down to very low-mass stars. The observed CMDs shown in Fig.~\ref{fig:CMD} indicate the presence of two populations. Here, after the accurate determination of the stellar parameters using the MUSE spectra (see previous Section) we reevaluate this using the extinction values derived for each individual star.

\begin{figure}
\includegraphics[width=\columnwidth]{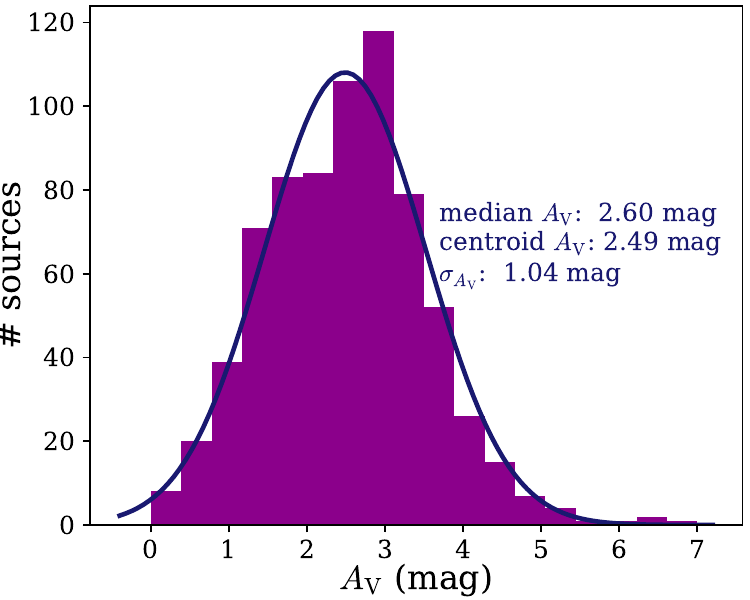}
\caption{Distribution of \Av\ estimated for \tr\ stars. Additionally, we fitted the Gaussian function to estimate the centroid of the distribution. Centroid, width of the distribution and median value in our sample is indicated in the upper right part of the figure. 
}
\label{fig:Av}
\end{figure}

Based on measurements toward individual stars, we estimate the visual extinction towards the \tr. The medium value of \Av\ is 2.60 
~mag. In Fig. \ref{fig:Av} we show the distribution of visual extinctions estimated in the previous sections. The distribution has a Gaussian-like shape, the fitted profile raises a centroid of 2.49~mag, 
consistent with the findings of \cite{ascenso2007} and slightly lower with respect to the value reported by \cite{beccari2015}. Our distribution of \Av\ is quite broad with Gaussian width of 1.04~mag. On average, the uncertainties of individual \Av\ estimates are $\sim0.5$~mag. We conclude that our measurements are in line with the literature values within uncertainties. 

\begin{figure*}%[h!]
\includegraphics[width=.65\columnwidth, trim={0 0 0.1cm .1cm}, clip]{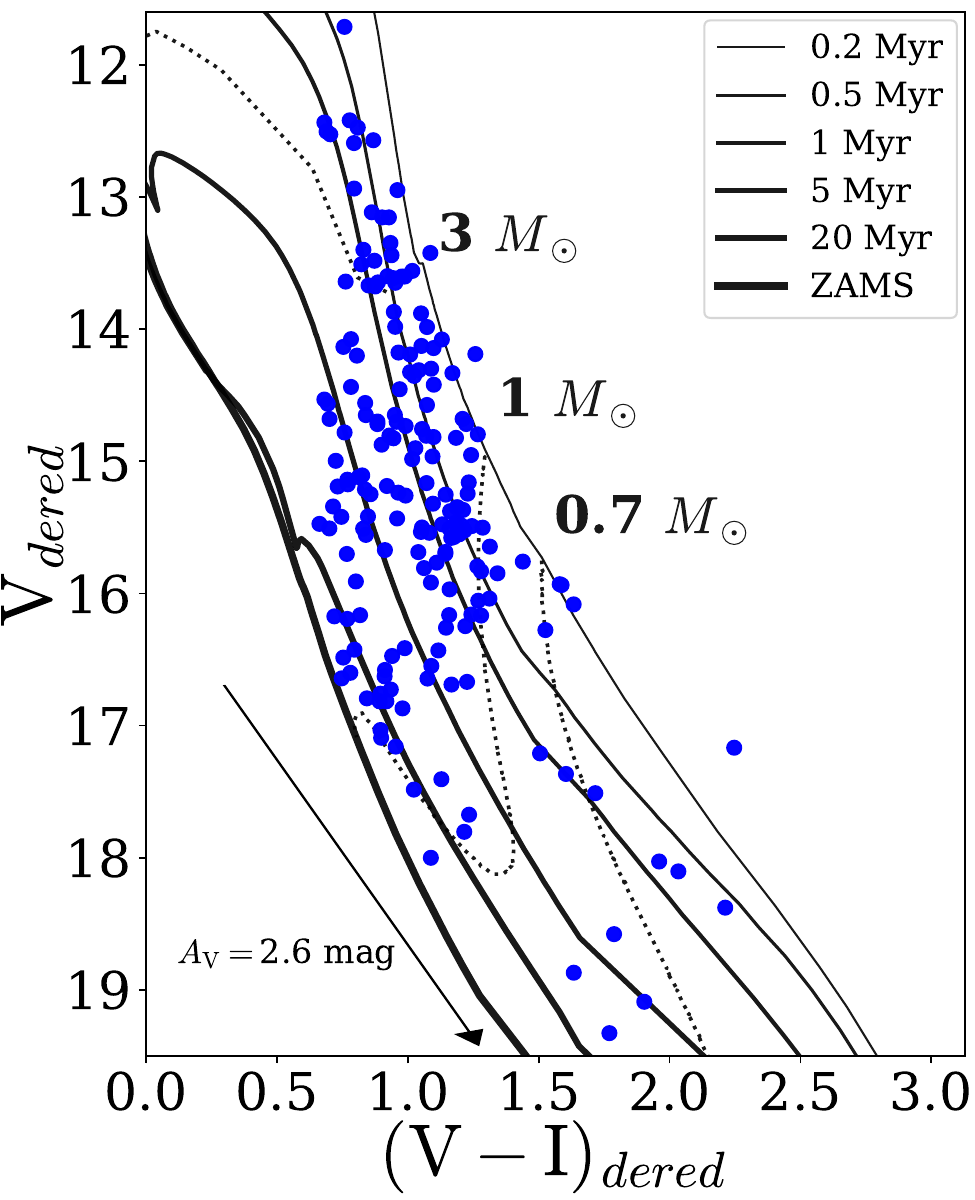} 
\includegraphics[width=.65\columnwidth, trim={0 0 .1cm .1cm}, clip ]{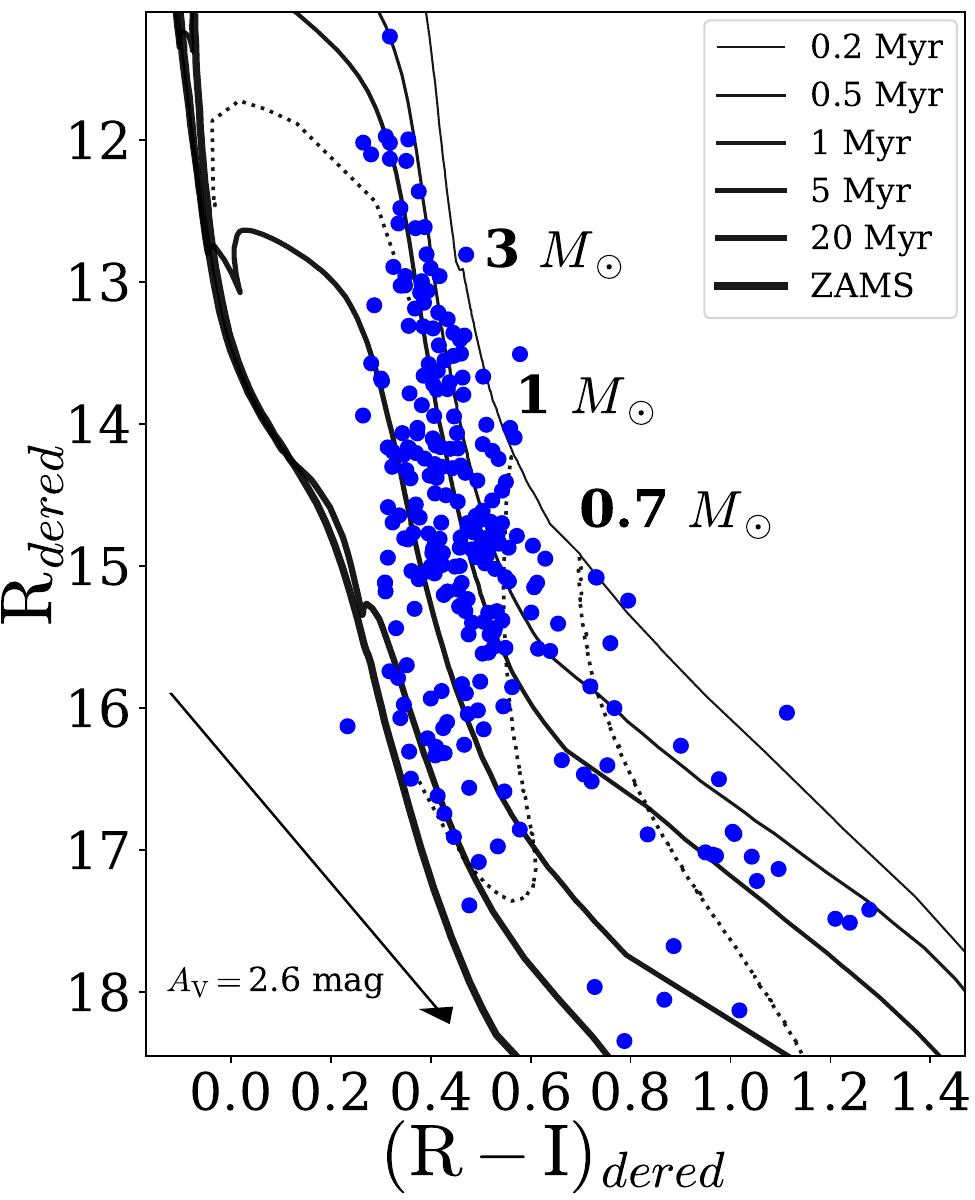}
\includegraphics[width=.65\columnwidth, trim={0 0 .1cm .1cm}, clip ]{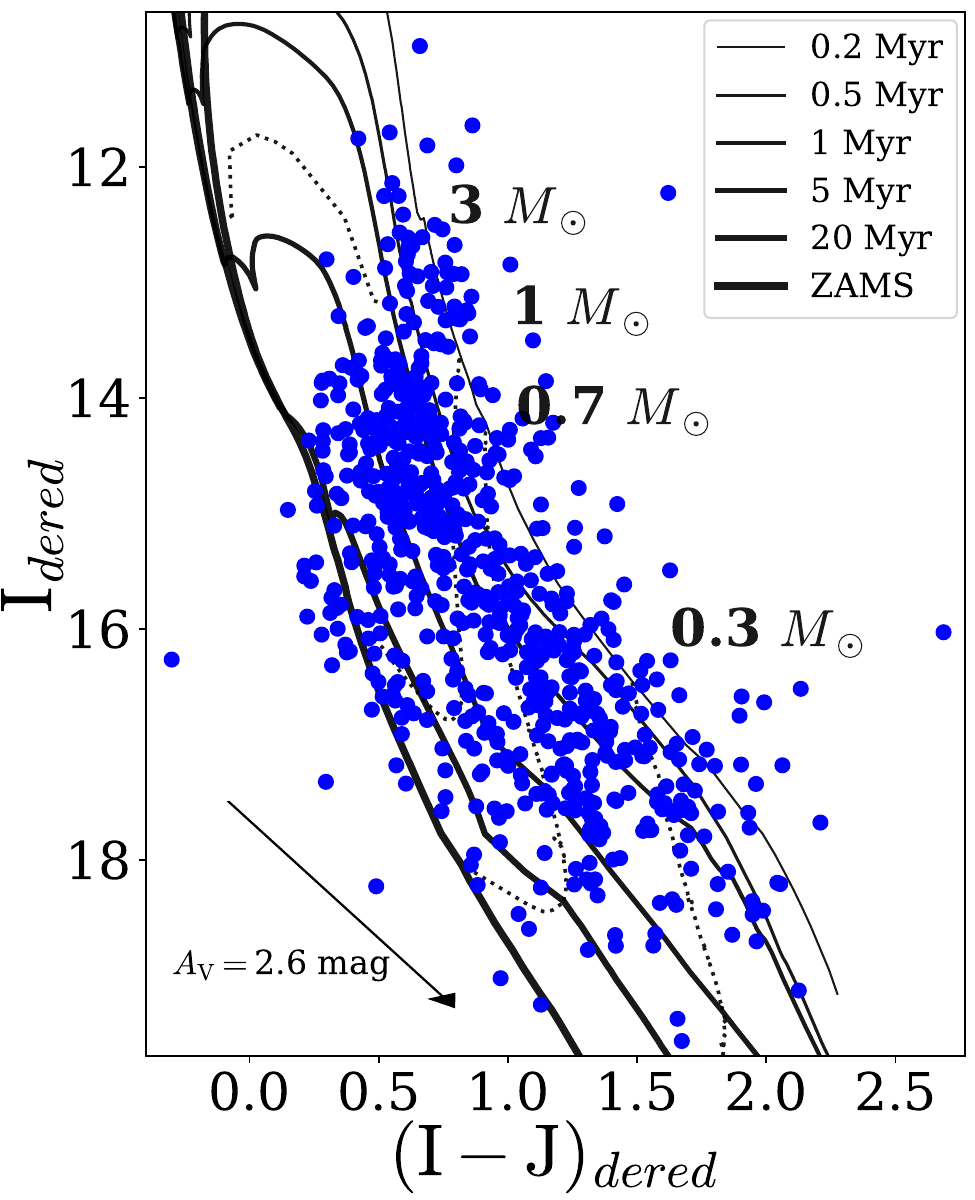}
\caption{Colour--magnitude diagrams from MUSE broad-band filters images in $V$ and $I$ (left panel), $R$ and $I$ magnitudes (middle panel), and $I$ and $J$ magnitudes (right panel) corrected for individual extinction. $J$-band magnitudes are from the VISTA and HAWK-I observations. Reddening vectors in the lower left corners show reddening by a median value of \Av\ estimated for Tr~14. {Solid lines show isochrones from 0.2 to 20~Myr (and ZAMS), dotted lines show isomasses of 0.3, 0.7, 1, and 3~$M_\odot$, as labeled \citep{bressan2012}.}
}
\label{fig:CMDdered}
\end{figure*}

We use the estimated \Av\ to correct the observed magnitudes. We show de-reddened colour-magnitude diagrams in Fig. \ref{fig:CMDdered} together with the isochrones {from the PARSEC v1.2S models \citep{bressan2012}}. 
The previously seen two populations {are no longer apparent} when the new extinction correction is applied, as expected for differently obscured populations \citep{ascenso2007}. This reassures us about the correctness of our procedure. The large scatter of points remains, we will discuss possible reasons in the following sections.

%%%%%%%%%%%%%%%%%%%%%%%%%%%
%%%%%%% discussion %%%%%%%%
\section{Physical parameters of the stars}

%%%%%%%
\subsection{Effective temperature and stellar luminosity}
\label{subsec:Lbol}
\indent

We derive the effective temperatures (\teff) of our stars based on their spectral types. For M-type stars we use the SpT -- \teff\ scale from \cite{luhman2003b} and for earlier types, we apply the scaling from \cite{kenyon1995} and interpolate linearly between the sub-classes. Newer scales, like e.g., from \cite{herczeg2014}, agree well for low-temperature stars (types later K5). The scale adopted here deviates for the hotter stars up to 380~K in case of K0 stars in comparison to \cite{herczeg2014}. 

It has been shown in the literature that the $J$-band photometry is most suitable in deriving bolometric correction for young stars. The spectral energy distribution in these objects can be strongly affected by the presence of NIR excess due to the ongoing mass accretion from a protoplanetary disk or intrinsic differential extinction. Such effects can not be fully avoided but are minimised using the $J$-band filter \citep[e.g.,][]{kenyon1995,luhman1999}.
Our bolometric luminosities are hence calculated using the $J$-band photometry from VISTA \citep{preibisch2014} and HAWK-I \citep{preibisch2011a,preibisch2011b}. Whenever magnitudes from both catalogues are available for a given star, we choose the one with smaller uncertainty. We first calculate the bolometric magnitude ($M_{\mathrm{bol}}$) dereddening the observed magnitudes by individual visual extinction determined from our spectral classification (Sec. \ref{subsec:spt}) and estimating its value in $J$-band using extinction law from \cite{cardelli1989}, subtracting distance modulus and adding bolometric correction with colours, as indicated by the equation: 
\begin{equation}
M_{\mathrm{bol}} = J - A_{\mathrm{J}} - {\rm DM} + ({\rm BC}_V + (V-K) - (H-K) - (J-H) )
\end{equation}
Values of the corrections and colours were taken from \cite{kenyon1995}. Finally, to obtain the bolometric luminosity in $L_\odot$, we subtract from the previously estimated $M_{\mathrm{bol}}$ the solar bolometric magnitude $M_{\mathrm{bol},\odot}=4.74$ \citep{cox2000}:
\begin{equation}
\log{(L_{\rm bol}/L_\odot)} = -0.4 \cdot (M_{\mathrm{bol}} - M_{\mathrm{bol},\odot})
\end{equation}
The $L_{\rm bol}$ values thus calculated are listed in Table \ref{tab:cat}. There is only $\sim$1\% of spectroscopically classified stars in our catalogue which were not matched with any source from the NIR catalogues and therefore do not have estimated stellar parameters. This might be due to the fact that NIR catalogues are not 100\% complete. 

The uncertainty of the stellar luminosity in our estimations is mostly driven by two factors: uncertainty of $J$-band photometry adopted from the VISTA and HAWK-I catalogs, and uncertainty of the extinction measured by us while performing the spectral classification of the each star (Sec. \ref{subsec:spt}). The latter has significantly greater impact: typical uncertainty of the $J$-band magnitudes used in this work is $\sim$0.03-0.05~mag, while the average \Av\ error is $\sim0.5$~mag, corresponding to $\Delta A_{\rm J}\sim0.16$~mag. 

\subsection{HR diagram and stellar parameters}
\label{subsec:HR}
\indent

\begin{figure}
\includegraphics[width=\columnwidth]{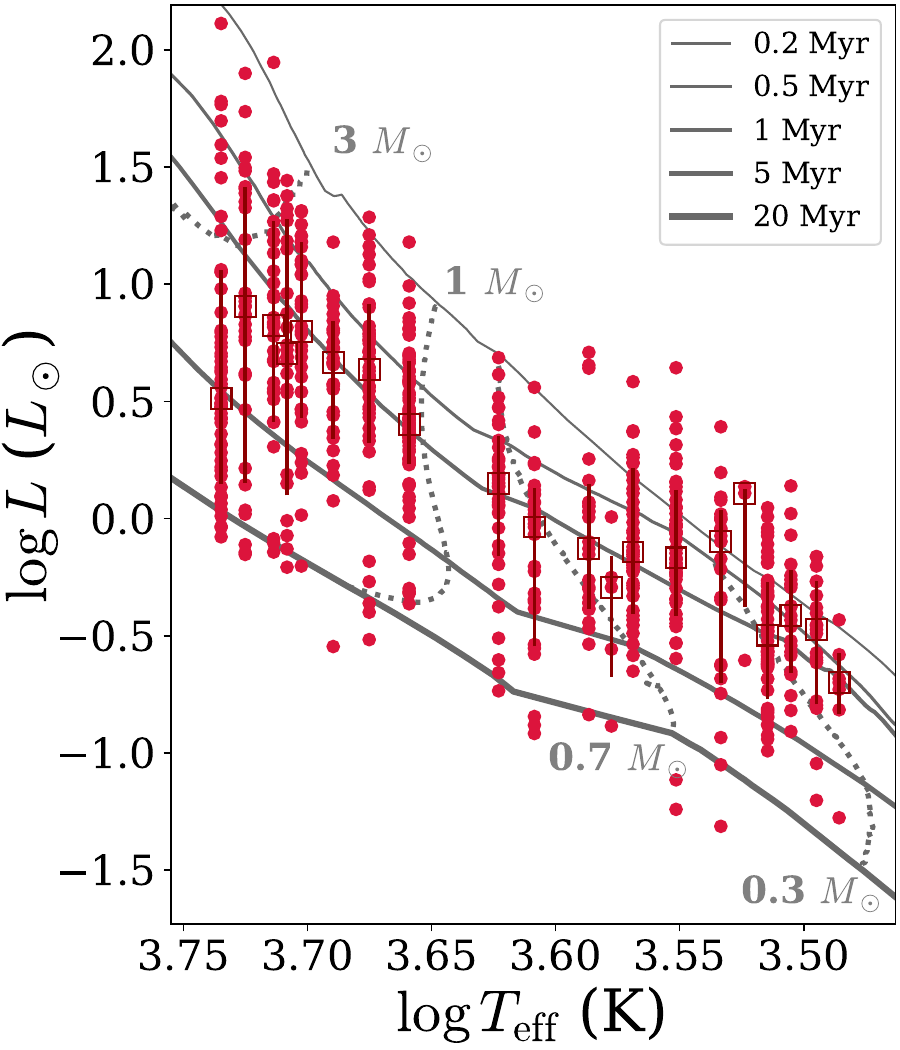}
\caption{HR diagram for low-mass stars of \tr. {Filled} circles show data points, {empty} squares are median values of the bolometric luminosity for each spectral subclass with errorbars indicating 1-$\sigma$ percentiles. Theoretical tracks from {\cite{bressan2012} are shown as solid black lines. Grey dotted lines show tracks for 0.3, 0.7, 1, and 3~$M_\odot$ stars.}
}
\label{fig:HR}
\end{figure}

In Figure \ref{fig:HR} we show the bolometric luminosity as function of effective temperature. The stars detected with MUSE are shown with {filled} 
circles. The {open} 
squares represent the median luminosities for each spectral type. We show on the HR diagram the {PARSEC v1.2S} theoretical isochrones {\citep{bressan2012}}. 
We assign the stellar masses and ages by performing linear interpolation between the tracks and isochrones. The resulting values are listed in Tab. \ref{tab:cat}. 
{For stars more luminous than predicted by the lowest age isochrone we assign the  boundary value of 0.1~Myr as a stellar age.}

The HR diagram (Fig. \ref{fig:HR}) shows the presence of a large spread of luminosities for sources with the same spectral type. Depending on the spectral type, the spread ranges from 0.5 to 2.0~dex. {Possible explanations of this behavior are two-fold: observational and physical. Observational reasons for the spread cover} 
uncertainties in estimations of stellar luminosity and / or effective temperature, {as well as} contamination from foreground sources. As we discussed in Sec. \ref{subsec:Lbol}, the main source of luminosity uncertainty is the extinction, closely linked to the uncertainty of the spectral type (and thus \teff) and veiling. On average, the luminosity values are uncertain by {$\sim$0.3~dex}, 
temperatures by {300~K}, and veiling by 0.1-0.3. Given the large distance to \tr\ \citep[2.35$\pm$0.05~kpc,][]{goeppl2022}, we expect a significant  contamination by foreground stars. 
In Sec. \ref{subsec:frg} we therefore used the {\it Gaia} DR3 catalogue to minimise this effect and remove objects in the foreground of Carina Nebula. However, the limited number of good astrometric measurements does not allow to identify all non-cluster members. It is then not trivial to estimate the contribution of foreground contamination to our results. This effect, combined with uncertainties in our measurements, can explain {a large part} 
of the observed luminosity spread in our HR diagram.

{The physical} sources of the luminosity spread include intrinsic age spread, 
variability, binarity, dispersion in distance, and accretion history. 
Episodic but vigorous accretion of low-mass objects at the very early stages of their formation (Class 0 -- Class I) can leave its imprint on their evolution for the next few Myr \citep{baraffe2009}. If most of the accreting kinetic energy is radiated away, the structure of the stars will be more compact (i.e., stellar radius will be smaller) than of the non-accreting star of the same age and mass. Short, intense and numerous accretion episodes do not leave enough time for the object to relax to a larger radius for the newly acquired mass. As a result, object has lower luminosity and seems to be older than non-accreting of the same \teff. \cite{baraffe2009} found that episodic accretion at early stages of stellar evolution can well reproduce luminosity spread equivalent to an age spread of $\sim$10~Myr observed in Orion Molecular Cloud \citep{peterson2008}. 
Moreover, the intrinsic spread of accretion rates in the cluster might add to the luminosity spread. In their estimates, \cite{hartmann2001} adopted arbitrarily an error of 0.1 in $\log{L}$ due to accretion (ignoring the effect of disk inclination). 
We use the $J$-band photometry to minimise the excess luminosity caused by the accretion \citep[following][]{kenyon1995}. We also included veiling in our spectral classification. However, we made a very simplistic approach, where the veiling is independent of the wavelength. 

Another physical process, that has a great impact on the luminosity of young stars is the photometric variability and to lesser extend, accretion variability. Usually, photometric variability is relatively small \citep[e.g., $\sim$0.2~mag in J, H, K$_{\rm s}$ bands, see][]{carpenter2001} and has a timescale of less than a few days. It is very often assigned to the rotational modulation of cool or hot spots. 
{Variability related to accretion can span wide range of amplitudes and timescales. Typically, changes in brightness are lower than 1-2~mag and last few days \citep{fischer2023}.} However, some extreme cases were also spotted. 
{For example,} \cite{claes2022} recently reported change by $\gtrsim$1.4~dex on accretion rate of XX~Cha measured on UV excess and by $\sim$0.5~dex measured on lines (including Pa$\beta$ in $J$-band) over a period of 11 years. 
Previous studies of accretion variability from photometry \citep[e.g.,][]{venuti2014} or spectroscopy \citep{costigan2012, costigan2014} recorded variability $<$0.5~dex at different-time scales (years, days, and minutes). %\lesssim
If behaviour of XX~Cha is more common for young stars than thought so far, 
it could explain a significant fraction of observed luminosity spread. The authors note also high photometric variability of the star in optical bands ($>2$~mag in $B$-band to $\sim0.5$~mag in $I$-band). \cite{hartmann2001} in their Taurus study adopted a variability of 0.1~mag to explain an observed luminosity scatter. While not negligible, this value alone cannot explain our observations. 

A different source of uncertainty, which impact is difficult to predict, is the inclination of the accretion disk: stars with edge-on disks will appear significantly redder than face-on ones. 
For instance, \cite{alcala2014} suggested highly inclined disks as an explanation of the sub-luminosity of four young stars in Lupus. After correction for disk obscuration by a factor of 4--25 (corresponding to 0.4--1.4~dex) their accretion properties were well in line with those from other sources in the region. 

Unresolved multiplicity has potentially a large impact on luminosity distribution, especially for young clusters, where the multiplicity fraction is observed to be higher than among the more evolved field stars \citep{duchene2013,zurlo2023}. Multiplicity also scales with stellar mass, from $\sim$25\% for M-type stars to almost 100\% for OB stars \citep{duchene2013,zurlo2023}. \cite{zagaria2022} noted, that {at lest} 20\% of all stellar systems in Lupus, Chameleon I and Upper Scorpius with measured disk masses and accretion rates are multiples. They all also have higher observed accretion rates than isolated stars. Similarly, \cite{zurlo2020} found the fraction of binaries in Ophiuchus with separations from 9 to 1200~au to be 18\%, whereas in Coronae Australis that number was estimated to be 36.2$\pm$8.8\% \citep[separations between 17 and 780~au,][]{koehler2008} and in Taurus is 37.4$\pm$4.6\% \citep[the same separation range,][]{leinert1993}. 
Unresolved multiples appear brighter with respect to the single stars in the HR diagram mimicking younger age.  \cite{hartmann2001} estimated this potential shift of luminosities to be $\sim0.2~\log{(L_\odot)}$. 

{The individual distances to the cluster members might also add to the observed spread. }
Here, we use the distance estimate of $2.35\pm0.05$~kpc based on {\it Gaia} EDR3 catalogue \citep{goeppl2022} for all sources in the field. We expect to include in that way both members of the \tr\ and young stars from the dispersed population of CNC. \tr\ has a compact core of radius of $\sim$0.6--0.7~pc with extended halo up to $\sim$3.4--5.3~pc \citep{ascenso2007,kharchenko2013}, much less than the distance uncertainty of 50~pc (note, that \cite{ascenso2007} assumed distance of 2.8~kpc to \tr, here we re-scaled their results to 2.35~kpc). This error corresponds to an uncertainty of 0.05~dex in luminosity and cannot explain the scatter of estimated values. Similarly, the dispersion in distances to the different clusters in Carina of 2\% \citep{goeppl2022} is too small to explain observed scatter. Therefore, we neglect any impact from the distance spread 
on the luminosity dispersion.

Summarising, we conclude 
that the luminosity spread is mostly caused by large uncertainties of photospheric parameters, contamination of non-cluster members, accretion and photospheric variability, and unresolved multiplicity. Other parameters, like internal spread of stellar ages, accretion properties, individual distances might play a role, but their impact is smaller.

\subsection{Age of Trumpler 14} 
\label{subsec:age}
\indent

\begin{figure}
\includegraphics[width=\columnwidth, trim={.2cm 0.1cm .cm .cm}, clip]{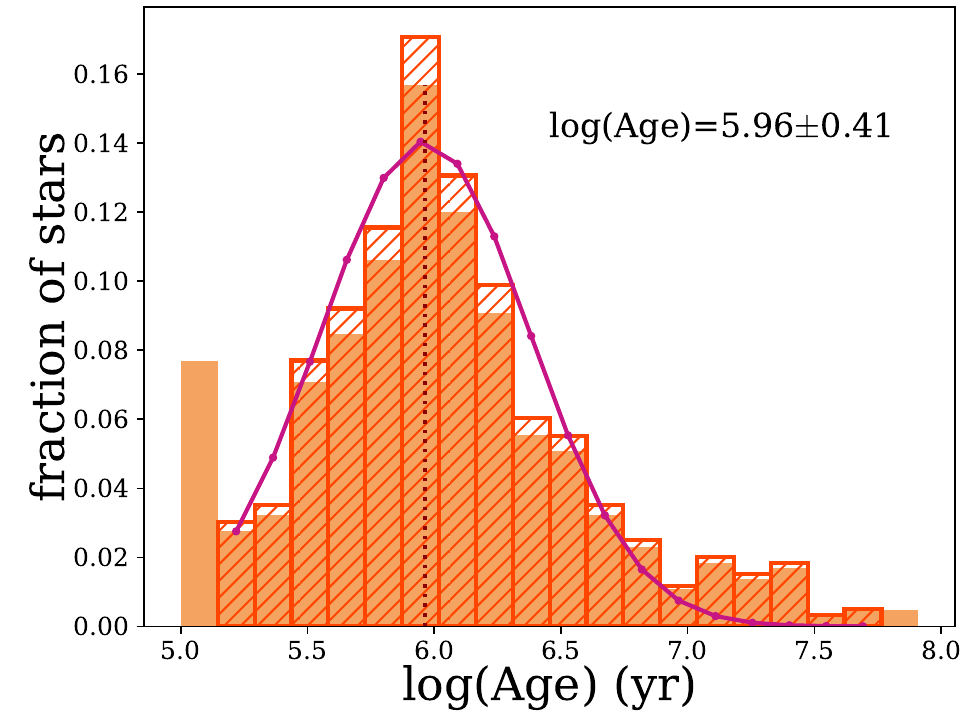}
\caption{Fraction of stellar ages derived from HR diagram {for stars with $\log{(T_{\rm eff})}<3.73$}. Filled orange histogram shows distribution for the whole sample, while hatched red histogram represents the fraction distribution after removing the extreme bars with respect to the total number of stars within the new age range. The normal fit to the probability density distribution converted into the fraction distribution for the visual purposes is shown as a dark violet curve with {mean value of $\log{\rm{(age)}}=5.96\pm0.41$} 
corresponding to the {0.9$^{+1.4}_{-0.6}$~Myr}. 
} 
\label{fig:age}
\end{figure}

The YSOs plotted in Fig. \ref{fig:HR} concentrate near the 1~Myr isochrone strongly suggesting a young age of the cluster. Here, we look more closely into the distribution of stellar ages in \tr.

Figure \ref{fig:age} presents the 
distribution of fraction of the stars within each age bin 
in logarithmic scale. Only measurements for stars with {$\log{(T_{\rm eff})}<$3.73} are included in the distribution. Ages were estimated based on {PARSEC evolutionary tracks \citep[see Sec. \ref{subsec:HR}]{bressan2012}}. 
{The lowest stellar ages provided by the models are 0.1~Myr. Some of our sources lay above this isochrone on HR diagram. Since we do not extrapolate stellar parameters beyond theoretical models, those sources have a fixed age of 0.1~Myr causing an artificial overdensity in the first bin of the age distribution in Fig. \ref{fig:age}. Therefore, }
to estimate the cluster age, we excluded {from this analysis} the boundary bars. 
We fit the lognormal profile to the remaining distribution, as shown in Fig. \ref{fig:age}. The fit peaks at the logarithm of {5.96$\pm$0.03}, 
which we interpret as a cluster age, with the width of {0.41$\pm$0.02}, 
which we adopt as an uncertainty. In linear scale that corresponds to the age of \tr\ of {0.9$^{+1.4}_{-0.6}$~Myr}. 

{We check how conclusion on the cluster age is impacted when using different set of models. Therefore, we employ tracks from \cite{baraffe2015} for low-mass stars (spectral types later than K5), and since they are limited to the solar-mass stars, for hotter stars we employ tracks from \cite{siess2000}. We present HR diagram, stellar properties, and comparison between those two set of tracks in Appendix \ref{app:tracks}. Those stellar parameters ($M_*$ and age) are also listed in Table \ref{tab:cat} and \ref{tab:highbkgcat}. Use of different models changes values of parameters for individual stars, but do not affect our conclusion on cluster age. We perform the same exercise as described above for a distribution of stellar ages based on \cite{baraffe2015} / \cite{siess2000} evolutionary tracks. The normal fit indicates cluster age of $\log{\rm{(age)}}=6.16\pm0.31$, corresponding to 1.4$^{+1.5}_{-0.7}$~Myr, consistent within errors to the previous estimate. Stellar, and hence cluster, ages around 1~Myr are difficult to infer precisely. We {adopt} 
that the age of \tr\ is $\sim$1~Myr. This is a robust result (given uncertainties related to differences in models and internal uncertainties of observations) since the estimate is not affected by the choice of evolutionary tracks.}

The large 
spread {in the HR diagram seen} 
in \tr\ 
lead {in the past} to conclusions of long, continuous star formation over last 10~Myr \citep{degioia2001,povich2019}, 1--6~Myr \citep{tapia2003}, or 5~Myr \citep{ascenso2007}. While comparing different clusters in Carina, \cite{damiani2017} stated that \tr\ is younger than Trumpler~16, similarly to \cite{smith2008} who found the age difference of 1-2~Myr between these two clusters. More precise estimates indicate a Trumpler~14's age of 2$\pm$1~Myr \citep{preibisch2011b}. \cite{rochau2011} found a recent (1.0$\pm$0.5~Myr) starburst-like event and hint of the presence of an older (3~Myr) population in \tr, which might be part of the dispersed population of CNC. Overall, our estimate is in line with general findings in literature. Our measurements also show a large spread of isochronal 
ages, which is a direct consequence of the luminosity spread. In the previous section (\ref{subsec:HR}) we listed several possible sources responsible for the spread in luminosity within the stellar population of \tr, with the uncertainty of the parameters estimated during spectral classification expected to have the strongest impact. It is important to note, that the aforementioned studies were mostly focused on massive and intermediate-mass stars ($\gtrsim1 M_\odot$), while here we do not analyse stars hotter than $\sim$5500~K. 

Additionally to the spread, Fig. \ref{fig:HR} shows also decrease in median luminosities toward hotter stars {and deviation from $\sim$1~Myr isochrone with the last temperature bin above $\log{(T_{\rm eff})}\sim$3.73 exhibiting significant drop in luminosity.} This behaviour can be caused by our selection bias {as we focus on low-mass objects and} do not identify in this work stars with spectral classes earlier than G8. This {might} result in the apparent \lq older\rq\ population of hotter stars. 
\cite{hartmann2003} noted a similar trend in Taurus star-forming region. Stars colder than 4350~K (corresponding to the masses below $\sim$1~$M_\odot$) had an age distribution strongly pointing to the values $<$2~Myr, while hotter stars exhibited flat distribution spanning up to 22~Myr \citep[][see their Fig. 1]{hartmann2003}. 
If we divide the age distributions into the same \teff\ ranges, {we do not see such a strong behavior, both sub-samples peak around 1~Myr, although we note that the youngest stars ($\leqslant$0.15~Myr) are among those with \teff$<$4350~K.} 
\cite{hartmann2003} argued, that {the flat distribution of hotter Taurus members} is due to the highly inaccurate positions of birth line for the more massive stars, as well as non-member contamination. 
{Similarly, \cite{fang2017} showed that cluster ages are higher when derived from luminosities and temperatures of hotter stars. The behaviour hold for different theoretical models and different young clusters.}
{This effect can also impact our results.} 

Our observations span prominent, dense and compact cluster core \citep[$r\sim$0.5\arcmin--0.9\arcmin,][]{ascenso2007,kharchenko2013} and extended area around. {The widely}
dispersed population of Carina Nebula Complex members \citep{feigelson2011,zeidler2016} is mixed in our observations with the \tr\ members causing the {apparent} age spread. {We investigate that possibility below.}

\begin{figure}
\includegraphics[width=\columnwidth, trim={.cm 0 0cm 0cm}, clip]{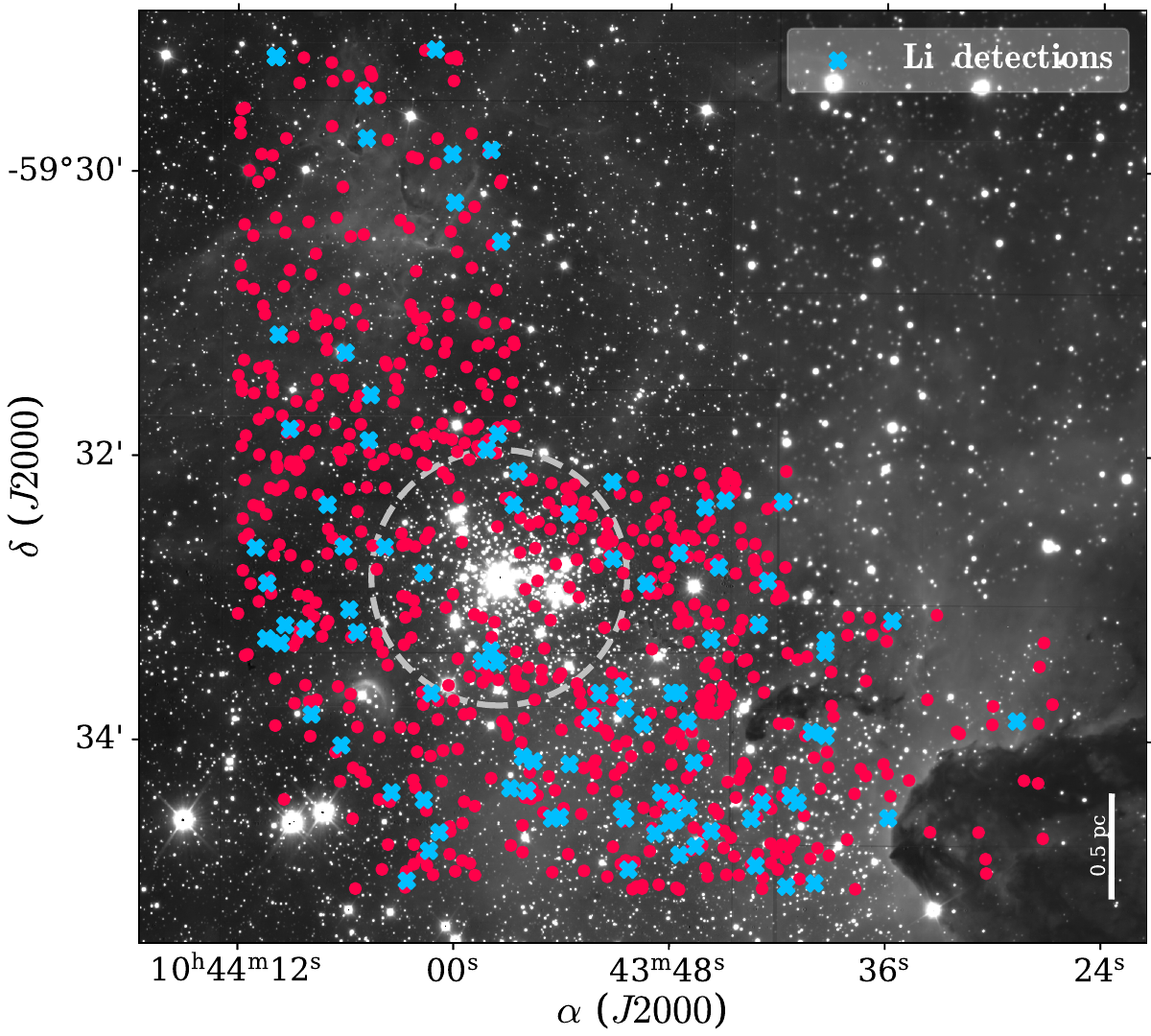}
\caption{Locations of the Li 6708\AA\ detections {in the MUSE field} 
blue crosses). All stars studied here are marked with red dots, as in Fig. \ref{fig:Tr14detect}. The dashed circle with radius of 0.9\arcmin\ shows the core of the \tr, as defined by \cite{kharchenko2013}. The background image in grey scale is the $H$-band image from HAWK-I \citep{preibisch2011a,preibisch2011b}.} 
\label{fig:sky-Li}
\end{figure}

\begin{figure}
\includegraphics[width=\columnwidth, trim={.5cm 0.5cm 0.4cm 0.3cm}, clip]{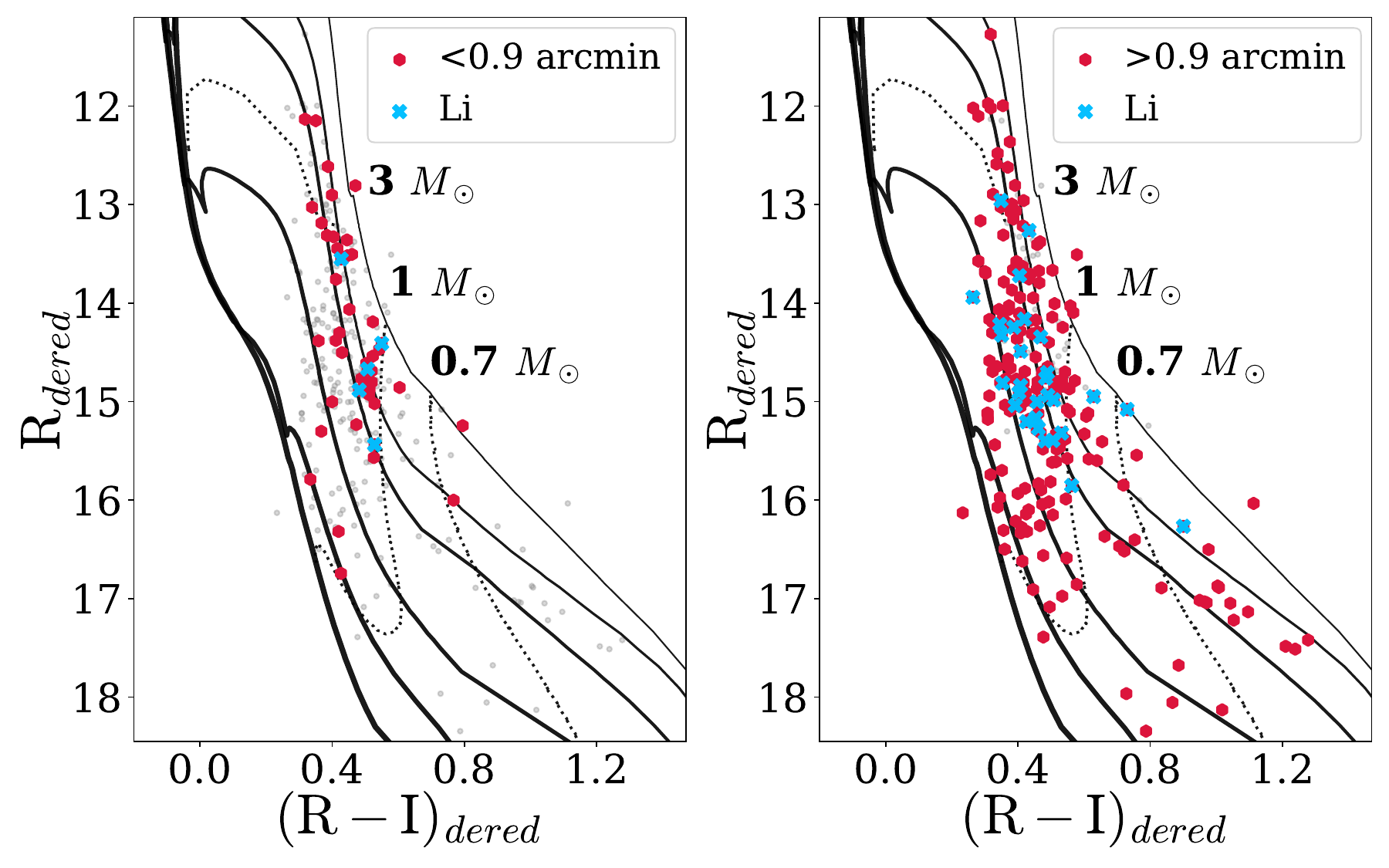}
\caption{Colour-magnitude diagrams for {de-reddened} $R$ and $I$-band magnitudes from MUSE. Red hexagons mark stars within the core of \tr\ ({left}, 0.9\arcmin, \cite{kharchenko2013}) or outside ({right}). The blue crosses indicate location of the Lithium-bearing stars on the CMDs within and outside the core radius, respectively. {Plotted are the same tracks as in Fig. \ref{fig:CMDdered}.}} 
\label{fig:CMD-pop-Li}
\end{figure}

In Fig. \ref{fig:sky-Li} we mark the \lq\lq core\rq\rq\ area with radius 0.9\arcmin\ and compare it to the location of our sources. We do not detect many sources in the most central area due to the spectral contamination. We investigate whether the stars inside the core have different properties than the population at larger radii from the cluster center. Figure \ref{fig:CMD-pop-Li} shows {de-reddened} CMDs where sources inside (left panel) and outside (right panel) the radius of 0.9\arcmin\ are marked with red hexagons. {The core population of \tr\ is mostly concentrated around 1~Myr isochrone. Although the extended, \lq\lq halo\rq\rq, population exhibits larger spread in colours and ages, most of the stars are also located around 1~Myr isochrone. There are more faint stars in the latter group which are affected by the higher observational uncertainties. 
It is likely that the \lq\lq halo\rq\rq\ population is a mixture of young \tr\ members and the older widely distributed population of the whole Carina Nebula Complex. Since the widely distributed population of young stars in the CNC exhibits a range of ages between $<$1~Myr and $\sim$8~Myr \citep{preibisch2011a}, it is not possible with the available data to distinguish between \tr\ members in the outer parts of the cluster and stars from the distributed population. %\lesssim
}

We additionally check the spatial distribution of young stars using the Li{\sc i} 6708~\AA\ absorption line. In the Figures \ref{fig:sky-Li} and \ref{fig:CMD-pop-Li} the stars where Lithium was detected are marked with blue crosses. We do not find any specific concentration in the cluster of those stars but they all follow the {$<$10}~Myr isochrones, as expected for Lithium-bearing stars. We note similar behaviour for NIR excess or X-ray emitting sources (see Appendix \ref{app:populations}). We conclude that in our dataset, where the most central core part is saturated and we can not characterise {most of the} stars located there, %\lesssim
the true, young members of \tr\ are distributed evenly across the cluster. However, our sample contains also number of stars from the widely distributed Carina population. We are not able to distinguish between the true and apparent \tr\ members nor to confirm the cluster membership of older stars.

\subsection{Mass distribution} 
\label{subsec:IMF}
\indent

\begin{figure}
\includegraphics[width=\columnwidth]{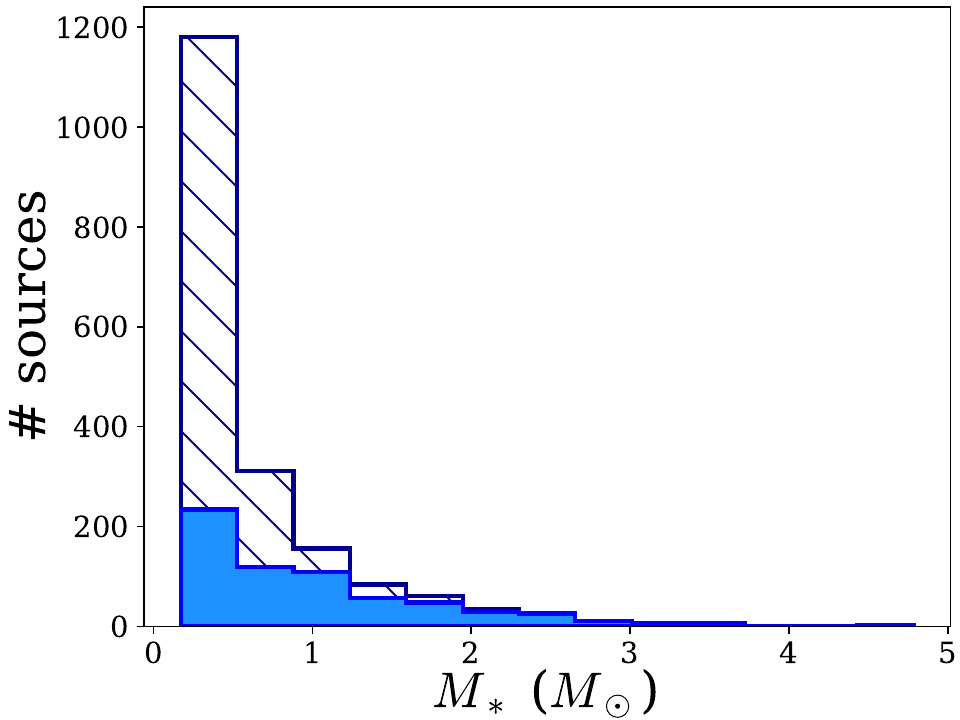}
\includegraphics[width=\columnwidth]{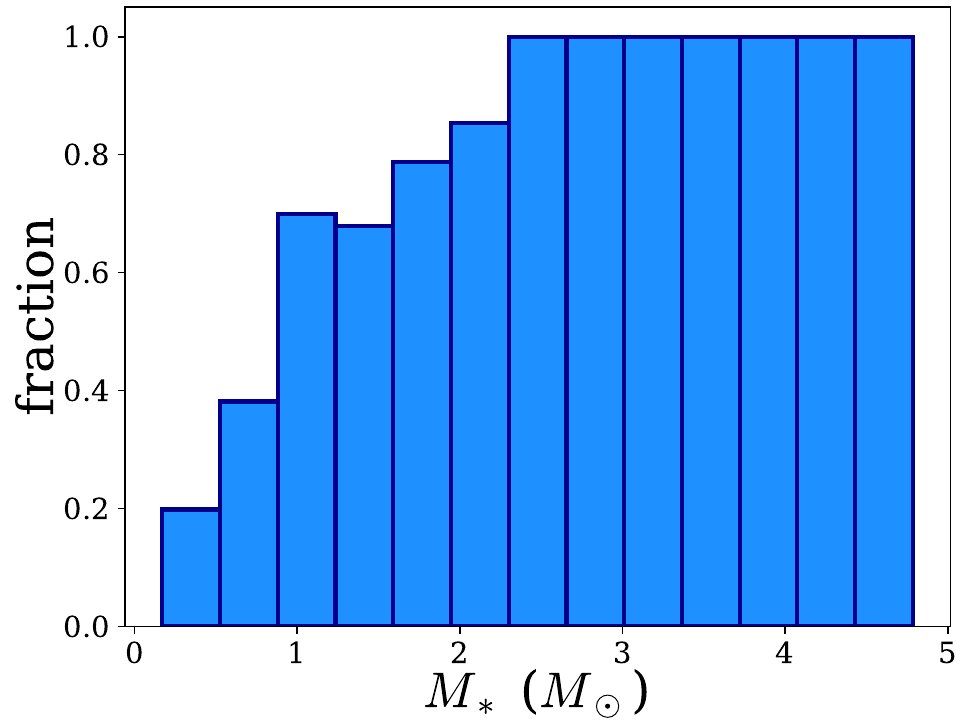}
\caption{Distribution of stellar masses in \tr\  {for stars with $\log{(T_{\rm eff})}<3.73$}. 
{Top:} filled histograms presents distribution of masses for stars analysed in this paper. On top of it (hatched histogram) we display the distribution of the probable members with uncertain photometry removed from the analysis due to the high variability of the background emission. As expected, most of the removed stars are faint, low-mass objects. {Bottom:} fraction of stars in the final spectroscopic catalogue within each mass bin relative to the combined catalogues of final sample and probable members with uncertain photometry {due to the variable background emission}. Bins are the same as in the upper panel.
}
\label{fig:IMF}
\end{figure}

%%%%%%%%%%%%%%%%%%%%%%%%%%%%%%%%%

Whether the environment can affect the initial mass function (IMF) of the stellar cluster was investigated in multiple studies. For example, \cite{damian2021} studied low-mass stars in eight young clusters ($\sim$2-3 Myr) observed in $J$ and $K$-band spanning wide range of FUV radiation levels, cluster densities, and galactocentric distances. Their log-normal IMFs \citep{chabrier2003} agreed well within each other peaking within the range 0.2–0.4~$M_\odot$ and not revealing any dependence on any of the three environmental properties. On the other hand, \cite{deMarchi2010} suggested that the present-day characteristic mass of the IMF is significantly correlated with the dynamical age of the cluster.

There is no study dedicated to investigate the impact of high FUV field on the IMF in the Carina Nebula. Only \cite{rochau2011} tried to look at the mass function in the closest vicinity of the massive stars in \tr, but they could not draw any binding conclusions on their impact onto neighbouring stars. Similarly, \cite{rainot2022} studied low-mass companions in the vicinity of seven O-type stars with VLT/SPHERE in $K$-band. Despite the found differences between their IMF and the one from \cite{rochau2011} or \cite{chabrier2003}, they could not robustly confirm if the presence of the massive stars impacts their neighbouring companions or if the noticed differences are due to the observational bias. Former IMF studies in \tr\ \citep[e.g.,][]{ascenso2007,hur2012} were also based mostly on NIR photometry, up to date there is no study employing spectroscopy in \tr\ to investigate the stellar mass distribution.

Our work focuses on low-mass stars with spectral type later than G8. Figure \ref{fig:IMF} presents the distribution of stellar masses estimated based on MUSE observations in \tr\ for stars with $\log{(T_{\rm eff})}$ below {3.73}.  
The shown masses range from {0.17 to 2.08}~$M_\odot$. The presented distribution is not an IMF as we did not correct for the photometric incompleteness. As we discussed in Sec. \ref{subsec:complete}, completeness of our catalogue is affected by crowdness in the cluster core, presence of bright stars, and highly variable nebular emission. According to the $J$-band magnitudes distribution in Fig. \ref{fig:JcomplHAWKI}, we reach 50\% completeness level at 15.5~mag corresponding to 0.8~$M_\odot$ at 1~Myr \citep{bressan2012,baraffe2015}. However, if we include detections excluded from our catalogue due to the variable background emission, the 50\% completeness level is already achieved at 18.5~mag (equivalent of {0.1}~$M_\odot$), 
demonstrating the depth and value of our MUSE data. 
In the second panel of Fig. \ref{fig:IMF} we show the fraction of stars in mass bins in the final (clean) catalogue in comparison to the sample without applying the background cut (Sec. \ref{subsec:bkg}). In the mass range below $\sim${0.8}~$M_\odot$ 
more than 50\% detections are missing due to our conservative approach to the background contamination. {At the same time, none of the stars with masses $\gtrsim$2.3~$M_\odot$ is removed due to the high background variation.} 
In Table \ref{tab:highbkgcat} we list targets with uncertain photometry and their stellar parameters, where available. Removing the lowest-mass stars from our catalogue prevents us from constructing an IMF estimate and identifying the characteristic mass of the \tr\ population. {More than half of the stars with masses $<$0.8~$M_\odot$ are removed from the final catalog due the highly variable background (Fig. \ref{fig:IMF}) but} the proper completeness analysis is beyond the goal of this work. {We note however, that when using combined set of \cite{baraffe2015} and \cite{siess2000} evolutionary tracks, the distribution of masses is similar, although stellar masses extend only up to 4.1~$M_\odot$. While the global picture of stellar mass distribution in \tr\ is not affected by the choice of evolutionary models, the individual values can be different even by a factor of few, especially for the least massive objects. This illustrate how challenging it is to derive accurate properties of young, low-mass stars.}
 
The (in)variance of the IMF in the high FUV environment is {also} outside the scope of this paper. Future work addressing this problem needs first to resolve the completeness issue. Including more massive stars will allow investigating the slope of the mass function in the high-mass end. More sophisticated approach to the estimation of the background emission may allow including significantly more low-mass stars and with that testing the breaking point of the Kroupa-like IMF or characteristic mass of log-normal IMF. High-spatial-resolution observations (e.g., with adaptive optics) of the very center of \tr\ can help resolving the inner core region. Brown dwarfs and very low-mass members of Tr 14 can only be well spectroscopically characterised by the NIR IFU instruments, like VLT/KMOS, VLT/ERIS, JWST/NIRspec, or future ELT/HARMONI.

%%%%%%% conclusions %%%%%%%
\section{Summary and conclusions}
\label{sec:conc}

\indent

In this work, we presented the first optical spectroscopic study of low-mass stars in \tr\ based on IFU observations from VLT/MUSE. We identified targets and extracted photometry and spectra using {\tt SExtractor}. We excluded from the catalogue all sources with uncertain photometry. Specifically, the most significant cut (of 1868 sources) was related to the highly variable in spatial dimension emission of the H{\sc ii} region in Carina Nebula. At the end, our catalogue consists of 780 stars. We make available both catalogs, with robust and uncertain photometry for possible future follow-up studies.

Most of our sources have photometric measurements from NIR (99\%) and optical (76\%) catalogs, and almost 40\% were detected in X-rays. 
We performed the spectral classification using spectra of Class III stars from \cite{manara2013,manara2017} as templates. Together with spectral type we estimated visual reddening and constant veiling of 717 stars. We converted the spectral types to the effective temperatures and used $J$-band photometry to calculate bolometric luminosities. We placed our stars in the HR diagram and by the comparison to the theoretical evolutionary tracks {\citep{bressan2012}} 
we estimated the stellar masses and ages. Based on the distribution of stellar ages we {estimate} 
the cluster age of $\sim$1~Myr. {This result is maintained even when using different evolutionary models \citep{siess2000, baraffe2015}.} Majority of our stars (51\%) 
have mass below 1~$M_\odot$, while the least massive object has estimated mass of {0.17~$M_\odot$.}

Massive star-forming regions represent the most common environment in which the stars form in the Galaxy. This environment differs from those seen in the solar neighbourhood and therefore necessitates for extension of the studies of the more distant regions. Those examinations are however challenging due to the large distance to the cluster, often high extinction, and high stellar crowding. This is particularly difficult for spectroscopic observations. Here, we presented that those kind of explorations are feasible with IFU spectrographs like VLT/MUSE. Low-mass stars are the most common in the Galaxy but at the same time the most vulnerable to the environmental conditions. Presented here stellar characteristics of the few hundreds of low-mass stars provide a step into the better understanding of formation and early evolution of low-mass stars in the massive cluster. This study set up a base for the follow-up investigation of the protoplanetary disk population response to the high FUV field in the cluster that will be presented in the consecutive paper. 

%%%%%%%%%%%%%%%%%%%%%%%%%%%%%%%%
%%%%%%%%%%%%%%%%%%%%%%%%%%%%%%%%

%%%%%%% acknowledgments %%%%%%%
\begin{acknowledgements}
{We thank the referee, Guido De Marchi, for a careful reading and for insightful comments helping improving this manuscript.} 
This work was partly supported by the Italian Ministero dell’Istruzione, Universit\`{a} e Ricerca through the grant Progetti Premiali 2012-iALMA (CUP C52I13000140001), by the Deutsche Forschungsgemeinschaft (DFG, German Research Foundation) - Ref no. 325594231 FOR 2634/2 TE 1024/2-1, by the DFG Cluster of Excellence Origins (www.origins-cluster.de), by the Fondazione Cariplo, grant no. 2022-1217. This project has received funding from the European Union’s Horizon 2020 research and innovation program under the Marie Skłodowska- Curie grant agreement No 823823 (DUSTBUSTERS), from the European Research Council (ERC) via the ERC Synergy Grant ECOGAL (grant 855130), and by the European Union (ERC, WANDA, 101039452 and ERC, DiscEvol, 101039651). Views and opinions expressed are however those of the author(s) only and do not necessarily reflect those of the European Union or the European Research Council Executive Agency. Neither the European Union nor the granting authority can be held responsible for them.

This work has made use of data from the European Space Agency (ESA) mission {\it Gaia} (\url{https://www.cosmos.esa.int/gaia}), processed by the {\it Gaia} Data Processing and Analysis Consortium (DPAC, \url{https://www.cosmos.esa.int/web/gaia/dpac/consortium}). Funding for the DPAC has been provided by national institutions, in particular the institutions participating in the {\it Gaia} Multilateral Agreement.

\end{acknowledgements}

%%%%%%% bibliography %%%%%%%
\bibliographystyle{aa}
\bibliography{references-Tr14-MUSE}

%%%%%%% appendix %%%%%%%
\begin{appendix} 
\section{Observational log}
\label{app:weather}

In Table \ref{tab:weather} we provide basic information 
about weather conditions during observations with \lq\lq long\rq\rq\ integration times ({3$\times$}13~min) and name of the standard star used for flux calibration for each set. Due to the bad atmospheric conditions some observations were repeated. We checked all datasets and used those, which were described as better quality by the Observatory. We include in the Table all observations made for the information of the astronomical community.

\begin{table}[h!]
\caption{Observational log} 
\label{tab:weather} 
\resizebox{\columnwidth}{!}{
\begin{tabular}{cccccc}
\hline
\hline
pointing & coordinates & date & seeing & grade & calibration \\ 
  & (h:m:s ~d:m:s) & & (\arcsec) &  & standard \\ 
\hline
 1 & 10:44:08.4 -59:29:39.9 & 28.02.2016 & 0.75 & A & GD108\\ 
 2 & 10:44:00.7 -59:29:39.6 & 25.02.2016 & 0.92 & B & GD71 \\ 
 3 & 10:44:00.7 -59:30:39.3 & 25.02.2016 & 0.85 & B & GD71 \\
 4 & 10:44:08.3 -59:30:39.7 & 25.02.2016 & 0.86 & B & GD71 \\
 5 & 10:44:08.3 -59:31:39.0 & 25.02.2016 & 0.92 & A & GD153 \\ 
 6 & 10:44:00.6 -59:31:38.6 & 28.02.2016 & 0.90 & B & GD71 \\ 
 7 & 10:44:00.6 -59:32:38.4 & 25.02.2016 & 1.24 & B & GD108\\
 8 & 10:44:08.2 -59:32:38.8 & 02.03.2016 & 0.89 & A & GD108\\ 
 ~~9*& 10:44:08.2 -59:33:38.1 & 25.03.2016 & -- &  C & -- \\
 9  & 10:44:08.2 -59:33:38.1 & 30.03.2016 & 1.01 &  A & GD108\\
 ~10* & 10:44:00.5 -59:33:37.7 & 25.03.2016 & -- &  C & -- \\
 10 & 10:44:00.5 -59:33:37.7 & 30.03.2016 & 1.07 & A & GD108\\
 11 & 10:44:00.5 -59:34:37.3 & 28.03.2016 & 1.10 & A & GD108 \\
 12 & 10:44:08.1 -59:34:37.8 & 27.03.2016 & 1.58 & B & GD108 \\ 
 ~13* & 10:43:52.9 -59:32:37.8 & 27.03.2016 & 2.71 & C & -- \\ 
 13 & 10:43:52.9 -59:32:37.8 & 04.04.2016 & 0.53 & A & GD108 \\ 
 14* & 10:43:45.3 -59:32:37.3 & 27.03.2016 & 2.80 & C & -- \\
 14 & 10:43:45.3 -59:32:37.3 & 04.04.2016 & 0.50 & A & GD108 \\ 
 15* & 10:43:45.3 -59:33:36.8 & 27.03.2016 & 2.06 & C & -- \\
 15* & 10:43:45.3 -59:33:36.8 & 16.04.2016 & 0.66 & C  & -- \\
 15 & 10:43:45.3 -59:33:36.8 & 17.04.2016 & 0.68 & B & GD108 \\ 
 16 & 10:43:52.9 -59:33:37.2 & 30.03.2016 & 1.48 & B & GD108 \\ 
 17 & 10:43:52.9 -59:34:36.9 & 30.03.2016 & 1.38 & B & GD108 \\
 18 & 10:43:45.2 -59:34:36.4 & 01.04.2016 & 0.79 & A & GD108 \\
 19 & 10:43:37.7 -59:33:36.2 & 01.03.2016 & 1.47 & B & LTT3218 \\ 
 20 & 10:43:30.0 -59:33:35.7 & 02.03.2016 & 1.33 & A & GD108 \\ 
 21 & 10:43:30.0 -59:34:35.3 & 02.03.2016 & 0.99 & A & GD108 \\
 22 & 10:43:37.6 -59:34:35.9 & 02.03.2016 & 1.15 & A & GD108 \\ 
\hline
\end{tabular}}
\tablefoot{
Each of the pointings were observed three times with a 90$^o$ dither pattern, listed seeing is a mean value for each Observational Block (OB). 
Due to bad weather conditions some pointings were repeated, those not used in this work due to the low quality are marked with an asterisk (*).
OB's grades refer to: A -- fully within constraints, OB completed; B -- mostly within constraints, some constraint is $\sim$10\% violated, OB completed; C - out of constraints, OB must be repeated.}
\end{table}
%OB Grades
 %A : Fully within constraints - OB completed
 %B : Mostly within constraints, some constraint is ~10\% violated - OB completed
 %C : Out of constraints - OB must be repeated
 %D : Out of constraints - OB considered completed
 %X : Uncertain / Not Applicable
 %_ : no grade assigned
%-1.0 changed to -- in seeing

\section{Coordinates correction and cross-match with photometric catalogs}
\label{app:coord}
\indent

\begin{figure*}\hspace{-0.2cm}
\includegraphics[width=0.5\columnwidth, trim={0 0.cm 1.45cm 0.6cm}, clip ]{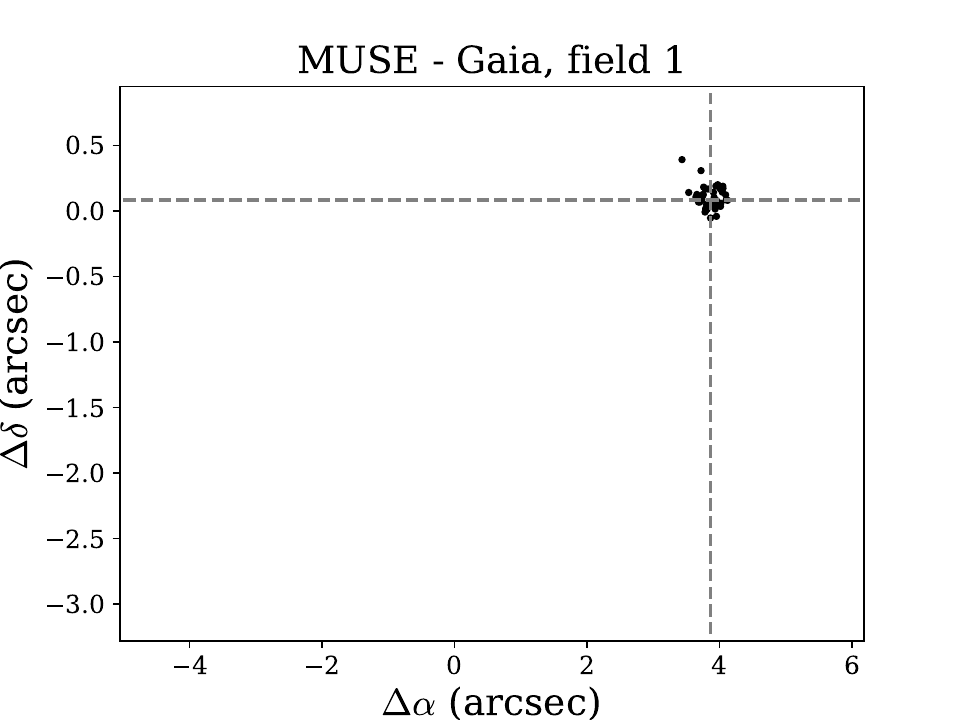}\hspace{-0.1cm}
\includegraphics[width=0.5\columnwidth, trim={0 0.cm 1.45cm 0.6cm}, clip ]{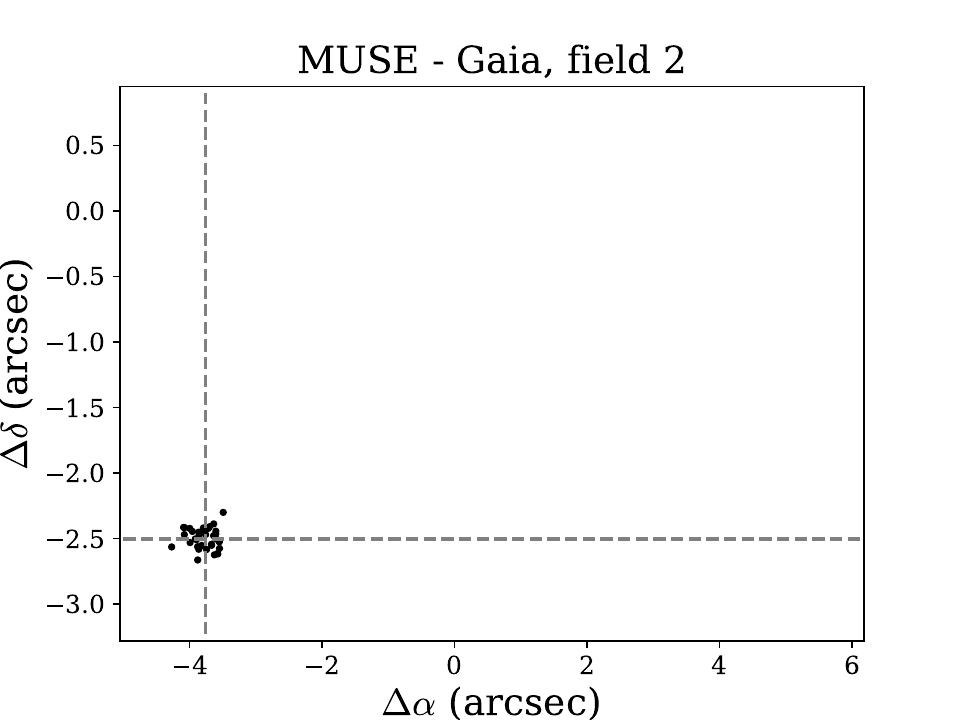}\hspace{-0.1cm}
\includegraphics[width=0.5\columnwidth, trim={0 0.cm 1.45cm 0.6cm}, clip ]{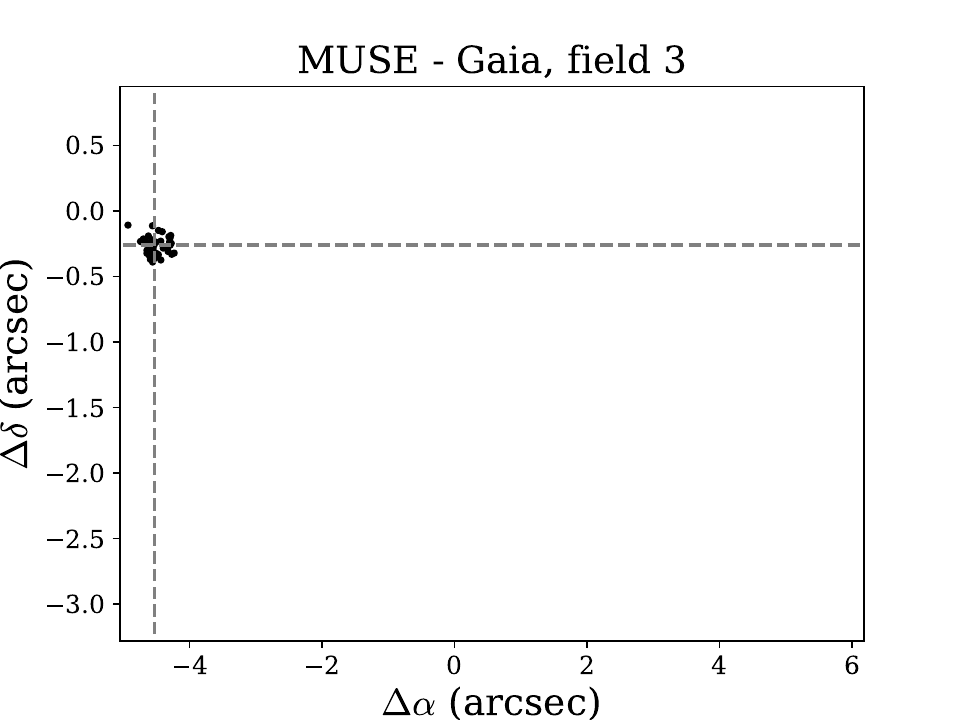}\hspace{-0.1cm}
\includegraphics[width=0.5\columnwidth, trim={0 0.cm 1.45cm 0.6cm}, clip ]{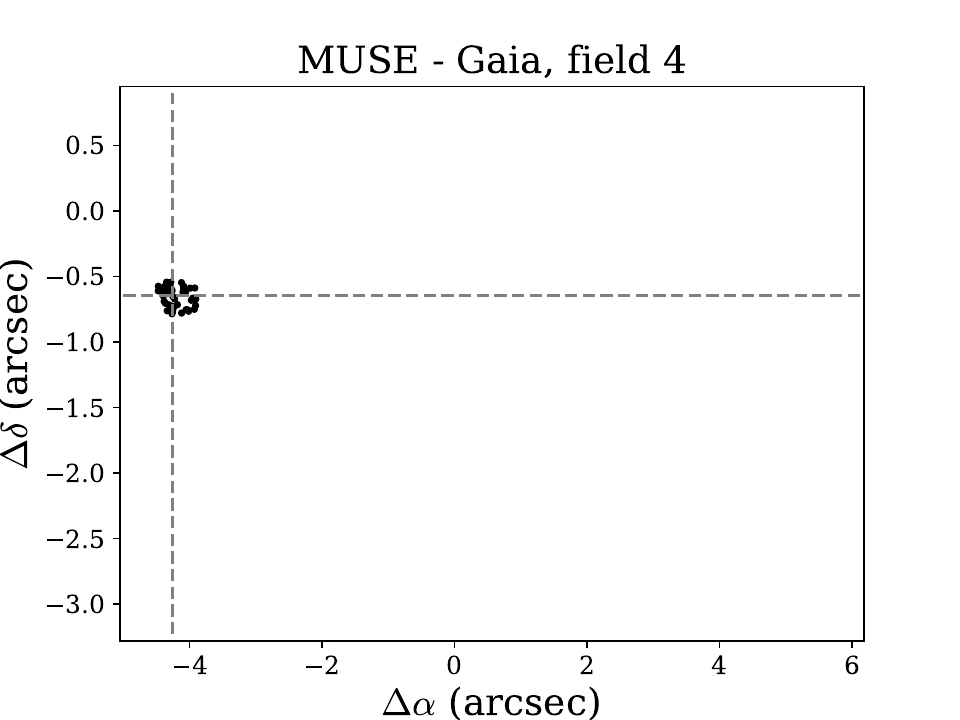}\hspace{-0.1cm}
%%%%%%%%%%%%%%%%%%%%%%%%%%
\hspace{-0.2cm}\includegraphics[width=0.5\columnwidth, trim={0 0.cm 1.45cm 0.6cm}, clip ]{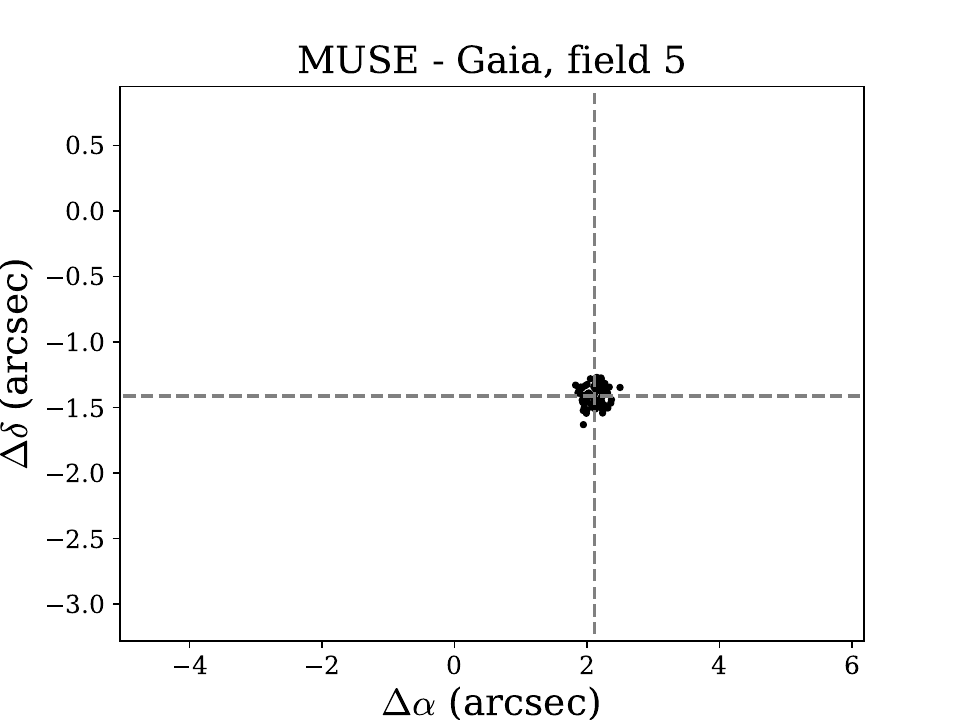}\hspace{-0.1cm}
\includegraphics[width=0.5\columnwidth, trim={0 0.cm 1.45cm 0.6cm}, clip ]{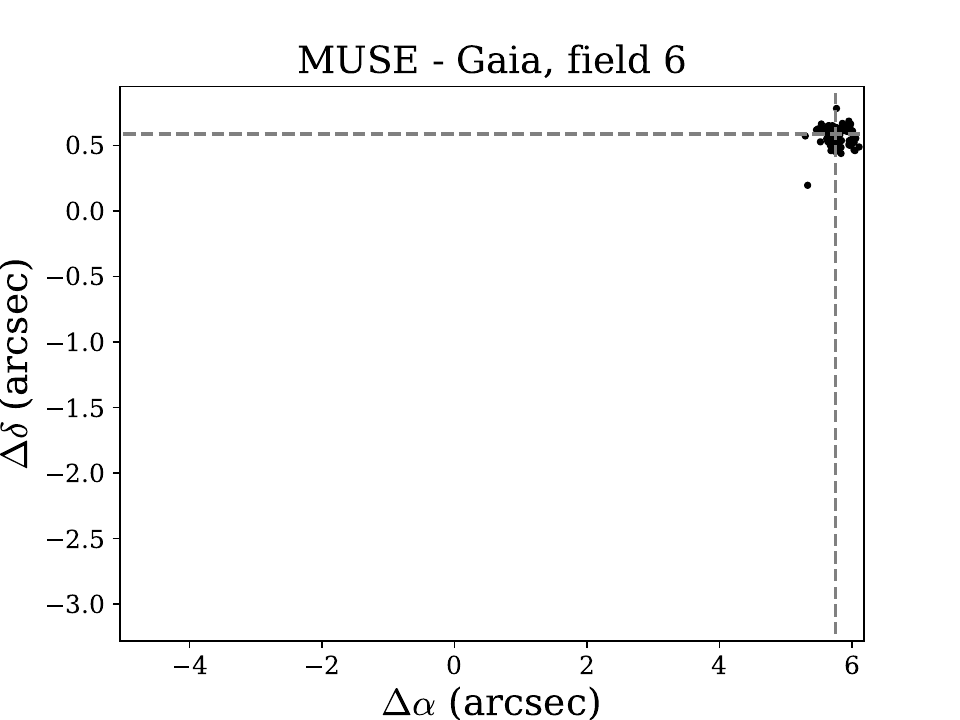}\hspace{-0.1cm}
\includegraphics[width=0.5\columnwidth, trim={0 0.cm 1.45cm 0.6cm}, clip ]{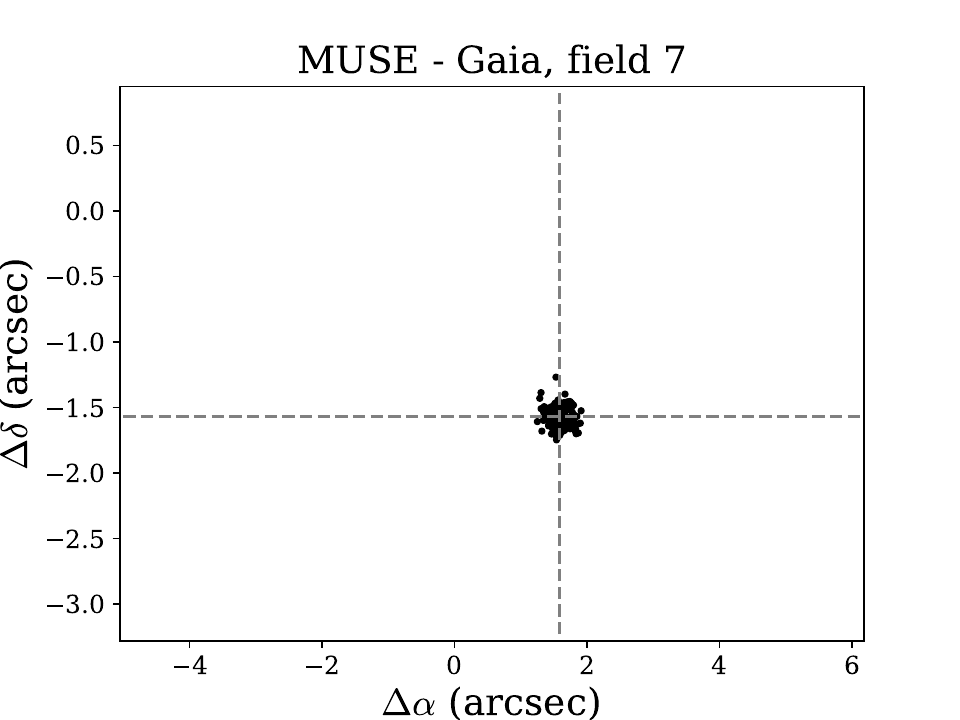}\hspace{-0.1cm}
\includegraphics[width=0.5\columnwidth, trim={0 0.cm 1.45cm 0.6cm}, clip ]{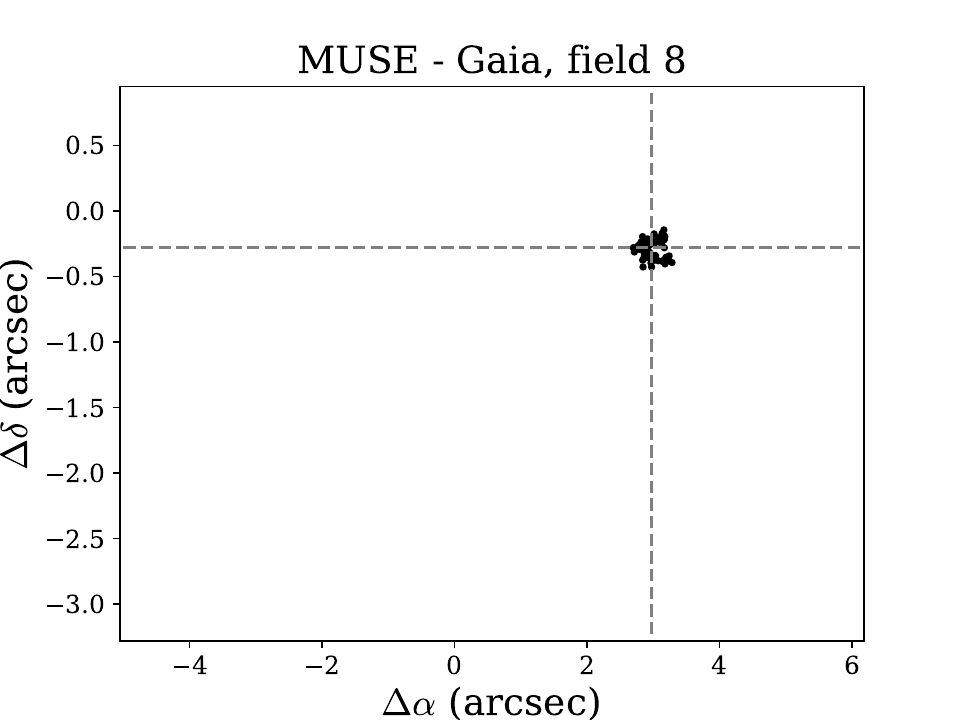}\hspace{-0.1cm}
%%%%%%%%%%%%%%%%%%%%%%%%%
\hspace{-0.2cm}\includegraphics[width=0.5\columnwidth, trim={0 0.cm 1.45cm 0.6cm}, clip ]{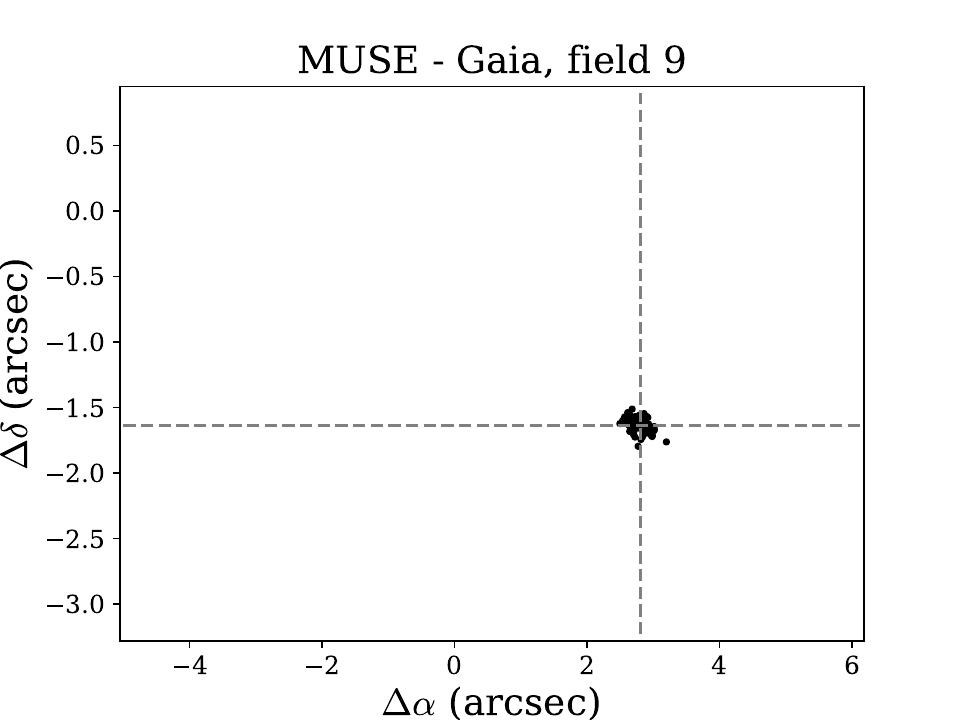}\hspace{-0.1cm}
\includegraphics[width=0.5\columnwidth, trim={0 0.cm 1.45cm 0.6cm}, clip ]{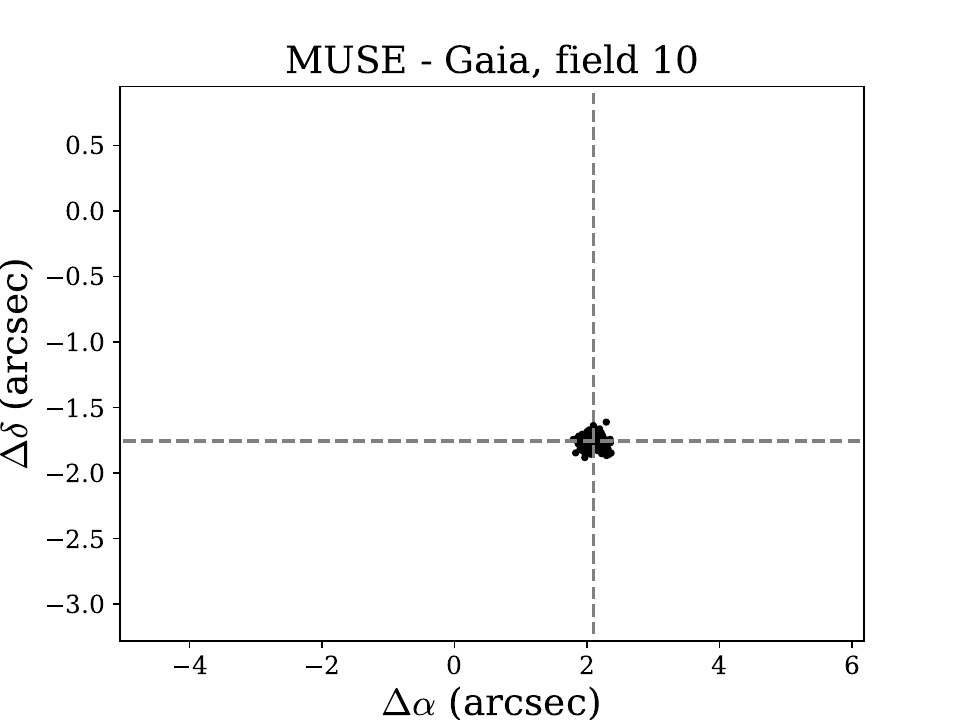}\hspace{-0.1cm}
\includegraphics[width=0.5\columnwidth, trim={0 0.cm 1.45cm 0.6cm}, clip ]{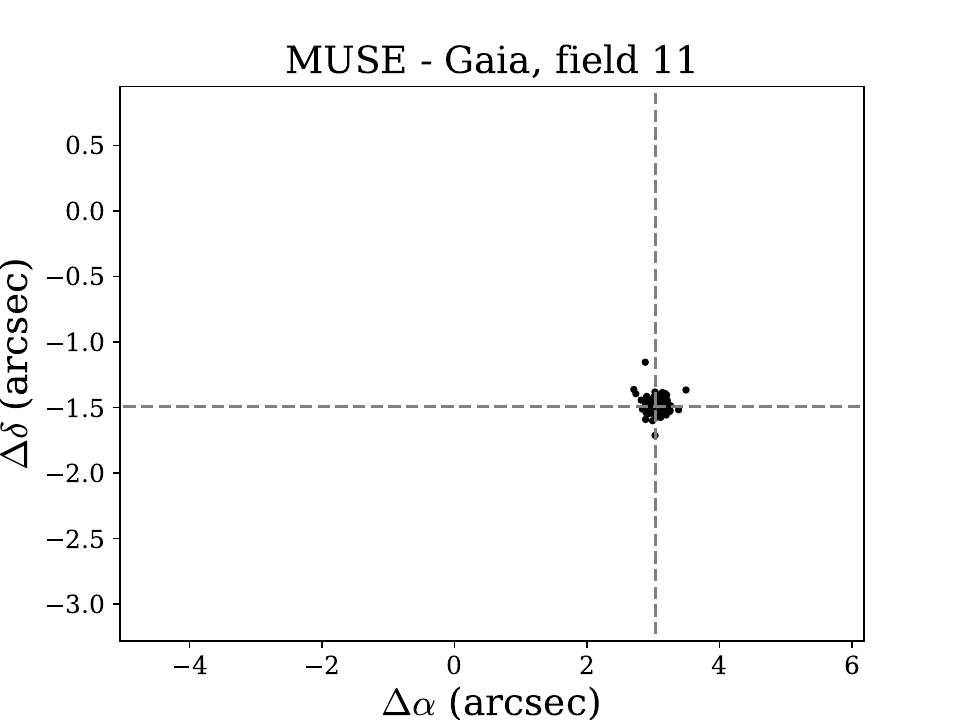}\hspace{-0.1cm}
\includegraphics[width=0.5\columnwidth, trim={0 0.cm 1.45cm 0.6cm}, clip ]{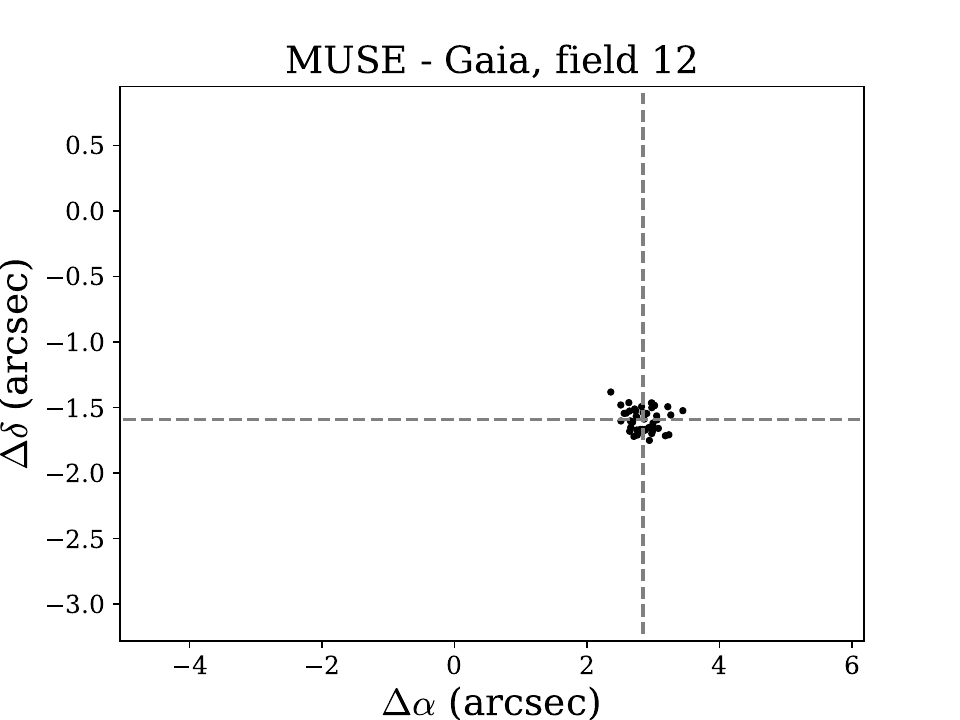}\hspace{-0.1cm}
%%%%%%%%%%%%%%%%%%%%%%%%%%%
\hspace{-0.2cm}\includegraphics[width=0.5\columnwidth, trim={0 0.cm 1.45cm 0.6cm}, clip ]{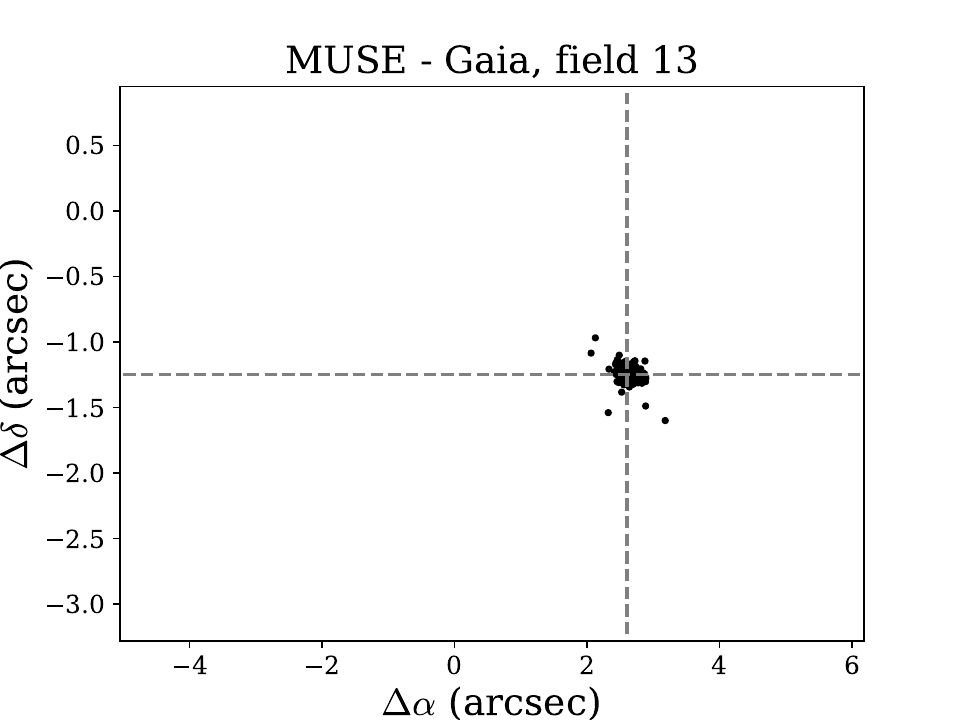}\hspace{-0.1cm}
\includegraphics[width=0.5\columnwidth, trim={0 0.cm 1.45cm 0.6cm}, clip ]{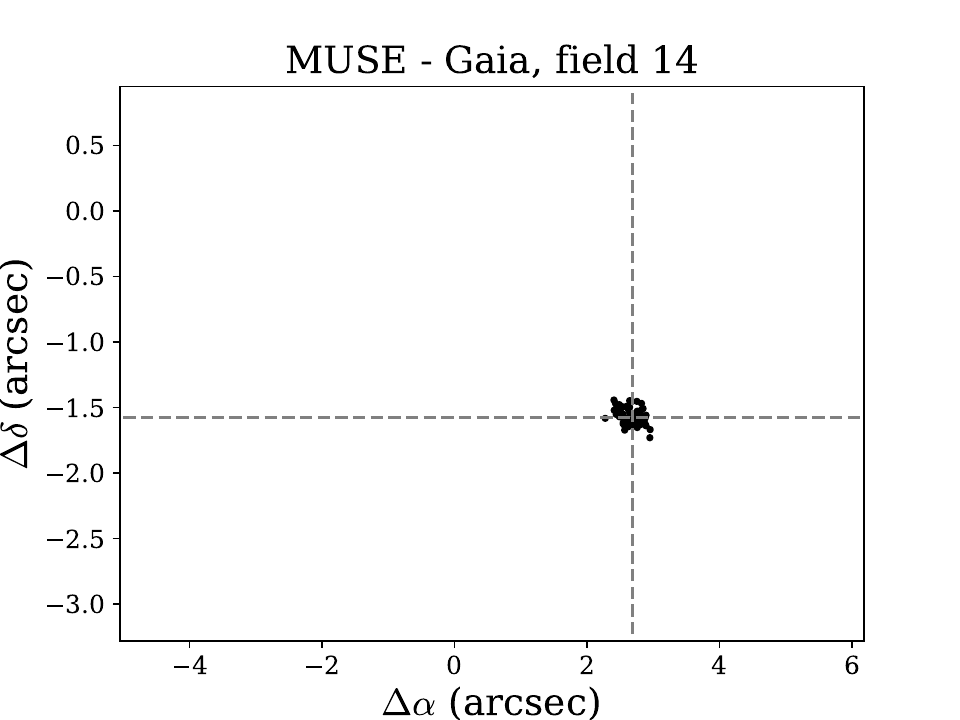}\hspace{-0.1cm}
\includegraphics[width=0.5\columnwidth, trim={0 0.cm 1.45cm 0.6cm}, clip ]{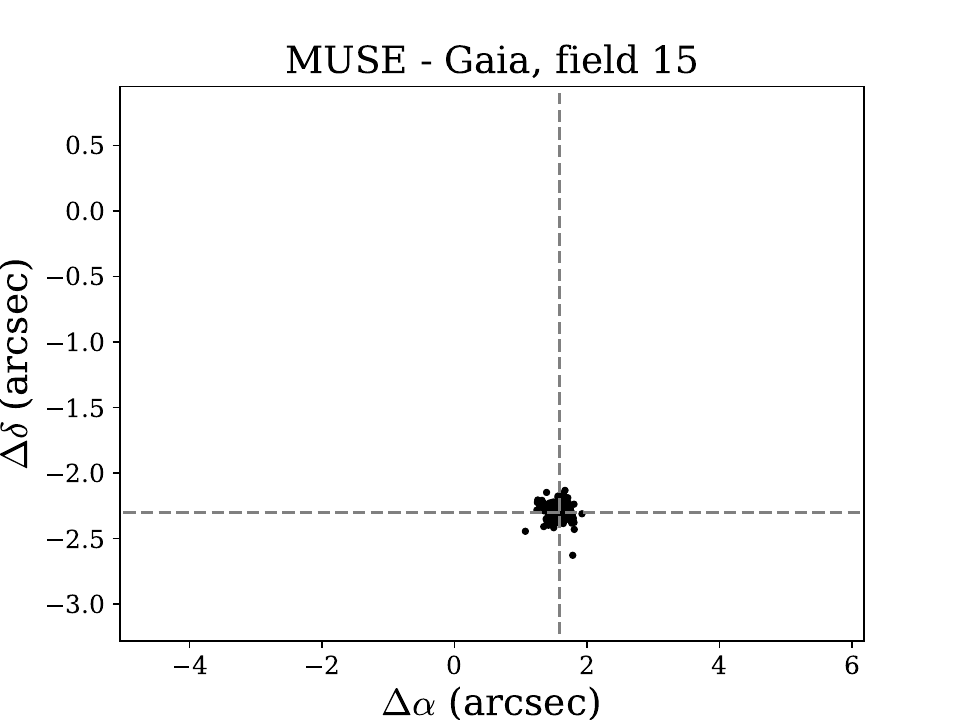}\hspace{-0.1cm}
\includegraphics[width=0.5\columnwidth, trim={0 0.cm 1.45cm 0.6cm}, clip ]{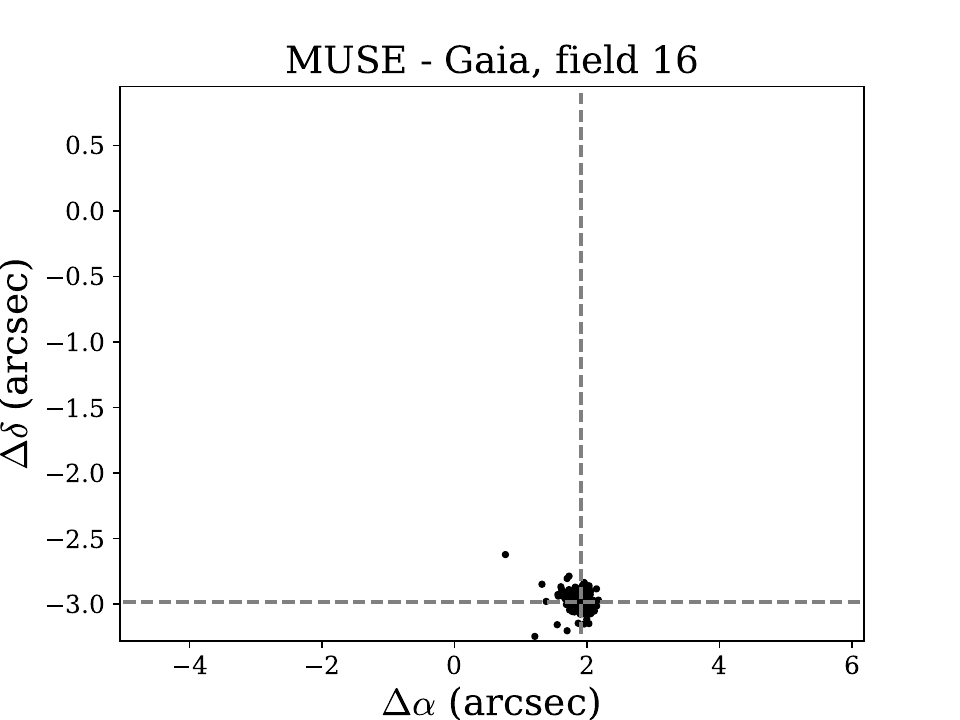}\hspace{-0.1cm}
%%%%%%%%%%%%%%%%%%%%%%%%%
\hspace{-0.2cm}\includegraphics[width=0.5\columnwidth, trim={0 0.cm 1.45cm 0.6cm}, clip ]{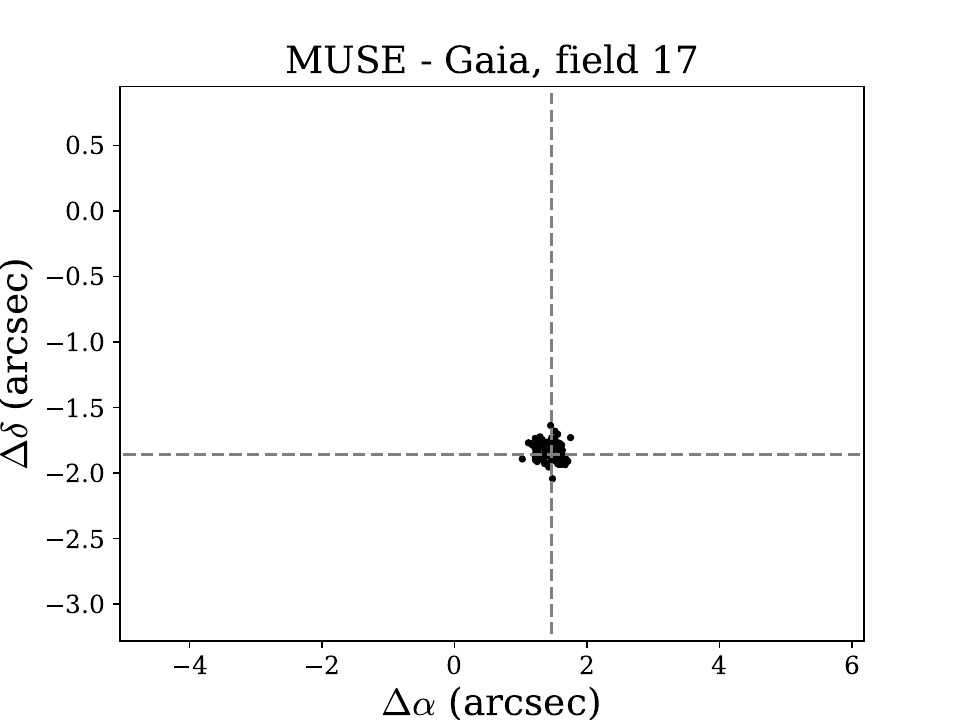}\hspace{-0.1cm}
\includegraphics[width=0.5\columnwidth, trim={0 0.cm 1.45cm 0.6cm}, clip ]{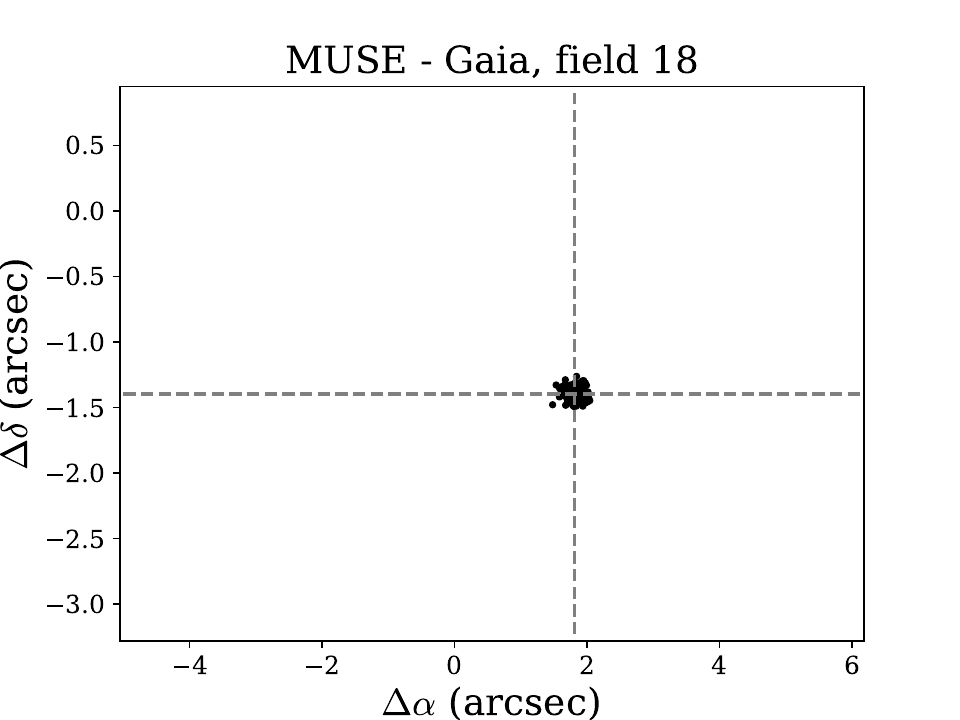}\hspace{-0.1cm}
\includegraphics[width=0.5\columnwidth, trim={0 0.cm 1.45cm 0.6cm}, clip ]{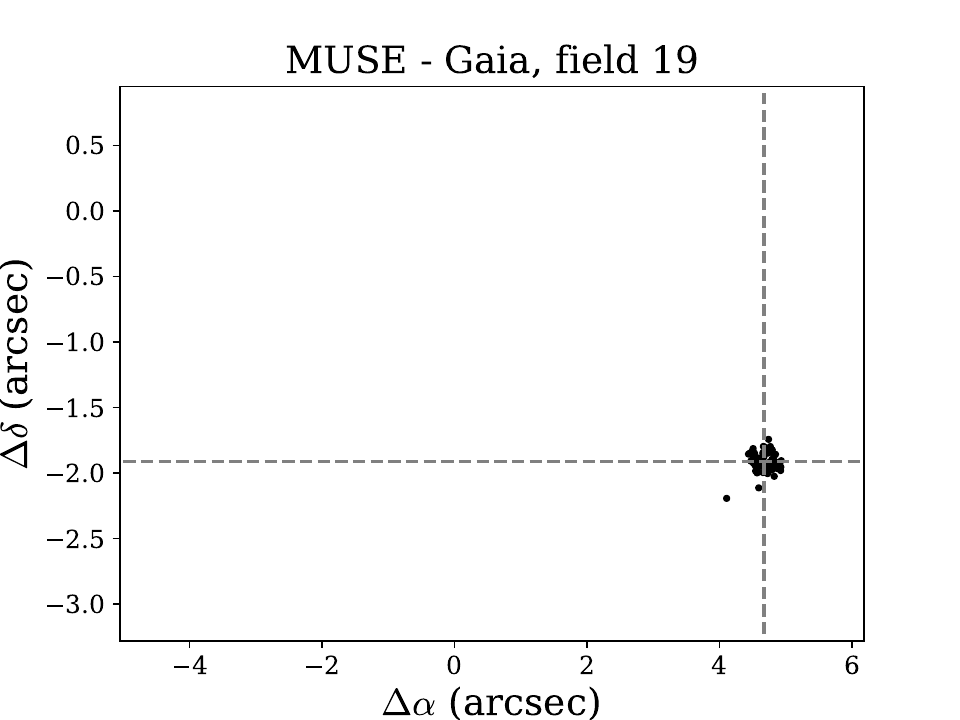}\hspace{-0.1cm}
\includegraphics[width=0.5\columnwidth, trim={0 0.cm 1.45cm 0.6cm}, clip ]{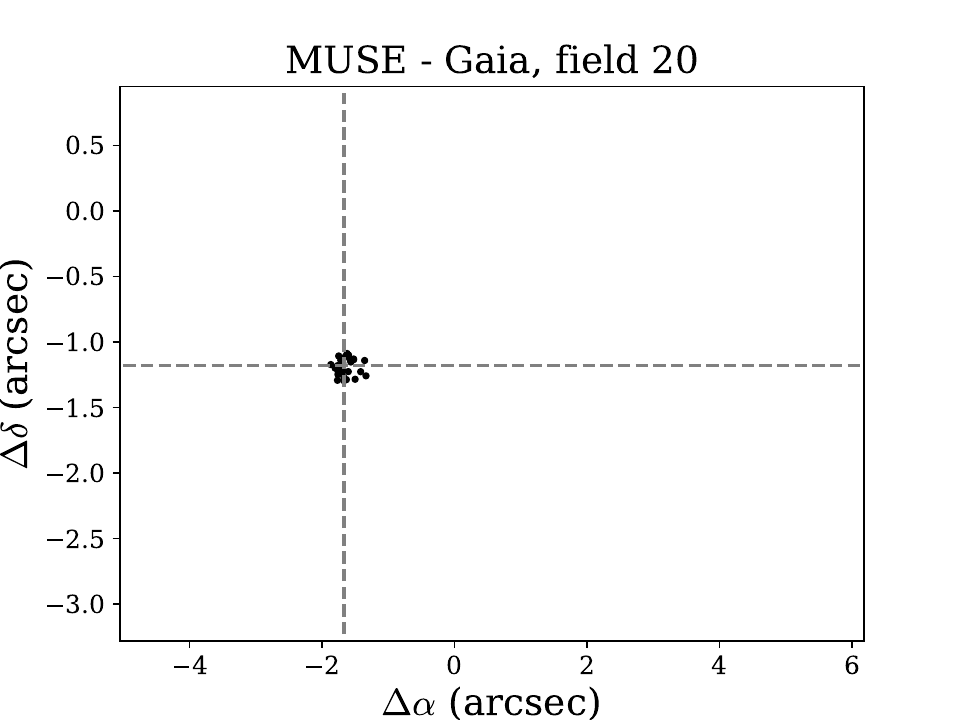}\hspace{-0.1cm}
%%%%%%%%%%%%%%%%%%%%%%
\hspace{-0.2cm}\includegraphics[width=0.5\columnwidth, trim={0 0.cm 1.45cm 0.6cm}, clip ]{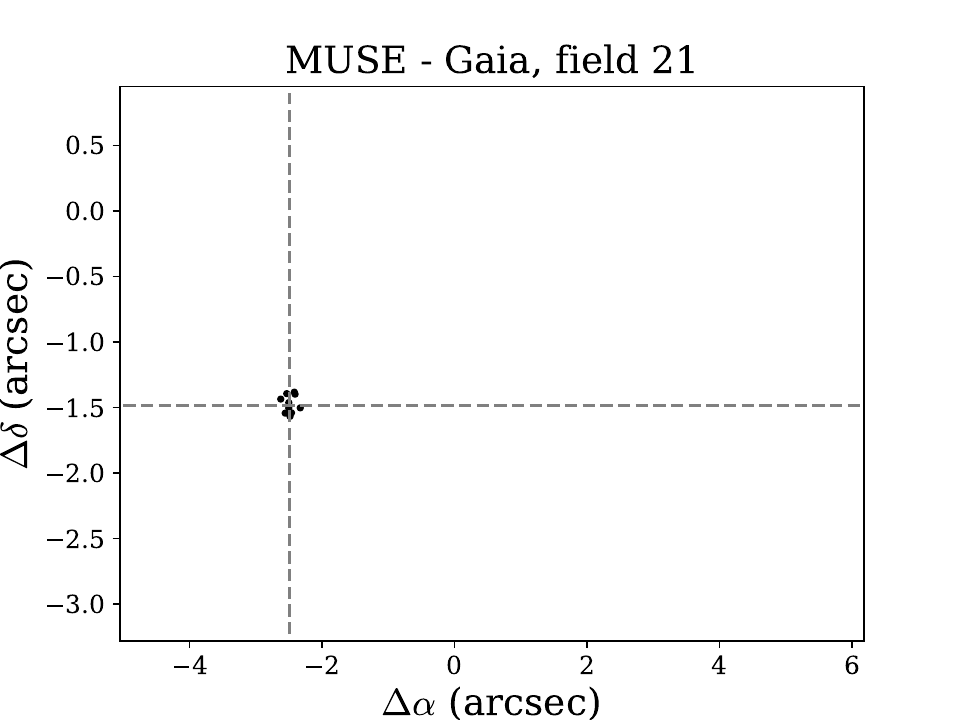}\hspace{-0.1cm}
\includegraphics[width=0.5\columnwidth, trim={0 0.cm 1.45cm 0.6cm}, clip ]{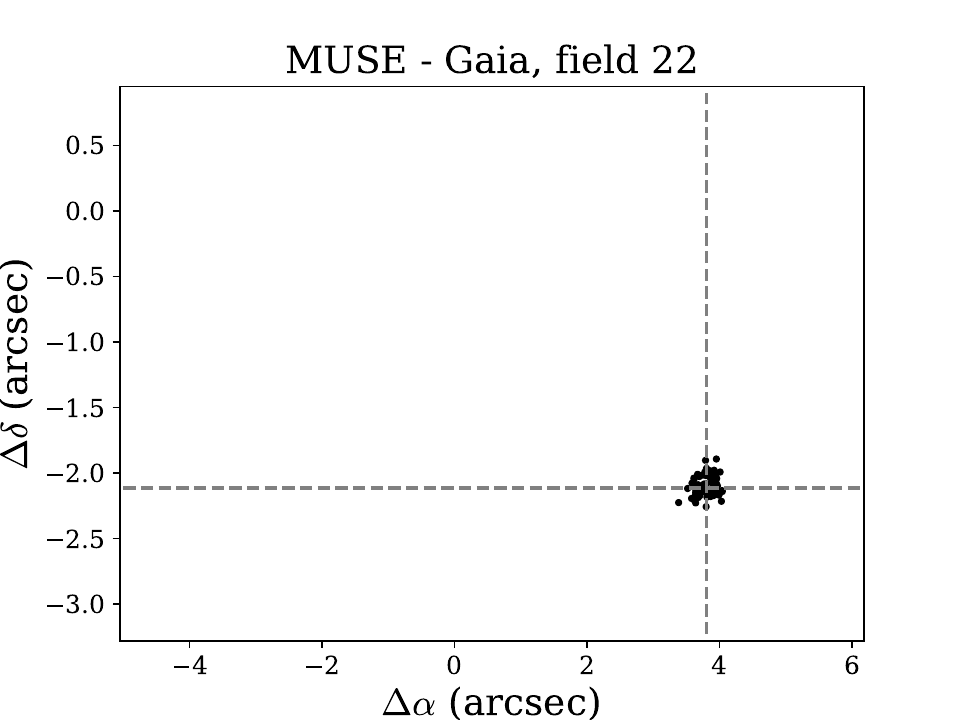}\hspace{-0.1cm}
\caption{Distribution of differences in right ascensions and declinations from MUSE and {\it Gaia} DR3 at each field.}
\label{fig:delta_coord}
\end{figure*}

We corrected MUSE coordinates using {\it Gaia} DR3 catalogue \citep{gaia-vallenari2022}. The applied corrections are median differences in right ascension and declination between MUSE and {\it Gaia}. We show distribution of those differences per each pointing in the Figure \ref{fig:delta_coord}. Corrections, which need to be added to the original MUSE coordinates, are listed in Table \ref{tab:coord-corr}. They range in absolute values between 1.46\arcsec and 5.75\arcsec for right ascension, and between 0.08\arcsec and 2.98\arcsec for declination with typical standard deviation of 0.15\arcsec for right ascension and 0.08\arcsec for declination.

\begin{table}%[h!]
\centering
\caption{Applied coordinate offsets.} 
\label{tab:coord-corr} 
%\resizebox{\columnwidth}{!}{
\begin{tabular}{ccc}
\hline
\hline
pointing & $\Delta\alpha$ & $\Delta\delta$ \\ 
  & (\arcsec) & (\arcsec) \\ 
\hline
 1 &  -3.86 $\pm$ 0.15 & -0.08 $\pm$ 0.08\\ 
 2 &  ~3.76 $\pm$ 0.19 & ~2.50 $\pm$ 0.08\\ 
 3 &  ~4.53 $\pm$ 0.15 & ~0.26 $\pm$ 0.07\\
 4 &  ~4.25 $\pm$ 0.17 & ~0.65 $\pm$ 0.07\\ 
 5 &  -2.12 $\pm$ 0.15 & ~1.41 $\pm$ 0.08\\ 
 6 &  -5.75 $\pm$ 0.16 & -0.59 $\pm$ 0.07\\ 
 7 &  -1.58 $\pm$ 0.14 & ~1.57 $\pm$ 0.07\\
 8 &  -2.98 $\pm$ 0.15 & ~0.28 $\pm$ 0.07\\ 
 9  & -2.80 $\pm$ 0.14 & ~1.64 $\pm$ 0.06\\
 10 & -2.09 $\pm$ 0.13 & ~1.75 $\pm$ 0.06\\
 11 & -3.04 $\pm$ 0.14 & ~1.49 $\pm$ 0.08\\
 12 & -2.85 $\pm$ 0.21 & ~1.59 $\pm$ 0.09\\
 13 & -2.60 $\pm$ 0.14 & ~1.25 $\pm$ 0.06\\ 
 14 & -2.68 $\pm$ 0.15 & ~1.58 $\pm$ 0.06\\ 
 15 & -1.58 $\pm$ 0.15 & ~2.30 $\pm$ 0.07\\ 
 16 & -1.91 $\pm$ 0.18 & ~2.98 $\pm$ 0.08\\ 
 17 & -1.46 $\pm$ 0.16 & ~1.86 $\pm$ 0.07\\
 18 & -1.81 $\pm$ 0.12 & ~1.40 $\pm$ 0.05\\
 19 & -4.67 $\pm$ 0.15 & ~1.91 $\pm$ 0.07\\ 
 20 & ~1.67 $\pm$ 0.13 & ~1.18 $\pm$ 0.06\\ 
 21 & ~2.49 $\pm$ 0.08 & ~1.48 $\pm$ 0.06\\
 22 & -3.80 $\pm$ 0.14 & ~2.11 $\pm$ 0.07\\ 
\hline
\end{tabular}%}
\tablefoot{Corrections calculated as a median difference between the {\it Gaia} and MUSE coordinates. Corrections were added to MUSE coordinates.}
\end{table}

\begin{figure}
\includegraphics[width=\columnwidth, trim={0.25cm 0.01cm 1.3cm 1.3cm}, clip]{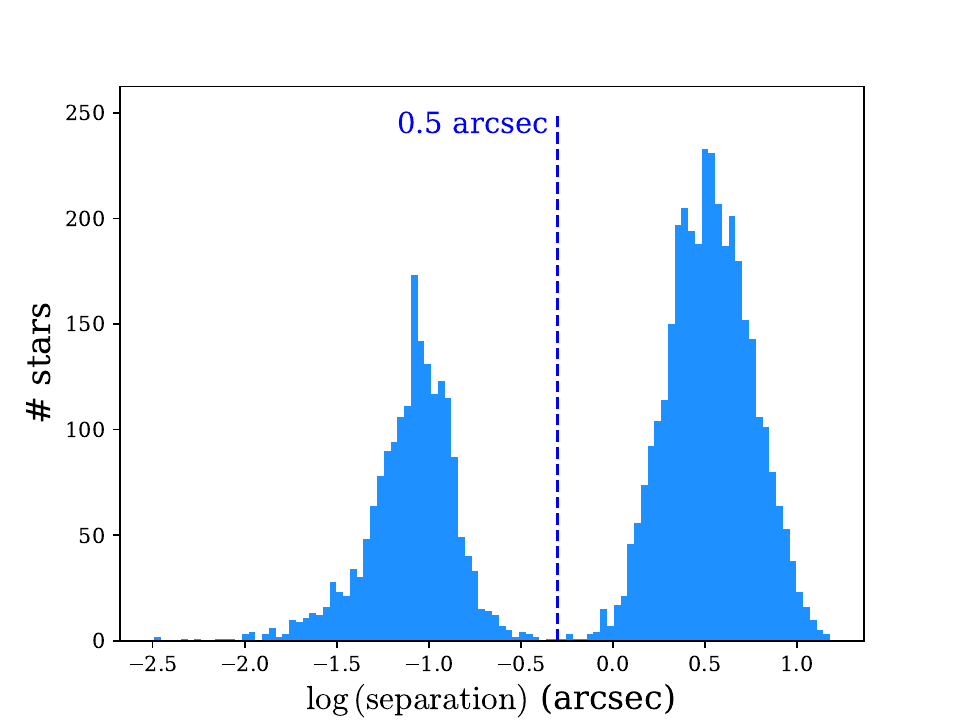}
\caption{Separations in arcsec between MUSE and {\it Gaia} DR3 stars in logarithmic scale. MUSE catalogue here contains data points as of Sec. \ref{subsec:coord}, before removing uncertain photometry. Overplotted is the threshold separation of 0.5$\arcsec$ used in this work.}
\label{fig:sep}
\end{figure}

After correcting MUSE coordinates we matched again our sources with {\it Gaia} to examine the goodness of our astrometry and define the best matching radius between different catalogs. Based on the bimodal distribution of separations between corrected MUSE and {\it Gaia} coordinates in Figure \ref{fig:sep}, we found that the best separation radius for cross-matching is 0.5\arcsec. Within it we find all true counterparts and do not include false matches with larger separations. False matches are caused by the crowding, especially large in the cluster center. The same distribution shows that the uncertainty of our astrometry is $\sim$0.1\arcsec.

\begin{figure}
\includegraphics[width=\columnwidth]{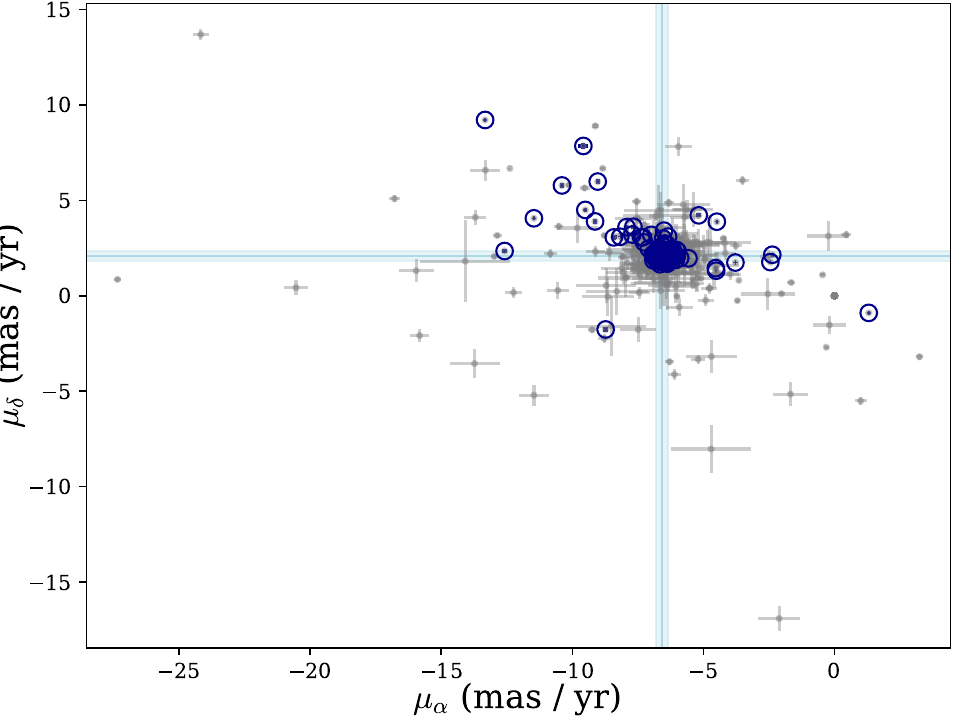}
\caption{Proper motions of {\it Gaia} counterparts. We show points with available proper motion measurements (grey dots) and with good astrometry (dark blue circles) (see Sec. \ref{subsec:cross-match} for details). Uncertainties of the latter points are smaller than the symbol sizes. Foreground and possibly contaminated stars are excluded. For reference, we also mark group proper motions from \cite{berlanas2023} derived from {\it Gaia} EDR3 astrometry of OB stars in \tr.}
\label{fig:pm}
\end{figure}

\section{Targets detected in more than one pointing}
\label{app:double}
\indent

The pointing of observations were designed to overlap. Therefore, some stars are detected multiple times in several pointings. After coordinates correction we identify those targets performing coordinate matching within the catalogue. We define the same separation limit of 0.5\arcsec, below which we assume that the two detections correspond to the same object. We find separations in the range from 0.09\arcsec\ to 0.26\arcsec\ with median of 0.20\arcsec\ corresponding to the size of one pixel. In Table \ref{tab:double} we list all the pairs of double detections.

\begin{figure}\hspace{-0.2cm}
\includegraphics[width=0.5\columnwidth]{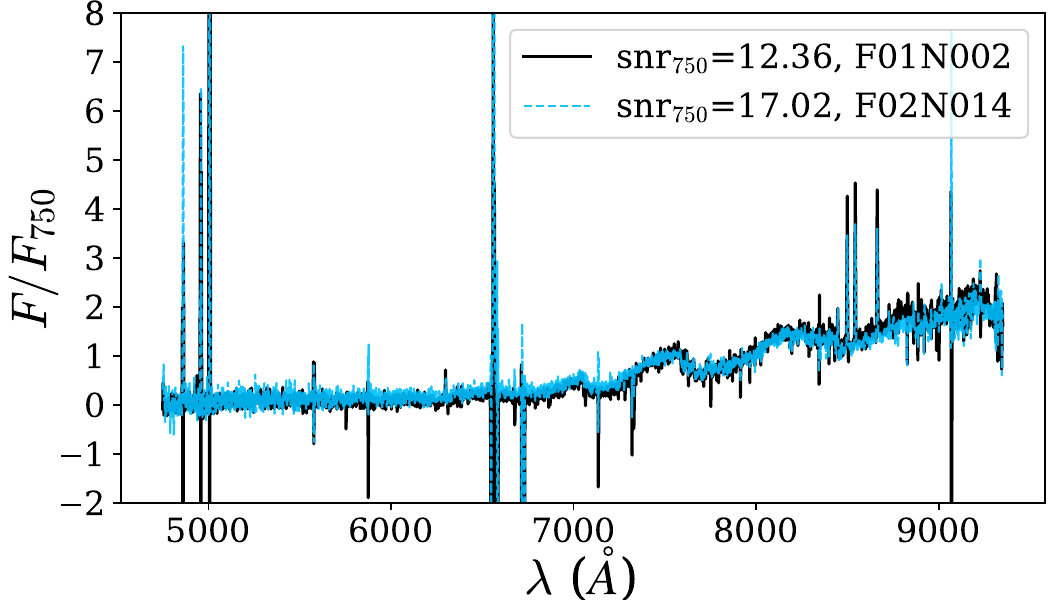}\hspace{-0.1cm}
\includegraphics[width=0.5\columnwidth]{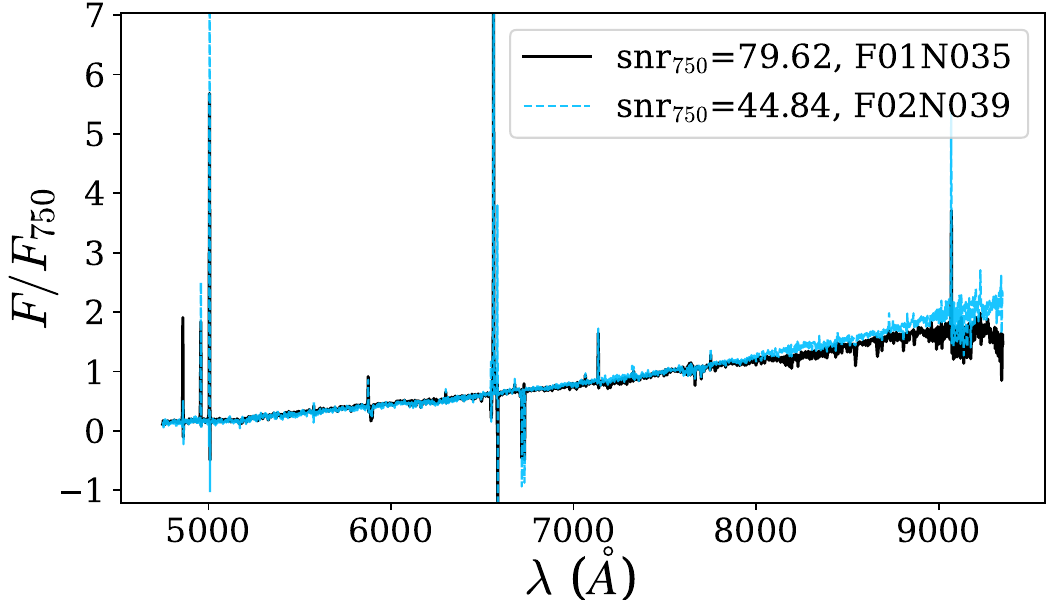}\hspace{-0.1cm}
\includegraphics[width=0.5\columnwidth]{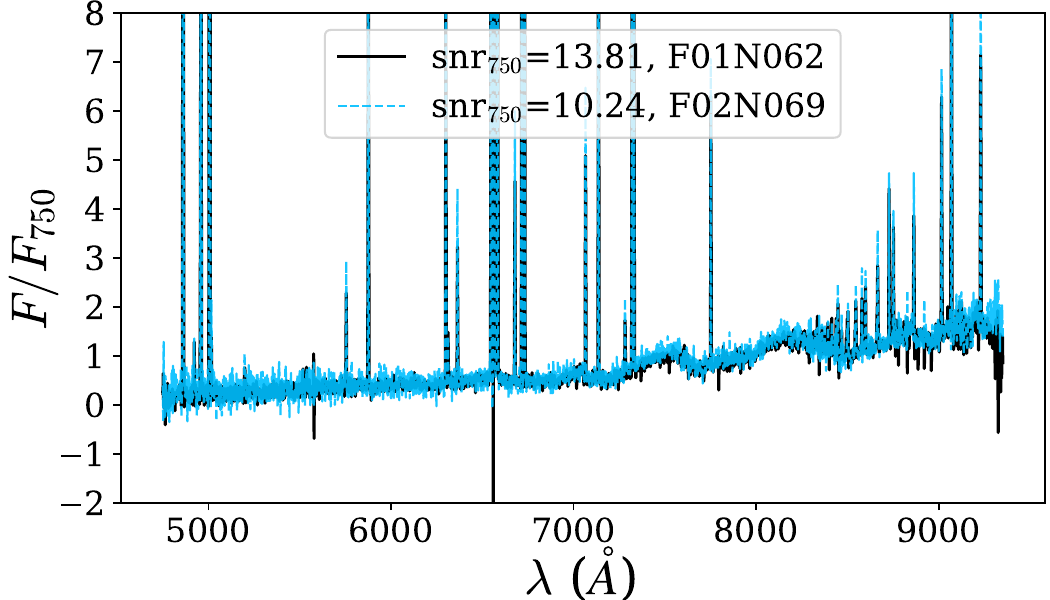}\hspace{-0.1cm}
\includegraphics[width=0.5\columnwidth]{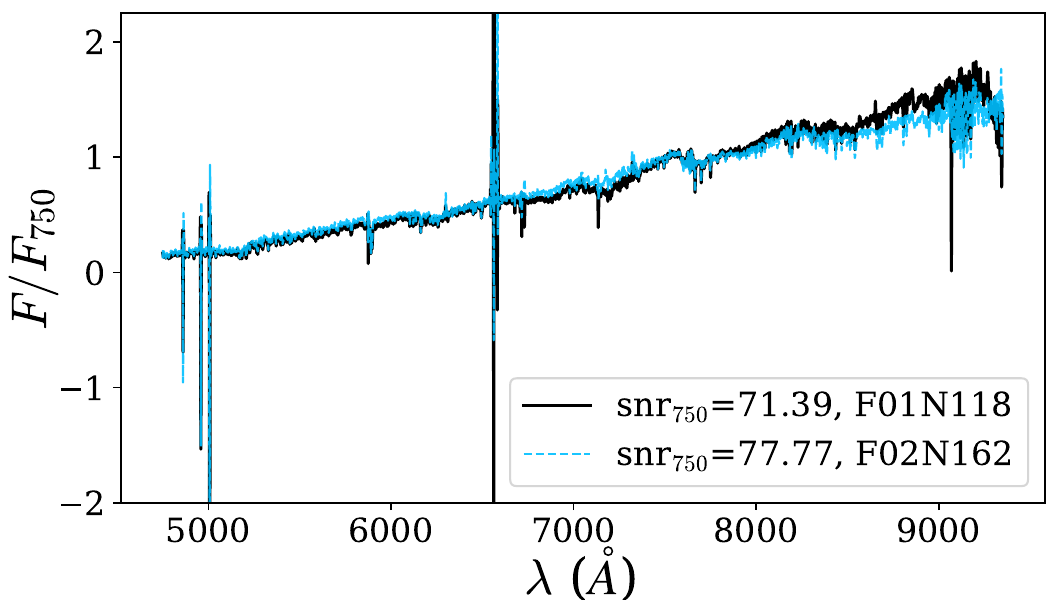}\hspace{-0.1cm}
\includegraphics[width=0.5\columnwidth]{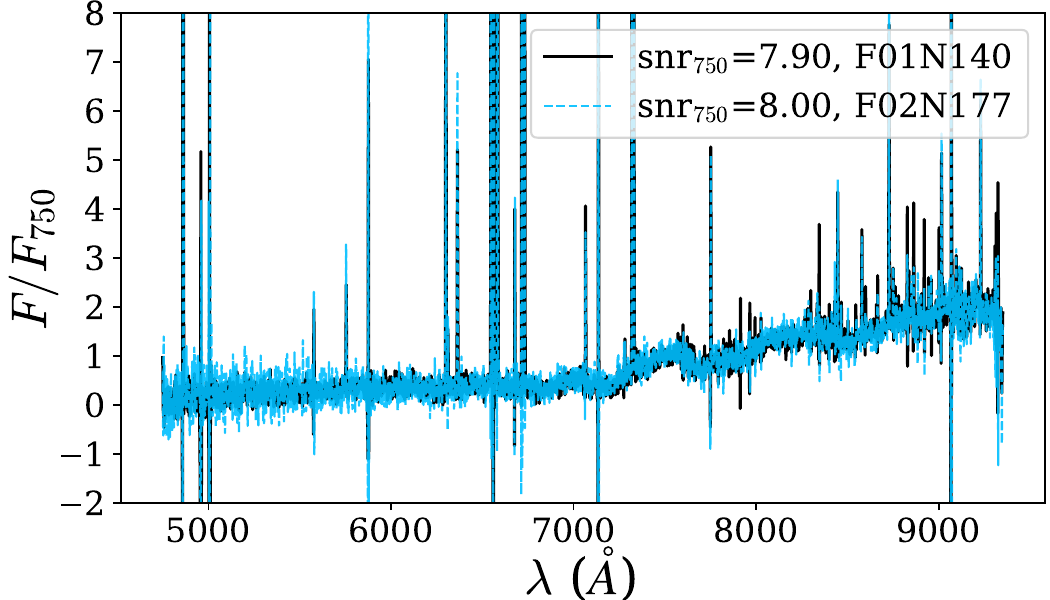}\hspace{-0.1cm}
\includegraphics[width=0.5\columnwidth]{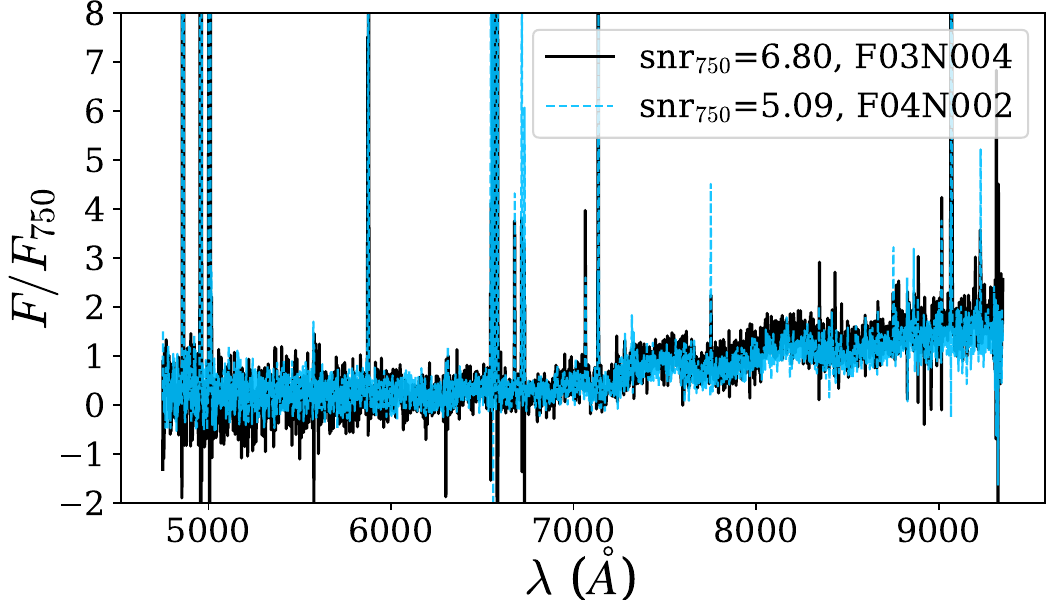}\hspace{-0.1cm}
\caption{Examples of spectra observed twice. Spectra were normalised to the flux at 7500~\AA. Label indicates source identifier and its snr around 7500~\AA\ used to decide which spectrum to keep for the analysis.}
\label{fig:doubles}
\end{figure}

In most of the cases the $I$-band magnitudes of both detections are consistent within the uncertainties. That reassures us about the overall good inter-frame calibration. 
We choose a signal-to-noise ratio in the vicinity of 7500~\AA\ (snr$_{750}$) as a measure of the quality and 
use in the analysis the spectrum with higher snr$_{750}$. In Figure \ref{fig:doubles} we show few examples of comparisons between the spectra of the two doubles. Corresponding snr$_{750}$ are indicated in the legend for each panel. The difference between the spectra is particularly visible in the case of late-type stars. There, the blue part of the spectrum is very vulnerable to the quality of the spectrum. In general, we see no trend between the chosen spectra and weather conditions, although there is a slight preference towards better seeing. We also note that stars located very close to the detector's edge were not detected by the {\tt SExtractor}.

\begin{table}
\centering
\caption{Double detected sources.} 
\label{tab:double} 
\resizebox{\columnwidth}{!}{
\begin{tabular}{ccc|ccc|c}
\hline
\hline
\multicolumn{3}{c|}{Selected double} & \multicolumn{3}{c|}{Discarded double} & \\
\hline
ID & snr$_{750}$ & $I$-band & ID & snr$_{750}$ & $I$-band & separation \\
  & & (mag) &  & & (mag) & (\arcsec) \\ 
\hline
2\_14 & 17.02 & 19.87 $\pm$ 0.02 & 1\_2 & 12.36 & 19.95 $\pm$ 0.02 & 0.22 \\ 
1\_29 & 63.08 & 17.808 $\pm$ 0.002 & 2\_33 & 62.55 & 17.784 $\pm$ 0.003 & 0.20 \\
1\_35 & 79.62 & 17.121 $\pm$ 0.001 & 2\_39 & 44.84 & 17.694 $\pm$ 0.003 & 0.22 \\
1\_55 & 6.53 & 21.09 $\pm$ 0.05 & 2\_66 & 3.64 & 21.07 $\pm$ 0.06 & 0.26 \\
1\_62 & 13.81 & 20.04 $\pm$ 0.02 & 2\_69 & 10.24 & 20.06 $\pm$ 0.02 & 0.20 \\
1\_90 & 8.61 & 20.65 $\pm$ 0.03 & 2\_135 & 7.14 & 20.61 $\pm$ 0.04 & 0.17 \\
2\_162 & 77.77 & 16.459 $\pm$ 0.001 & 1\_118 & 71.39 & 16.763 $\pm$ 0.001 & 0.25 \\
2\_177 & 8.00 & 20.43 $\pm$ 0.03 & 1\_140 & 7.90 & 20.27 $\pm$ 0.02 & 0.18 \\
3\_4 & 6.80 & 20.70 $\pm$ 0.05 & 4\_2 & 5.09 & 20.53 $\pm$ 0.04 & 0.22 \\
3\_84 & 5.18 & 20.75 $\pm$ 0.05 & 4\_45 & 4.27 & 20.78 $\pm$ 0.05 & 0.20 \\
3\_184 & 15.78 & 19.39 $\pm$ 0.02 & 4\_116 & 14.32 & 19.39 $\pm$ 0.01 & 0.22 \\
7\_12 & 18.22 & 19.69 $\pm$ 0.08 & 10\_280 & 11.33 & 19.77 $\pm$ 0.03 & 0.26 \\
8\_23 & 27.06 & 19.34 $\pm$ 0.01 & 7\_33 & 20.22 & 19.46 $\pm$ 0.06 & 0.20 \\
13\_140 & 51.00 & 17.94 $\pm$ 0.01 & 7\_164 & 21.99 & 18.41 $\pm$ 0.02 & 0.17 \\
7\_222 & 20.56 & 19.72 $\pm$ 0.08 & 8\_122 & 8.48 & 19.75 $\pm$ 0.01 & 0.25 \\
8\_127 & 23.12 & 19.11 $\pm$ 0.01 & 7\_237 & 15.69 & 19.12 $\pm$ 0.05 & 0.18 \\
8\_182 & 53.25 & 18.544 $\pm$ 0.004 & 7\_376 & 35.47 & 18.62 $\pm$ 0.03 & 0.22 \\
8\_1 & 6.40 & 20.79 $\pm$ 0.03 & 9\_7 & 5.29 & 20.75 $\pm$ 0.04 & 0.26 \\
10\_48 & 10.51 & 20.41 $\pm$ 0.05 & 9\_19 & 8.79 & 19.98 $\pm$ 0.02 & 0.26 \\
10\_86 & 47.456 & 18.16 $\pm$ 0.01 & 9\_43 & 42.22 & 18.235 $\pm$ 0.004 & 0.22 \\
10\_100 & 15.94 & 19.60 $\pm$ 0.03 & 9\_54 & 14.83 & 19.56 $\pm$ 0.01 & 0.26 \\
10\_126 & 18.29 & 19.45 $\pm$ 0.02 & 9\_67 & 16.14 & 19.45 $\pm$ 0.01 & 0.26 \\
10\_232 & 73.77 & 17.308 $\pm$ 0.003 & 9\_144 & 65.14 & 17.383 $\pm$ 0.002 & 0.17 \\
10\_266 & 15.93 & 19.66 $\pm$ 0.03 & 9\_180 & 11.07 & 19.65 $\pm$ 0.01 & 0.25 \\
16\_39 & 50.09 & 17.51 $\pm$ 0.01 & 10\_22 & 42.31 & 17.560 $\pm$ 0.004 & 0.15 \\
16\_43 & 73.70 & 16.008 $\pm$ 0.002 & 10\_27 & 73.43 & 15.965 $\pm$ 0.001 & 0.24 \\
16\_68 & 34.34 & 18.12 $\pm$ 0.01 & 10\_65 & 28.41 & 18.01 $\pm$ 0.01 & 0.25 \\
10\_72 & 79.62 & 17.696 $\pm$ 0.005 & 16\_79 & 35.85 & 17.84 $\pm$ 0.01 & 0.18 \\
10\_87 & 80.83 & 17.716 $\pm$ 0.005 & 16\_89 & 46.28 & 17.77 $\pm$ 0.01 & 0.20 \\
16\_98 & 20.65 & 18.99 $\pm$ 0.02 & 10\_102 & 19.55 & 18.96 $\pm$ 0.01 & 0.25 \\
10\_140 & 50.56 & 18.02 $\pm$ 0.01 & 16\_149 & 40.08 & 18.07 $\pm$ 0.01 & 0.24 \\
10\_165 & 92.15 & 16.245 $\pm$ 0.001 & 16\_183 & 73.15 & 16.267 $\pm$ 0.002 & 0.25 \\
10\_255 & 28.86 & 18.721 $\pm$ 0.012 & 16\_256 & 25.06 & 18.72 $\pm$ 0.02 & 0.18 \\
12\_67 & 89.56 & 14.873 $\pm$ 0.001 & 11\_76 & 85.15 & 14.9120 $\pm$ 0.0002 & 0.20 \\
11\_104 & 7.77 & 20.37 $\pm$0.03 & 17\_99 & 5.86 & 20.47 $\pm$ 0.06 & 0.18 \\
13\_3 & 28.83 & 18.50 $\pm$ 0.01 & 16\_284 & 27.22 & 18.11 $\pm$ 0.01 & 0.21 \\
13\_4 & 6.61 & 20.65 $\pm$ 0.07 & 16\_286 & 5.07 & 20.41 $\pm$ 0.09 & 0.23 \\
16\_6 & 96.93 & 16.546 $\pm$ 0.002 & 13\_6 & 88.40 & 16.39 $\pm$ 0.001 & 0.22 \\
13\_29 & 11.36 & 20.26 $\pm$ 0.05 & 14\_53 & 10.03 & 20.36 $\pm$ 0.03 & 0.24 \\
14\_140 & 11.75 & 19.98 $\pm$ 0.02 & 13\_144 & 10.61 & 20.12 $\pm$ 0.04 & 0.19 \\
14\_301 & 16.92 & 19.78 $\pm$ 0.02 & 13\_361 & 8.05 & 19.28 $\pm$ 0.02 & 0.14 \\
16\_2 & 69.07 & 16.325 $\pm$ 0.002 & 15\_5 & 47.33 & 16.138 $\pm$ 0.001 & 0.15 \\
16\_57 & 13.90 & 19.47 $\pm$ 0.04 & 15\_59 & 5.67 & 19.49 $\pm$ 0.03 & 0.18 \\
16\_241 & 45.60 & 18.11 $\pm$ 0.01 & 15\_350 & 23.01 & 17.86 $\pm$ 0.01 & 0.09 \\
16\_270 & 14.70 & 19.66 $\pm$ 0.04 & 15\_429 & 4.96 & 19.39 $\pm$ 0.02 & 0.15 \\
18\_138 & 11.89 & 20.08 $\pm$ 0.03 & 17\_87 & 10.84 & 20.02 $\pm$ 0.04 & 0.21 \\
18\_177 & 37.29 & 18.65 $\pm$ 0.01 & 17\_112 & 24.83 & 18.57 $\pm$ 0.01 & 0.15 \\
18\_246 & 10.48 & 20.34 $\pm$ 0.04 & 22\_142 & 6.51 & 20.10 $\pm$ 0.06 & 0.23 \\
20\_23 & 6.83 & 20.42 $\pm$ 0.05 & 19\_24 & 5.19 & 20.36 $\pm$ 0.09 & 0.17 \\
20\_35 & 23.49 & 19.61 $\pm$ 0.02 & 19\_33 & 10.86 & 19.54 $\pm$ 0.04 & 0.22 \\
20\_65 & 16.53 & 19.69 $\pm$ 0.02 & 19\_60 & 11.32 & 19.59 $\pm$ 0.05 & 0.16 \\
20\_80 & 15.67 & 19.55 $\pm$ 0.02 & 19\_76 & 11.60 & 19.54 $\pm$ 0.04 & 0.16 \\
19\_84 & 53.84 & 17.62 $\pm$ 0.01 & 20\_95 & 51.17 & 17.623 $\pm$ 0.004 & 0.24 \\
20\_113 & 68.79 & 17.776 $\pm$ 0.004 & 19\_100 & 40.30 & 17.82 $\pm$ 0.01 & 0.13 \\
21\_22 & 6.49 & 20.42 $\pm$ 0.02 & 22\_14 & 6.15 & 20.38 $\pm$ 0.08 & 0.19 \\
\hline
\end{tabular}}
\tablefoot{Table is separated into two part: left part lists ID, snr at 750~nm, and $I$-band magnitude for measurements used in the analysis, while right part lists the same properties of the discarded measurements. Magnitudes are corrected to match those from WFI catalogue \citep{beccari2015}. Additionally, last column shows separation between the two measurements. One MUSE pixel has width of 0.2\arcsec.}
\end{table}

\section{Assessment of the background variability}
\label{app:bkg}
\indent

\begin{figure}
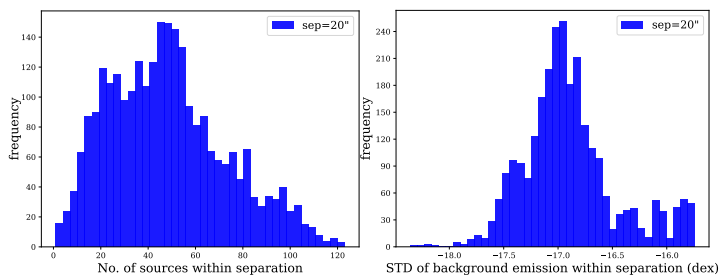
\hspace{-0.2cm}
\includegraphics[width=.252\textwidth, trim={0.1cm 0.cm 0.2cm 0.cm}, clip]{figures/No\_of\_sources-separation.pdf}%
\includegraphics[width=.256\textwidth, trim={0.cm 0.0cm 0.08cm 0.cm}, clip]{figures/STD\_bkg-separation.pdf}
\caption{{Left:} distribution of number of sources within a separation of 20\arcsec from a given star. {Right:} distribution of the standard deviation (std) of background emission estimated in $I$-band from the individual measurements within 20\arcsec\ from the given star.}
\label{fig:bkg-sep20}
\end{figure}

%%%%%%% Table 3 %%%%%%
\begin{table*}[ht]
\caption{Parameters of stars removed from the catalogue due to the high background variation.} 
\label{tab:highbkgcat} 
\resizebox{\textwidth}{!}{
\begin{tabular}{ccccccccccccccccccccccccc}
\hline
\hline
ID & coordinates & $I$-band & $R$-band & $V$-band & snr$_{I,\mathrm{bkg}}$ & snr$_{R,\mathrm{bkg}}$ & snr$_{V,\mathrm{bkg}}$ &  possible\_frg\_bkg & gaia\_flag & wfi\_flag & VISTA\_flag & hawki\_flag & spitzer\_flag & chandra\_flag & NIR excess & SpT & $T_{\mathrm{eff}}$ & \Av & \veil & $\log{(L_{\mathrm{bol}})}$ & $M_{\mathrm{*,PARSEC}}$ & $M_{\mathrm{*,B15S00}}$ & Age$_{\mathrm{PARSEC}}$ & Age$_{\mathrm{B15S00}}$ \\  
 & (h:m:s ~d:m:s) & (mag) & (mag) & (mag) & & & & & & & & & & & &  & (K) & (mag) &  & $L_\odot$ & ($M_\odot$) & ($M_\odot$) & (Myr) & (Myr) \\ 
\hline
 F01N008 & 10:44:07.40 -59:29:11.60 & 19.29$\pm$0.01 & 20.80$\pm$0.04 & 22.01$\pm$0.04 & 2.42 & 0.36 & 0.37 & False & Poor & True & True & True & False & False & False & M3.0$^{+0.6}_{-0.0}$ & 3415$_{-89}^{+0}$ & 0.00$_{-0.00}^{+0.17}$ & 0.02$_{-0.00}^{+0.05}$ & -1.09$_{-0.22}^{+0.17}$ & 0.60 & 0.33 & 23.6 & 6.8 \\ 
 F01N011 & 10:44:11.86 -59:29:12.11 & 20.94$\pm$0.03 & -- & -- &  1.10 & -- & -- & False & None  &  False & True & True & False & False & True & M4.0$^{+3.5}_{-1.7}$ & 3270$_{-475}^{+248}$ & 3.50$_{-0.95}^{+1.02}$ & 0.00$_{-0.00}^{+0.58}$ & -0.99$_{-1.75}^{+0.52}$ & 0.48 & 0.23 & 8.9 & 2.3 \\ 
\hline
\end{tabular}}
\tablefoot{The first column gives IDs of the detected sources, the second -- coordinates. The third, fourth, and fifth columns give apparent magnitudes in $I$, $R$, and $V$-band, respectively. The sixth, seventh, and eighth columns -- signal-to-noise of the flux with respect to the background variation in a given band, as indicated by the lower script (see Sec. \ref{subsec:bkg} for details).  
The ninth marks 
possible foreground or background stars (see Sec. \ref{subsec:frg} for definitions). 
The next five columns flag matches with other catalogs: {\it Gaia} \citep{gaia-vallenari2022}, WFI \citep{beccari2015}, VISTA \citep{preibisch2014}, HAWK-I \citep{preibisch2011a,preibisch2011b}, {\it Spitzer} \citep{povich2011}, and {\it Chandra} \citep{preibisch2011a,townsley2011}. The following indicates if the star has an NIR excess as defined by \cite{zeidler2016}. In the consecutive {nine} columns are given stellar parameters: spectral type, effective temperature, visual extinction, constant veiling at 7500~\AA, bolometric luminosity, {and stellar mass and stellar age estimated from PARSEC \citep{bressan2012} and \cite{baraffe2015} / \cite{siess2000} tracks, as indicated by the subscript}. 
A full version of this table will be available at the CDS upon publication. The first few rows are shown as an example.}
\end{table*}

{\tt SExtractor} provides estimates of stellar fluxes, magnitudes, and sky emission at the positions of the stars. It assumes smooth variation of the sky emission across the whole image. The tool does not provide uncertainty of the background estimation. Since some parts of the cluster covered by our observations are very crowded and sky emission exhibits prominent gaseous structures, local variation of the sky emission might not be smooth. Therefore, we employ another strategy to assess the quality of stellar magnitudes and spectra.

We look at the variation of the background emission estimated for each target by {\tt SExtractor} within the defined area around each star. We checked that the circle with the radius of 20\arcsec\ is a large area enough to cover satisfactory number of neighbouring sources, and at the same time, small enough to cover only \lq\lq local\rq\rq\ area. Left panel of Figure \ref{fig:bkg-sep20} shows how many neighbouring sources for each star is within the radius 20\arcsec. The distribution peaks at 40-50 neighbours giving a satisfactory large statistics. We calculate the standard deviation (std) of the sky emission within this area for every star in our catalogue. Right panel of the Fig. \ref{fig:bkg-sep20} shows the distribution of the std measured in flux units of erg s$^{-1}$ cm$^{-2}$ \AA$^{-1}$. We use such defined variability to select a robust photometry: we discard measurement of the stellar flux that is below the threshold of three sigma (here $\sigma$=std). We initially perform this exercise in $I$-band, as the presence of this magnitude is our definition of detection, and repeat for $R$ and $V$-bands. Since our stars have late spectroscopic types, they appear fainter in bluer bands, therefore, there are fewer photometric measurements in $R$ and $V$-bands than in the $I$-band. That can be noticed in our CMDs in Fig. \ref{fig:CMD} and \ref{fig:CMDdered}. We note that our approach is very conservative: spectra of some of the removed targets from the final catalogue have high enough snr for spectral classification. For this reason, we list in Table \ref{tab:highbkgcat} sources removed due to the uncertain photometry in $I$-band caused by the high background variation. We include in the Table uncertain photometry from all bands. The columns \lq\lq snr$_{I,\rm bkg}$\rq\rq\, \lq\lq snr$_{R,\rm bkg}$\rq\rq\, and \lq\lq snr$_{V,\rm bkg}$\rq\rq\ show ratios between stellar flux and background variation in a given band, and therefore can be used as indicators of photometric certainty.

\section{MUSE photometry}
\label{app:mag}
\indent

\begin{figure*}\hspace{-0.2cm}
\includegraphics[width=0.5\columnwidth, trim={0 0.2cm 0cm 0.cm}, clip ]{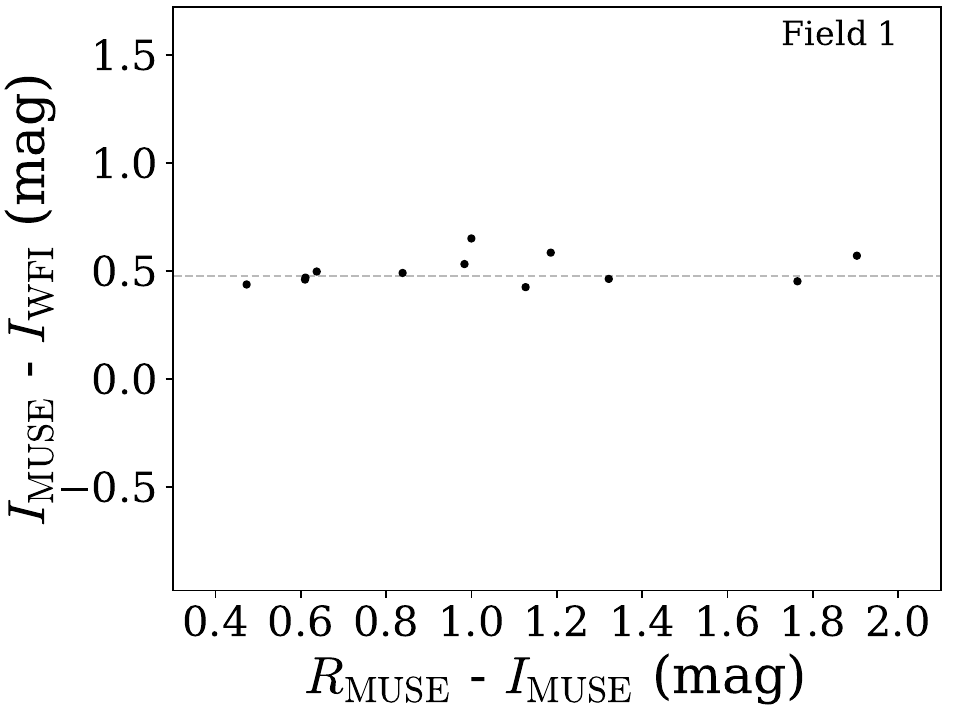}\hspace{-0.1cm}
\includegraphics[width=0.5\columnwidth, trim={0 0.2cm 0cm 0.cm}, clip ]{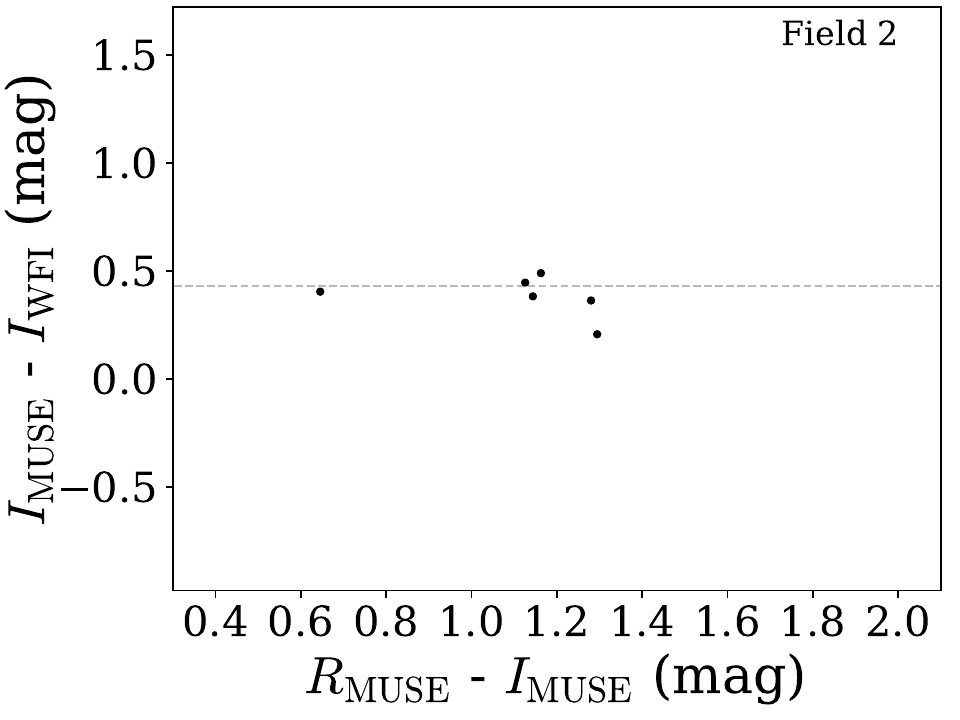}\hspace{-0.1cm}
\includegraphics[width=0.5\columnwidth, trim={0 0.2cm 0cm 0.cm}, clip ]{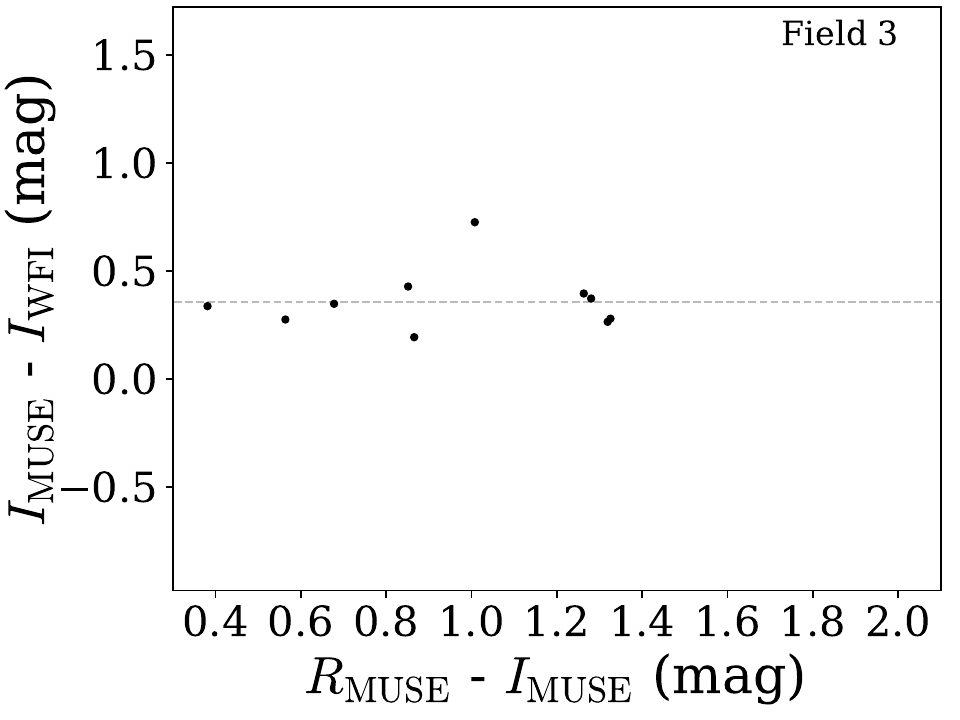}\hspace{-0.1cm}
\includegraphics[width=0.5\columnwidth, trim={0 0.2cm 0cm 0.cm}, clip ]{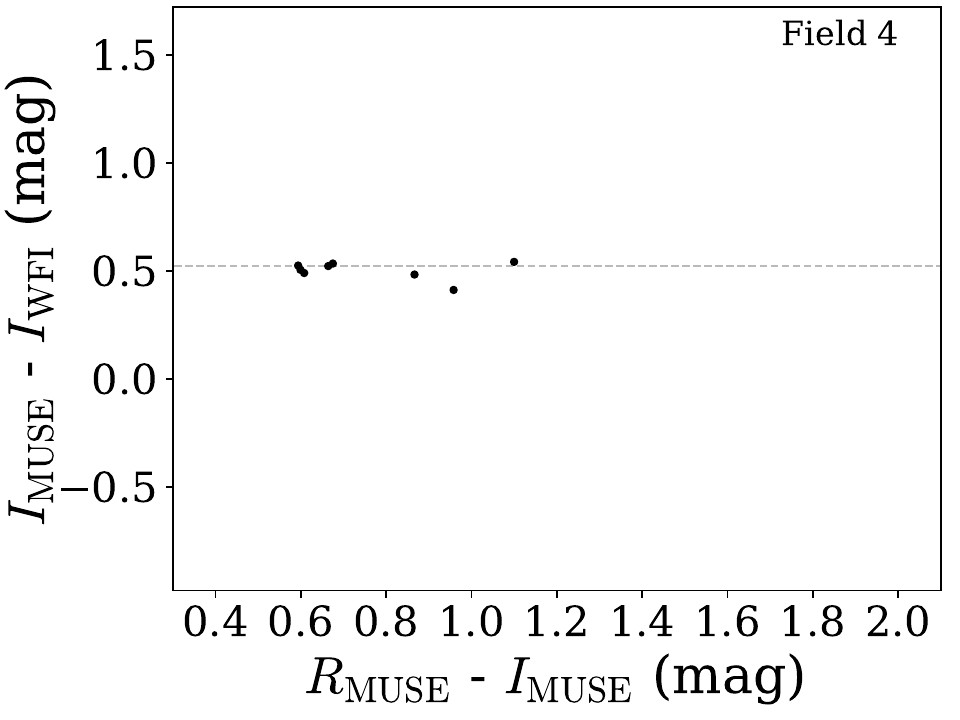}\hspace{-0.1cm}
%%%%%%%%%%%%%%%%%%%%%%%%%%
\hspace{-0.2cm}\includegraphics[width=0.5\columnwidth, trim={0 0.2cm 0cm 0.cm}, clip ]{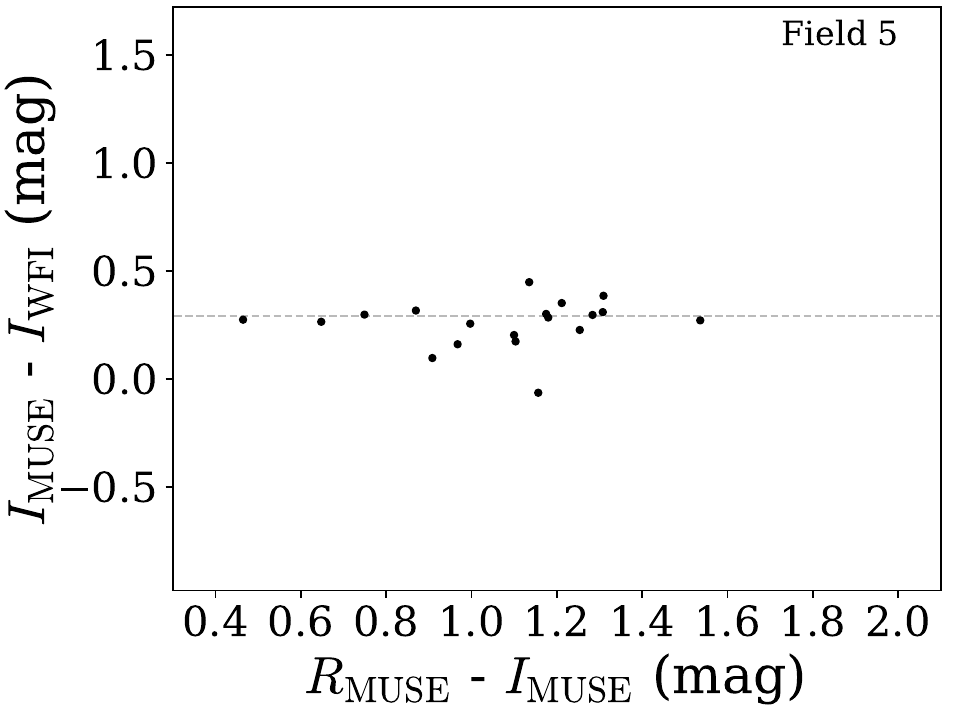}\hspace{-0.1cm}
\includegraphics[width=0.5\columnwidth, trim={0 0.2cm 0cm 0.cm}, clip ]{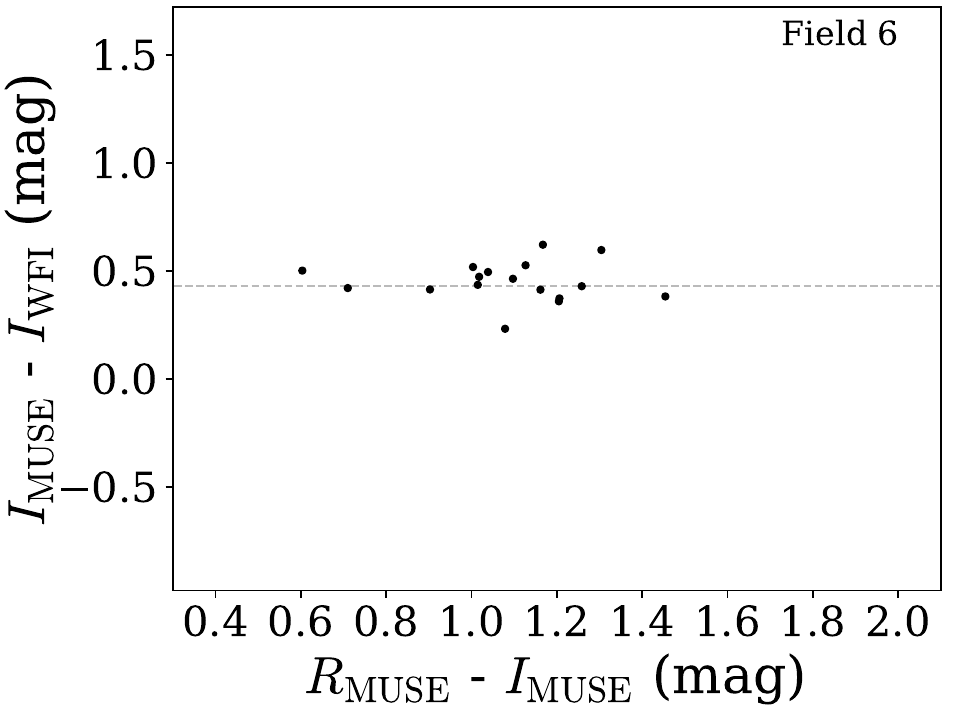}\hspace{-0.1cm}
\includegraphics[width=0.5\columnwidth, trim={0 0.2cm 0cm 0.cm}, clip ]{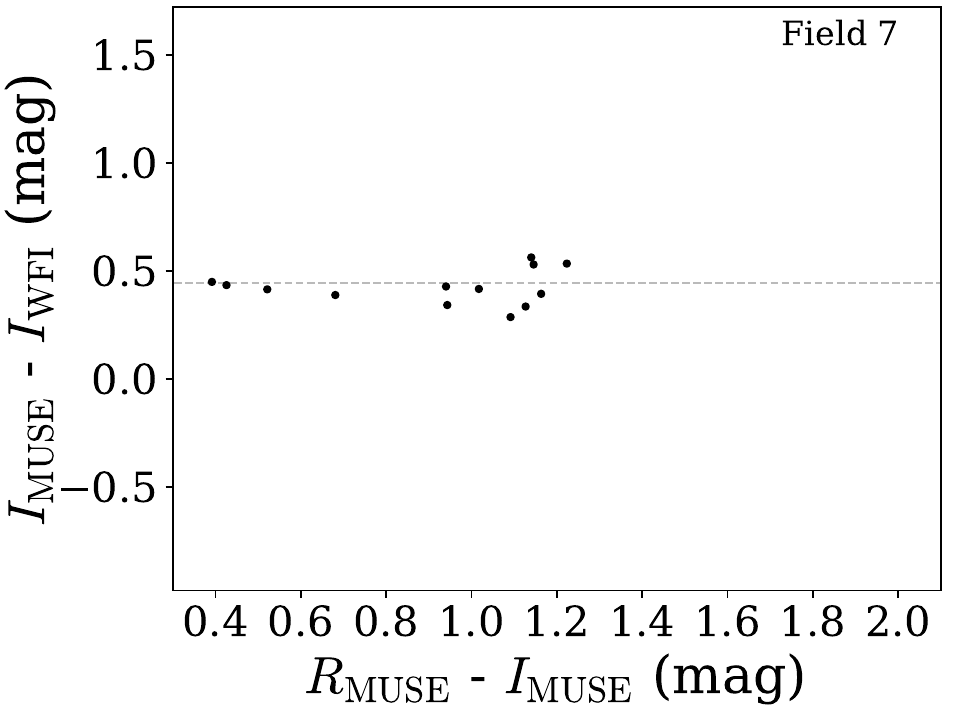}\hspace{-0.1cm}
\includegraphics[width=0.5\columnwidth, trim={0 0.2cm 0cm 0.cm}, clip ]{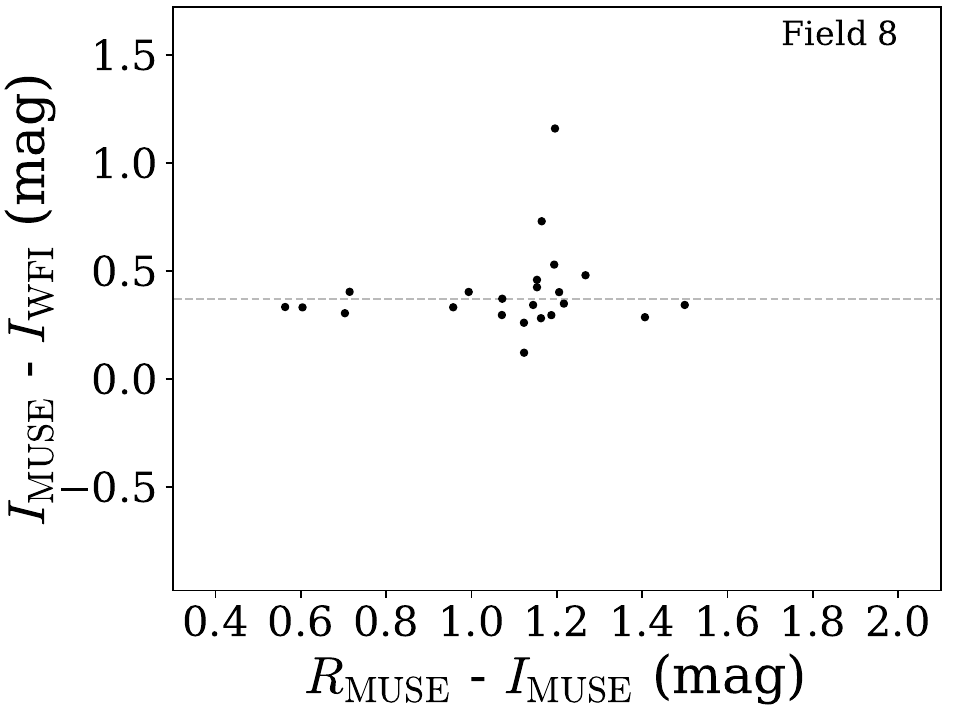}\hspace{-0.1cm}
%%%%%%%%%%%%%%%%%%%%%%%%%
\hspace{-0.2cm}\includegraphics[width=0.5\columnwidth, trim={0 0.2cm 0cm 0.cm}, clip ]{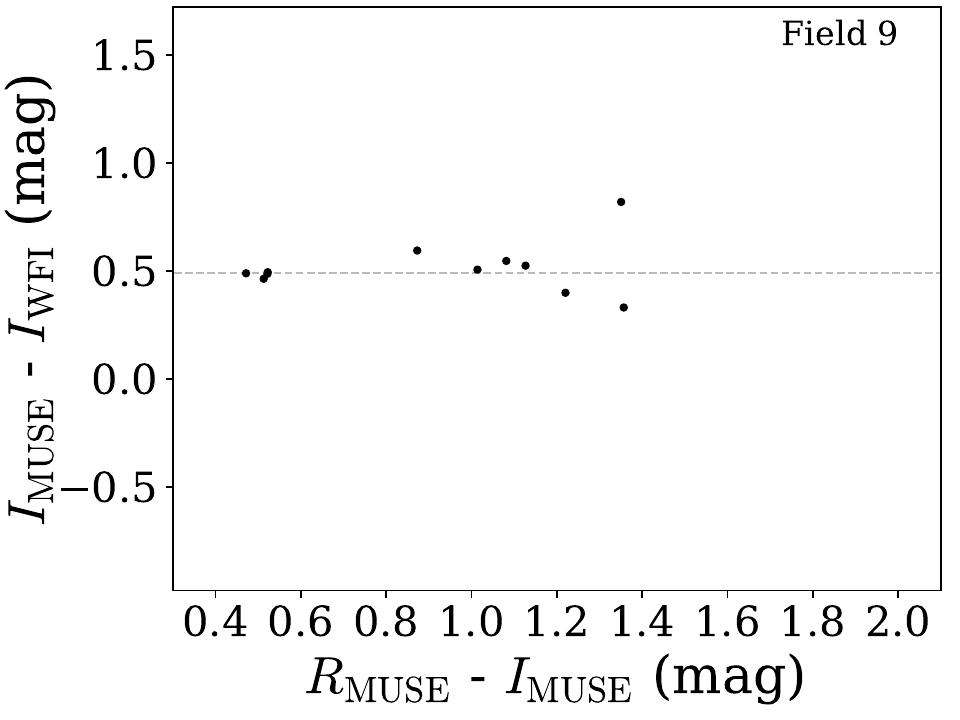}\hspace{-0.1cm}
\includegraphics[width=0.5\columnwidth, trim={0 0.2cm 0cm 0.cm}, clip ]{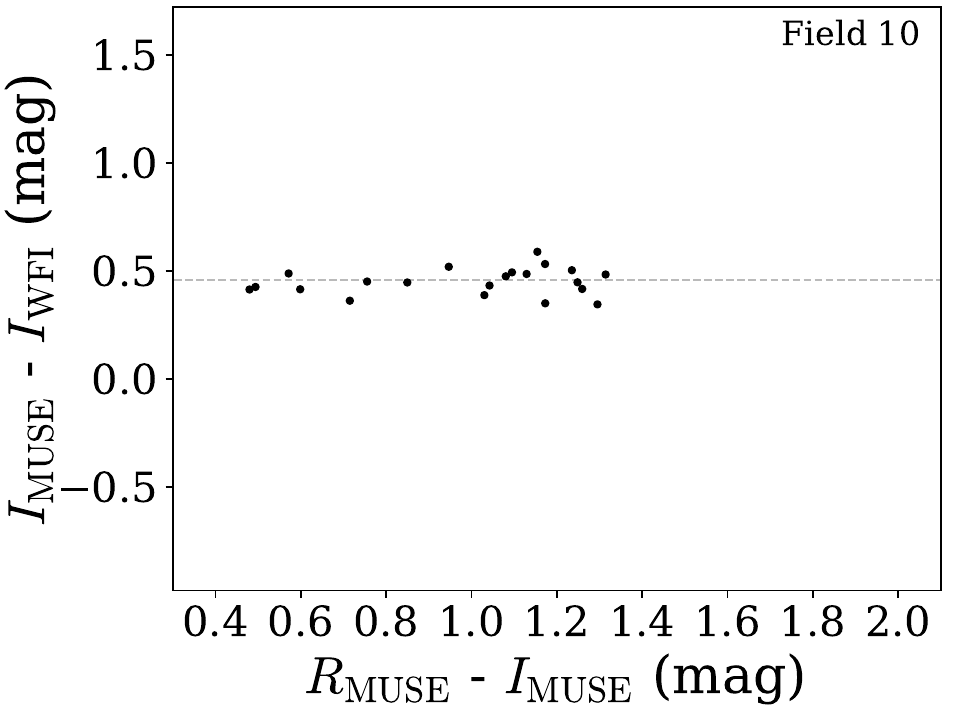}\hspace{-0.1cm}
\includegraphics[width=0.5\columnwidth, trim={0 0.2cm 0cm 0.cm}, clip ]{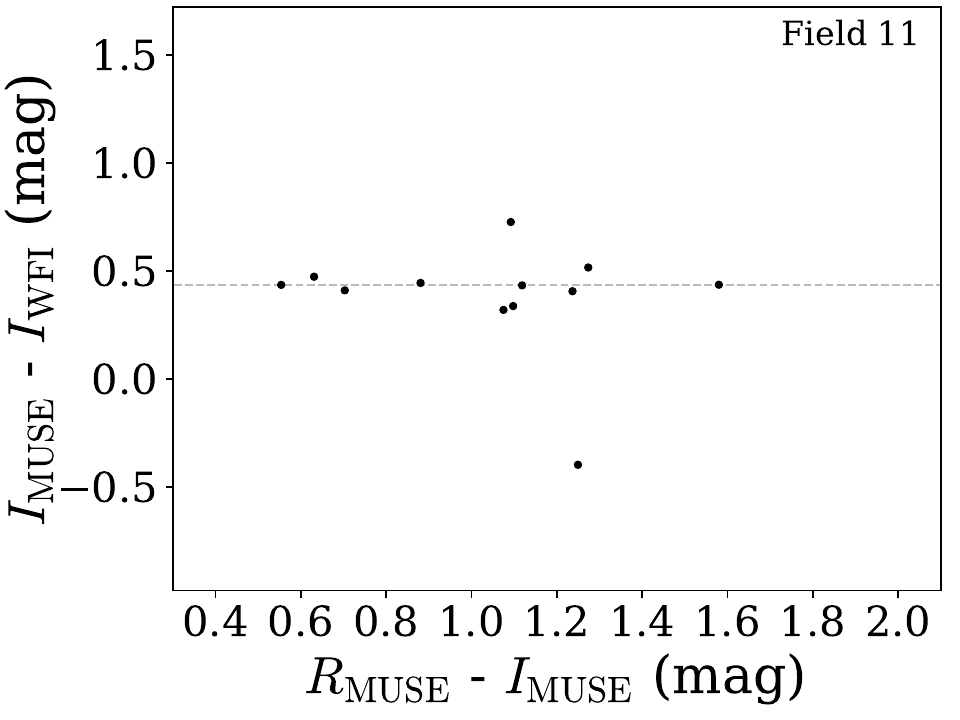}\hspace{-0.1cm}
\includegraphics[width=0.5\columnwidth, trim={0 0.2cm 0cm 0.cm}, clip ]{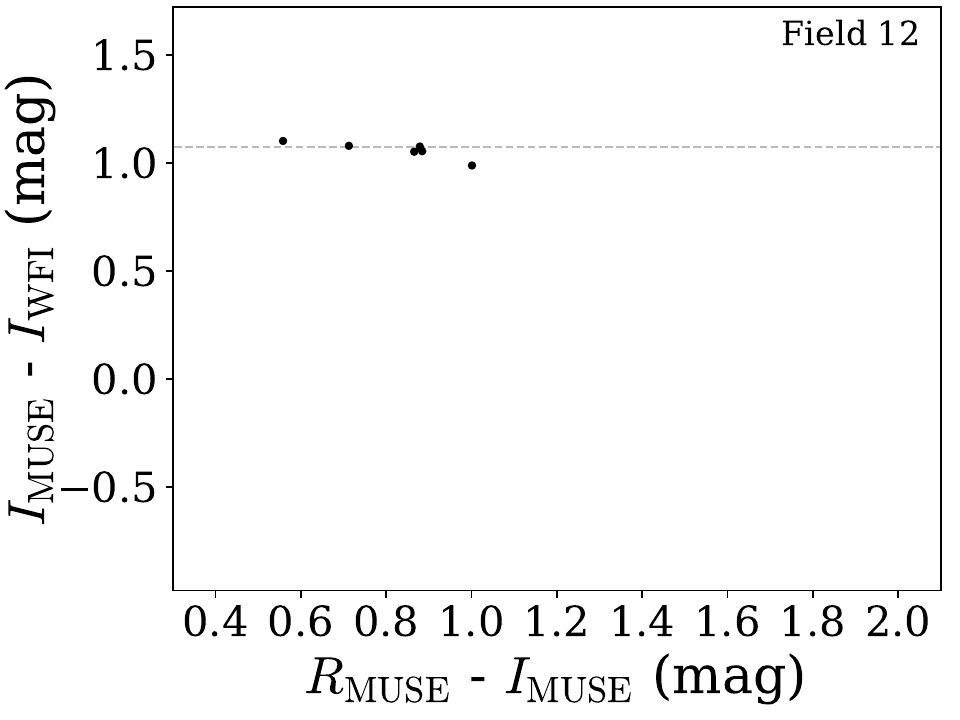}\hspace{-0.1cm}
%%%%%%%%%%%%%%%%%%%%%%%%%%%
\hspace{-0.2cm}\includegraphics[width=0.5\columnwidth, trim={0 0.2cm 0cm 0.cm}, clip ]{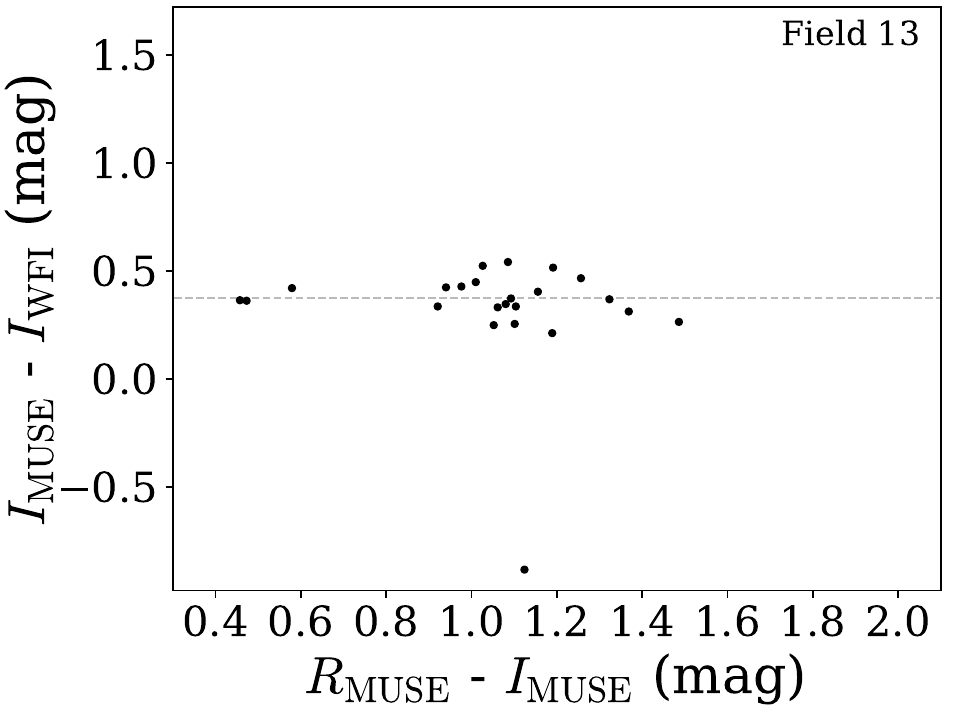}\hspace{-0.1cm}
\includegraphics[width=0.5\columnwidth, trim={0 0.2cm 0cm 0.cm}, clip ]{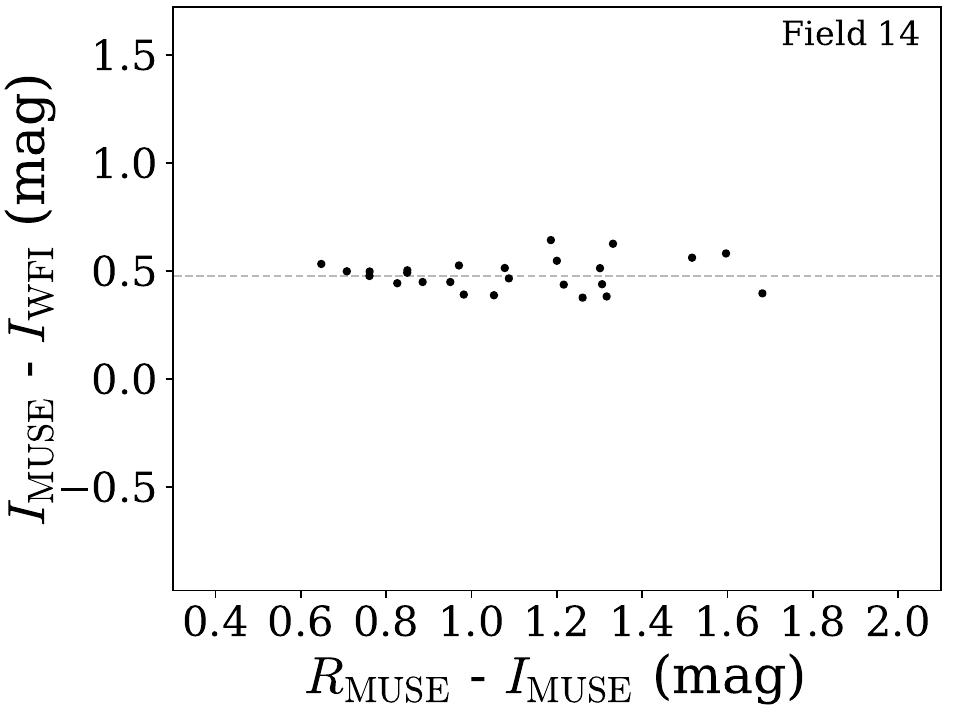}\hspace{-0.1cm}
\includegraphics[width=0.5\columnwidth, trim={0 0.2cm 0cm 0.cm}, clip ]{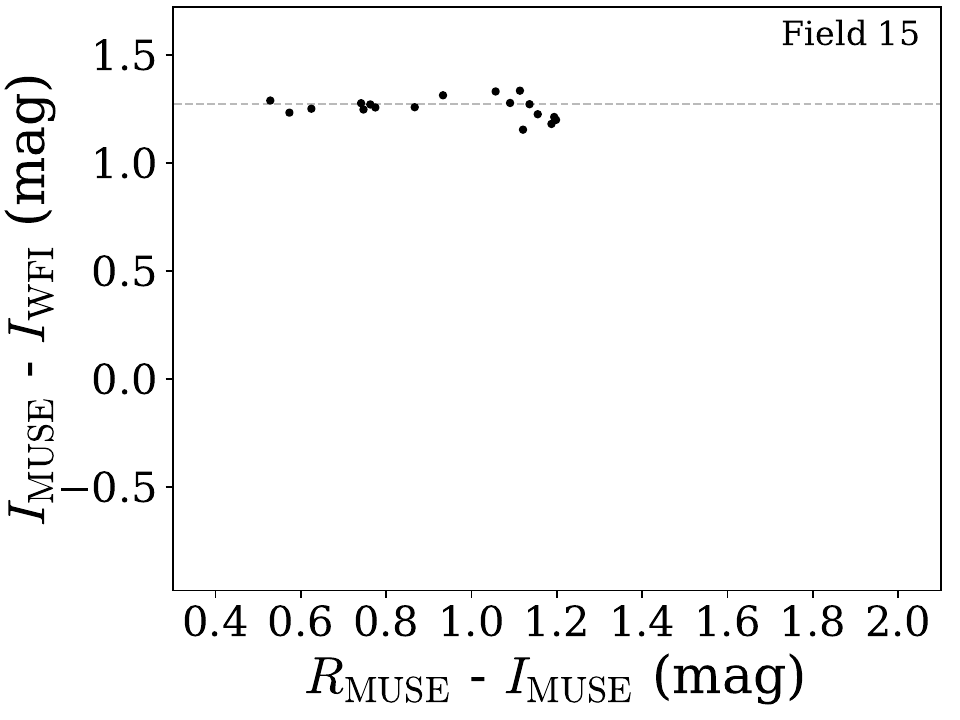}\hspace{-0.1cm}
\includegraphics[width=0.5\columnwidth, trim={0 0.2cm 0cm 0.cm}, clip ]{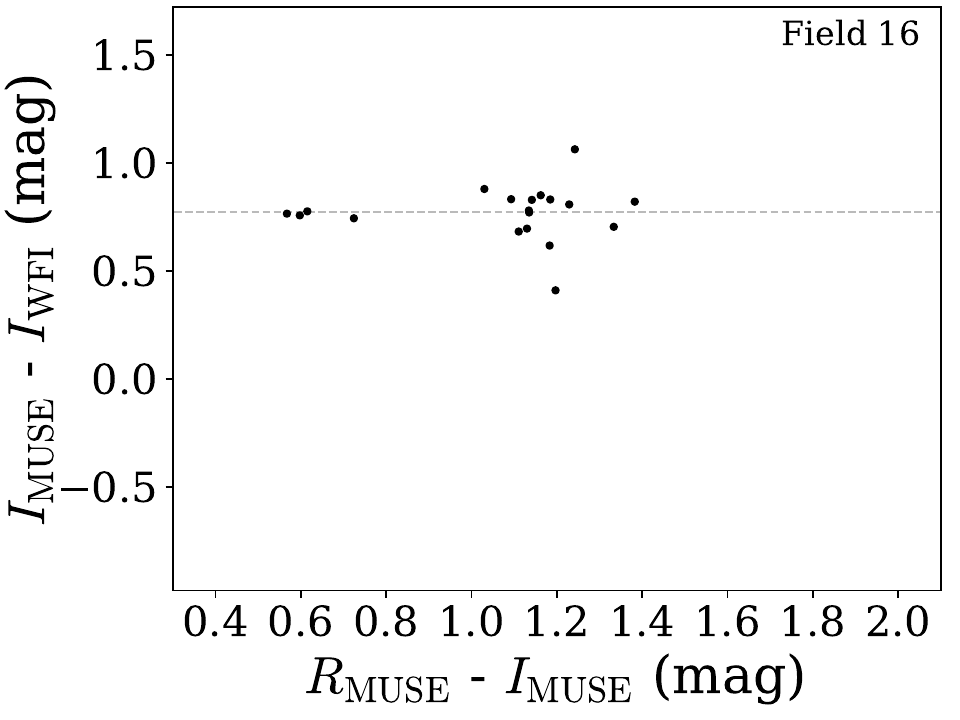}\hspace{-0.1cm}
%%%%%%%%%%%%%%%%%%%%%%%%%
\hspace{-0.2cm}\includegraphics[width=0.5\columnwidth, trim={0 0.2cm 0cm 0.cm}, clip ]{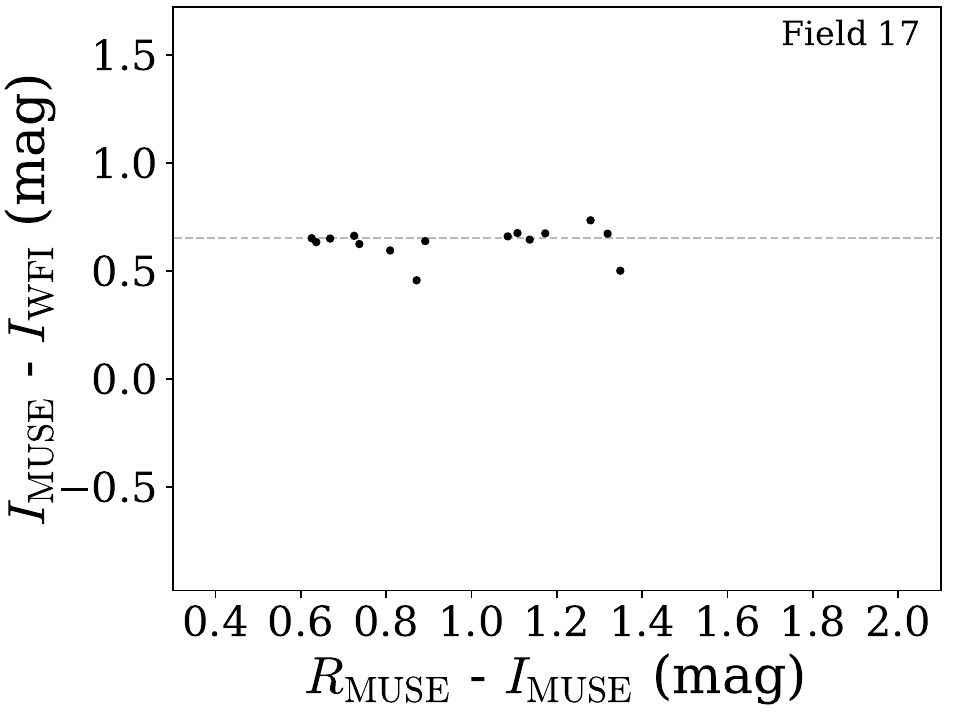}\hspace{-0.1cm}
\includegraphics[width=0.5\columnwidth, trim={0 0.2cm 0cm 0.cm}, clip ]{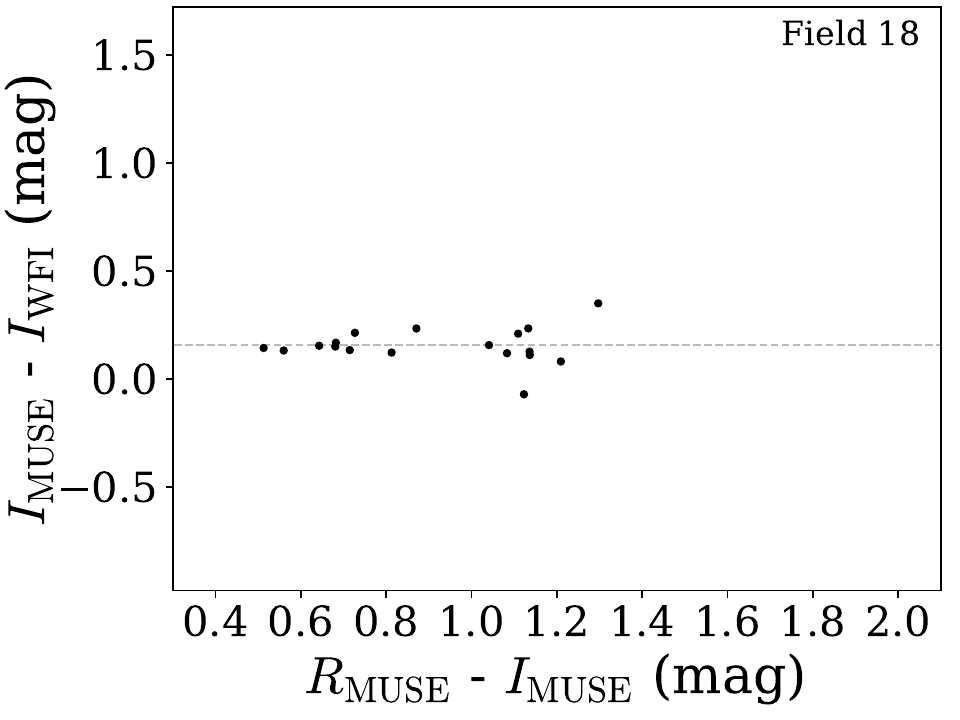}\hspace{-0.1cm}
\includegraphics[width=0.5\columnwidth, trim={0 0.2cm 0cm 0.cm}, clip ]{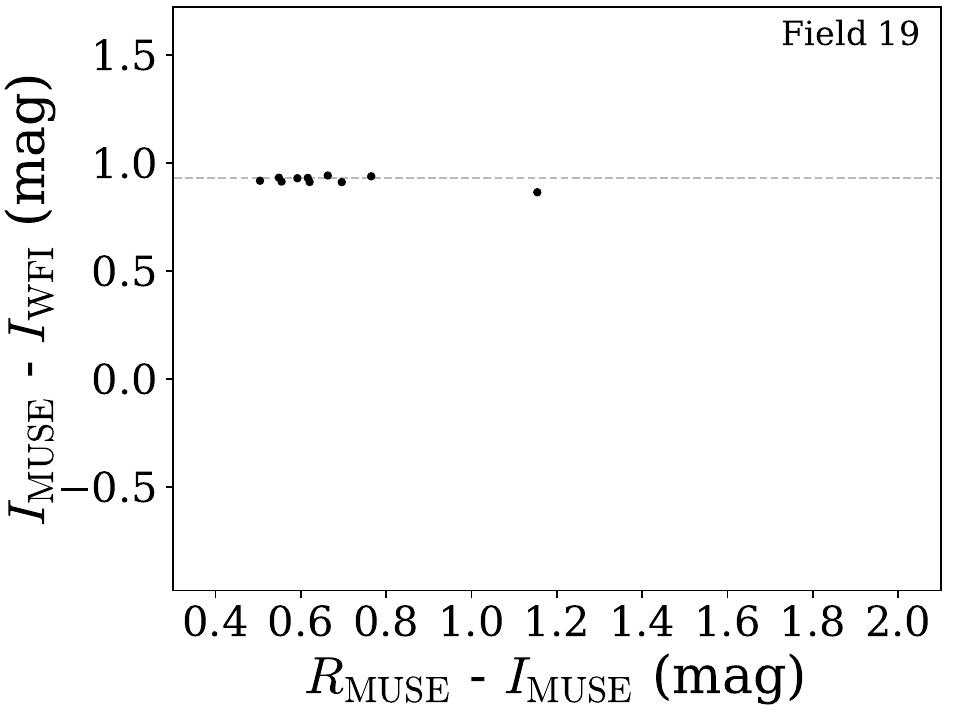}\hspace{-0.1cm}
\includegraphics[width=0.5\columnwidth, trim={0 0.2cm 0cm 0.cm}, clip ]{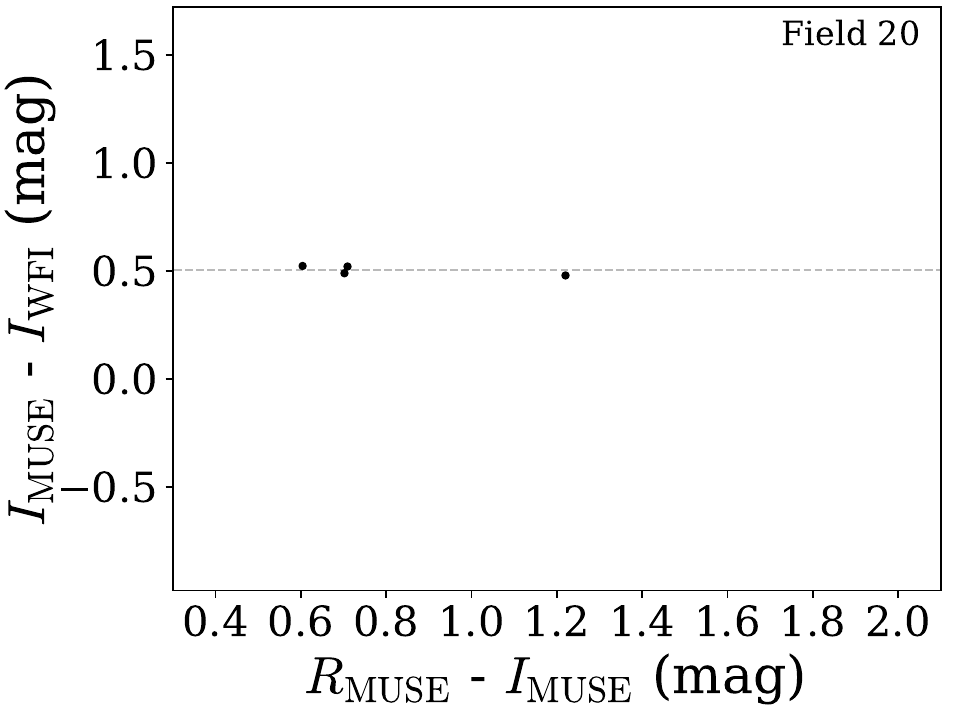}\hspace{-0.1cm}
%%%%%%%%%%%%%%%%%%%%%%
\hspace{-0.2cm}\includegraphics[width=0.5\columnwidth, trim={0 0.2cm 0cm 0.cm}, clip ]{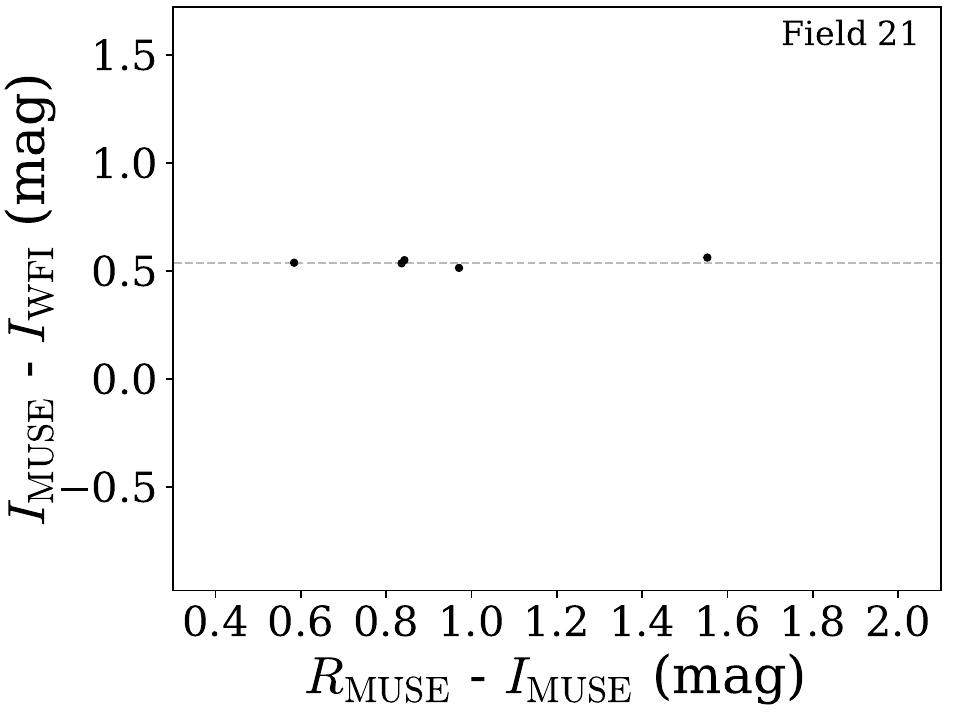}\hspace{-0.1cm}
\includegraphics[width=0.5\columnwidth, trim={0 0.2cm 0cm 0.cm}, clip ]{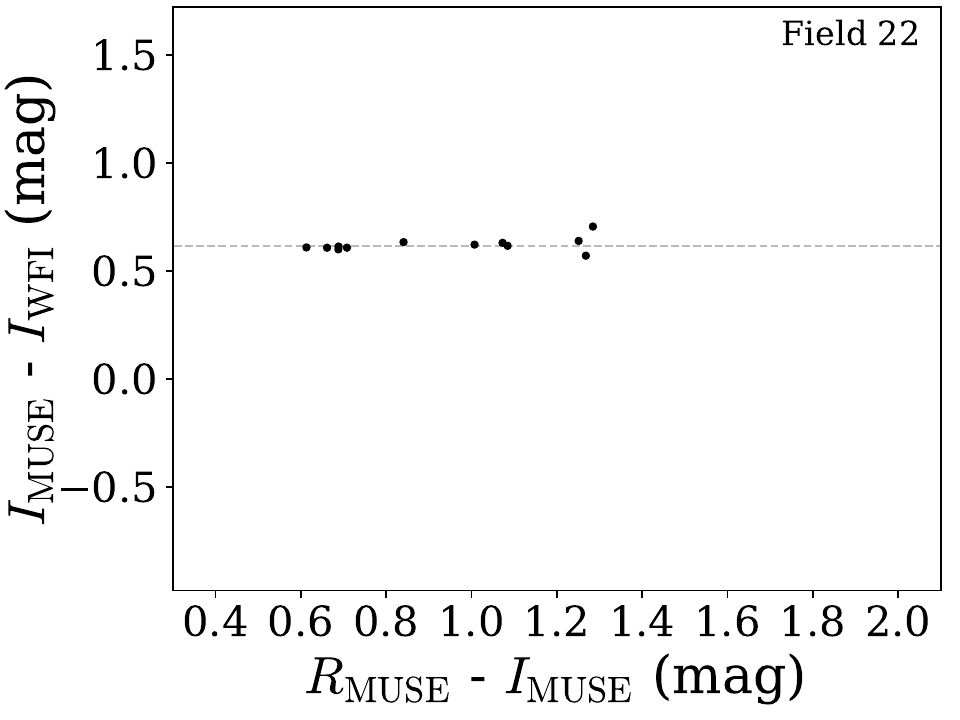}\hspace{-0.1cm}
\caption{Difference between MUSE and WFI \citep{beccari2015} $I$-band photometry as a function of MUSE colour $R-I$. Every panel represents separate field, as indicated in the upper right corner. Grey dashed line shows mean difference between $I$-band magnitudes applied to MUSE photometry as magnitude correction.}
\label{fig:colour-term}
\end{figure*}

The MUSE $I$, $R$, and $V$-band images in flux units were produced by collapsing the 3D MUSE cubes within the corresponding wavelength range and applying the filter transmission curve embedded in the ESO reduction pipeline \citep{weilbacher2020}. We performed the aperture photometry with {\tt SExtractor} extracting stellar fluxes from the images and converting them to magnitudes using Vega zero points written in the headers by the pipeline. Even though our observations were flux calibrated using standard stars, we found that our magnitudes deviate from those measured with Wide Field Imager by \cite{beccari2015} (described below). As those measurements are well calibrated, we correct MUSE magnitudes so they match those from WFI. We define a correction as a difference between MUSE and WFI magnitudes and subtract it from MUSE photometry. Corrections for each field and each band are listed in the Table  \ref{tab:phot-corr}. In $I$-band they range from 0.29~mag in field No. 5 to 1.27~mag in field No. 15. We also check whether a colour term is present in MUSE photometry. Fig. \ref{fig:colour-term} shows an example of this examination with the result of no colour term between $I$ and $R$-bands.   

The highest values of corrections are in fields No. 15, 12, and 19 ($\gtrsim1.0$~mag). Pointings No. 12 and 19 have the worst seeing from all used observations, which may explain the difference in the estimated flux. No. 12 suffers additionally from the presence of two very bright stars in the center of the field, whose brightness impact all neighbouring stars in the image, possibly to a larger extent than assumed in this work. No. 15 is one of the most crowded pointings, although it does not cover the very center of the \tr. The presence of few bright stars and prominent extincted feature in the lower right corner of the field might be another explanation of the large magnitude difference. If, as the result, the background estimation from the {\tt SExtractor} is incorrect, that would lead to the uncertain stellar photometry. Weather conditions were moderate (thin clouds) which might also have affected the observations. 

The uncertainty of magnitude corrections, measured as a standard deviation of magnitude differences, is $\sim$0.1~mag for most of the pointings. It also usually increases towards bluer bands. However, smaller number of available magnitudes in bluer bands due to the high background variability can cause underestimation of those uncertainties. 
Overall, the offset of the MUSE magnitudes seem to be relatively constant within the pointing. We add linearly correction uncertainties to the magnitude uncertainties of our sources and report them in the catalogues (Tab. \ref{tab:cat} \& \ref{tab:highbkgcat}).

\begin{table}%[h!]
\centering
\caption{Photometric corrections.} 
\label{tab:phot-corr} 
%\resizebox{\columnwidth}{!}{
\begin{tabular}{cccc}
\hline
\hline
\# & $I$ & $R$ & $V$  \\ 
  & (mag) & (mag) & (mag)  \\ 
\hline
1  &	0.47 $\pm$ 0.09 &	0.57 $\pm$ 0.08 &	0.58 $\pm$ 0.08	\\
2  &	0.43 $\pm$ 0.11 &	0.42 $\pm$ 0.12 &	0.41 $\pm$ 0.15	\\
3  &	0.36 $\pm$ 0.12 &	0.34 $\pm$ 0.18 &	0.37 $\pm$ 0.21	\\
4  &	0.52 $\pm$ 0.06 &   0.58 $\pm$ 0.07 &	0.55 $\pm$ 0.07	\\
5  &	0.29 $\pm$ 0.22 &	0.26 $\pm$ 0.12 &	0.22 $\pm$ 0.18	\\
6  &	0.43 $\pm$ 0.14 &	0.47 $\pm$ 0.11 &	0.46 $\pm$ 0.11	\\
7  &	0.44 $\pm$ 0.07 &	0.52 $\pm$ 0.11 &	0.56 $\pm$ 0.14	\\
8  &	0.37 $\pm$ 0.25 &	0.42 $\pm$ 0.25 &	0.42 $\pm$ 0.17	\\
9  &	0.49 $\pm$ 0.15 &	0.58 $\pm$ 0.14 &	0.62 $\pm$ 0.13	\\
10 &	0.46 $\pm$ 0.11 &	0.53 $\pm$ 0.08 &   0.55 $\pm$ 0.06 \\
11 &	0.44 $\pm$ 0.18 &	0.48 $\pm$ 0.36 &	0.48 $\pm$ 0.41	\\
12 &	1.07 $\pm$ 0.05 &	1.18 $\pm$ 0.05 &	1.17 $\pm$ 0.08	\\
13 &	0.37 $\pm$ 0.27 &	0.42 $\pm$ 0.31 &	0.41 $\pm$ 0.15	\\
14 &	0.47 $\pm$ 0.08 &	0.54 $\pm$ 0.10 &	0.49 $\pm$ 0.13	\\
15 &	1.27 $\pm$ 0.09 &	1.33 $\pm$ 0.06 &	1.31 $\pm$ 0.06	\\
16 &	0.77 $\pm$ 0.11 &	0.86 $\pm$ 0.14 &	0.91 $\pm$ 0.15	\\
17 &	0.65 $\pm$ 0.10 &	0.74 $\pm$ 0.13 &	0.75 $\pm$ 0.21	\\
18 &	0.16 $\pm$ 0.19 &	0.20 $\pm$ 0.10 &	0.18 $\pm$ 0.09	\\
19 &	0.93 $\pm$ 0.37 &	1.01 $\pm$ 0.02 &	1.05 $\pm$ 0.02	\\
20 &	0.50 $\pm$ 0.03 &   0.58 $\pm$ 0.04 &	0.62 $\pm$ 0.06	\\
21 &	0.54 $\pm$ 0.02 &	0.61 $\pm$ 0.05 &	0.65 $\pm$ 0.12	\\
22 &	0.62 $\pm$ 0.27 &	0.72 $\pm$ 0.36 &	0.75 $\pm$ 0.06 \\ 
\hline
\end{tabular}%}
\tablefoot{Corrections are defined as mean differences between MUSE and WFI \citep{beccari2015} magnitudes. Provided uncertainties are standard deviations of the difference between MUSE and WFI magnitudes.}
\end{table}

\begin{figure}
\includegraphics[width=\columnwidth]{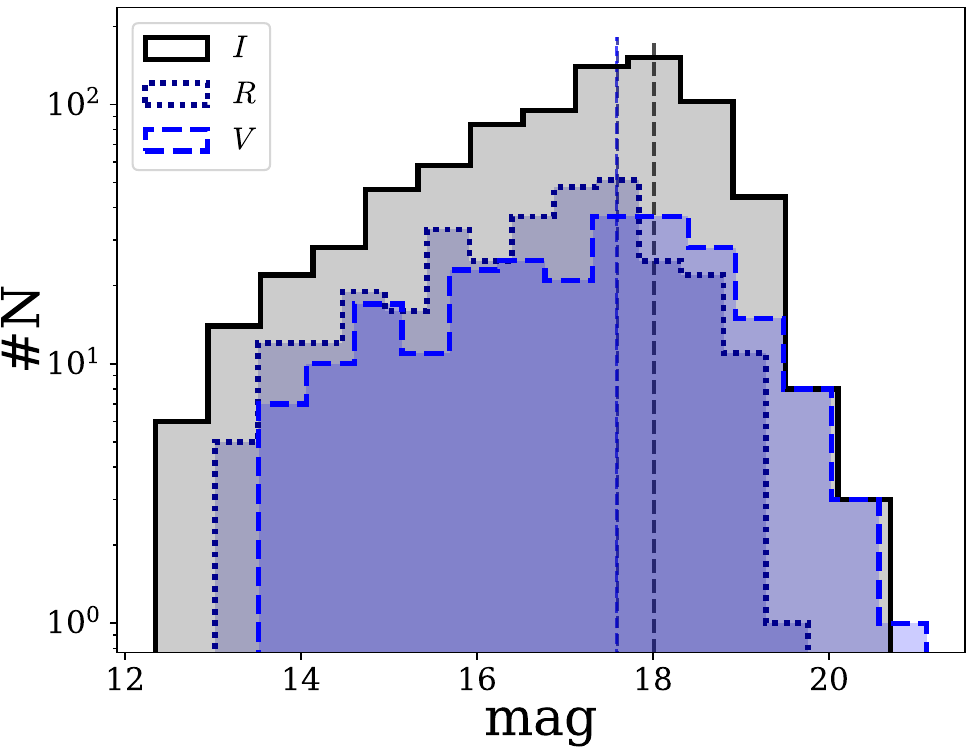} 
\caption{Distribution of corrected MUSE magnitudes. 
colours indicate different photometric bands as stated in the upper left corner of the figure. Dashed lines show the turn-over point of the distributions located at: 18.01~mag ($I$-band), 17.60~mag ($R$-band), 17.58~mag ($V$-band).
}
\label{fig:mag-sens}
\end{figure}

We show the resulting distributions of the corrected MUSE magnitudes in Figure \ref{fig:mag-sens}. The distribution of $I$-band magnitudes peaks at 18.01~mag, $R$-band -- 17.60~mag, and $V$-band -- 17.58~mag. The number of magnitudes in each band decreases bluewards from 780 in $I$-band, through 294 in $R$-band, to 223 in $V$-band.

\section{Spectral templates}
\label{app:templ}
\indent

Here, we list all Class III stars and their properties used in spectral classification as spectral templates (Table \ref{tab:templ}). 
Spectral types for those stars later than K5 were obtained based on depth of molecular absorption bands (TiO, VO and CaH) and few photospheric lines (e.g., NaI, CaI, MgI, etc.) present in optical part of the spectra \citep{manara2013}. Earlier $K$-type stars were identified using the spectral indices introduced by \cite{herczeg2014}, while $G$-type stars were identified based on the difference at 5150~\AA\ of continuum estimated between 4600 and 5400~\AA, and 4900 and 5150~\AA~\citep{herczeg2014}. Effective temperatures ($T_{\mathrm{eff}}$) were derived from spectral types using relations from \cite{luhman2003b} for $M$-type objects and \cite{kenyon1995} for $K$- and $G$-type stars. Most of the templates have none or negligible extinction \citep[$A_{\mathrm{V}}<0.5$~mag,][]{manara2017}; spectra were dereddened before analysis assuming the extinction law from \cite{cardelli1989} and $R_{\mathrm{V}}=3.1$. All the details of the data reduction, calibration, and spectral classification are provided in the original papers. 

\begin{table}[h!]
\centering
\caption{Properties of Class III stars used as spectral templates.} 
\label{tab:templ} 
%\resizebox{\columnwidth}{!}{
\begin{tabular}{cccc}
\hline
\hline
Object & SpT & $T_{\mathrm{eff}}$ (K) & Reference \\ 
\hline
RXJ1508.6-4423      & G8.0  &	5520 & 2 \\
RXJ1526.0-4501      & G9.0  &	5410 & 2 \\
HBC407              & K0.0  &	5200 & 2 \\
RXJ1515.8-3331      & K0.5  &	5050 & 2 \\
PZ99J160550.5-253313 & K1.0 &	5000 & 2 \\
RXJ0457.5+2014		& K1.0	&   5000 & 2  \\
RXJ0438.6+1546      & K2.0  &	4900 & 2 \\
RXJ1547.7-4018      & K3.0  &	4730 & 2 \\
RXJ1538.6-3916      & K4.0  &	4590 & 2 \\
RXJ1540.7-3756      & K6.0  &	4205 & 2 \\
RXJ1543.1-3920		& K6.0	&   4205 & 2 \\
SO879			    & K7.0	&   4060 & 1 \\	
TWA6                & K7.0  &	4060 & 1 \\	
Tyc7760283\_1       & M0.0  &	3850 & 1 \\
TWA14               & M0.5  &	3780 & 1 \\
RXJ1121.3-3447\_app2 & M1.0  &	3705 & 1 \\
RXJ1121.3-3447\_app1 & M1.0  &	3705 & 1 \\
CD\_29\_8887A       & M2.0  &	3560 & 1 \\
Sz122               & M2.0  &	3560 & 1 \\
TWA15\_app2         & M3.0  &	3415 & 1 \\
TWA7                & M3.0  &	3415 & 1 \\
TWA15\_app1         & M3.5  &	3340 & 1 \\
Sz94                & M4.0  &	3270 & 1 \\
Sz121               & M4.0  &	3270 & 1 \\ 
SO797               & M4.5  &	3200 & 1 \\
SO641               & M5.0  &	3125 & 1 \\
Par\_Lup3\_2        & M5.0  &	3125 & 1 \\
SO925               & M5.5  &	3060 & 1 \\
SO999               & M5.5  &	3060 & 1 \\
Sz107               & M5.5  &	3060 & 1 \\
Par\_Lup3\_1        & M6.5  &	2935 & 1 \\
LM717               & M6.5  &	2935 & 2 \\
J11195652-7504529   & M7.0  &	2880 & 2 \\
LM601               & M7.5  &	2795 & 2 \\
CHSM17173           & M8.0  &	2710 & 2 \\
TWA26               & M9.0  &	2400 & 1 \\
DENIS1245           & M9.5  &	2330 & 1 \\
\hline
\end{tabular}%}
\tablefoot{References: (1) \cite{manara2013}; (2) \cite{manara2017}.}
\end{table}

\section{Spectral classification}
\label{app:spt}
\indent

In this Appendix, we list spectral indices used for spectral classification of K- and late G-type stars, as well as we explain more in detail estimation of uncertainties based on $\chi^2_{\rm red}$ maps. 

Indices were defined, calibrated, and tested on Class III spectra listed in the Table \ref{tab:templ}. Each spectral index is defined as an equivalent width of a given line or a line ratio of the two. Table \ref{tab:ind} shows our indices together with their weights. The final spectral type is a weighted average of indices. Indices with values outside the applicable range of spectral types (indicating type earlier than G8 or later than M0) were excluded from the average to avoid extrapolation. We required at least 3 valid indices to estimate the spectral type. Additionally, we included in our list the index from \cite{oliveira2003, jeffries2007}, TiO~7140\AA. It is applicable only to the stars with spectral type later than K5. 

Extinction and veiling at 7500\AA, as well as spectral type for M-type stars, were estimated based on $\chi^2_{\rm red}$ maps. Figure \ref{fig:chi2-Mstar} shows the examples of such maps for an M-type star, and Figure \ref{fig:chi2-Kstar} for an K-type star. At each time, we examine the distribution of $\chi^2_{\rm red}$ in relation to a given two out of three variables in our problem (spectral type, SpT; visual extinction, \Av, and constant veiling at 7500~\AA, \veil). Hence, for M-type stars, there are three $\chi^2_{\rm red}$ maps for each star. For K-type stars, where uncertainties of the spectral type are assigned differently, there is only one $\chi^2_{\rm red}$ map constructed based on the Class III template closest with the SpT to the SpT of a given star. 
The best set of parameters' values is indicated by the minimum value of the $\chi^2_{\rm red}$. The 1-sigma contours drawn on top of the distributions are the basis for the uncertainty estimates. We adopt the projections of the contours onto the axis as the uncertainties of the given parameters. Fig. \ref{fig:chi2-Mstar} and \ref{fig:chi2-Kstar} show 1-sigma contours, and additionally also 2-, and 3-sigma ones, for reference.

Estimating uncertainties based on our $\chi^2_{\rm red}$ distributions is itself prone to the uncertainty. The sampling of spectral templates used for classification is not even causing discontinuous, step-like shape of 1-sigma contours. We do not propagate errors outside the range of adopted values for our parameters. As a result, when the best value is close to the edge if this range, one of the uncertainties will be underestimated. We observe high degeneracy between veiling and extinction, as well as high uncertainty of the value of the veiling, especially in hotter stars. Therefore, our veiling estimates might be inaccurate and we recommend to treat them as the rough indications of the presence of the veiling and its prominence.

\begin{table*}
\centering
\caption{Indices used for spectral classification of K- and late G-type stars.} 
\label{tab:ind} 
\resizebox{\textwidth}{!}{
\begin{tabular}{ccccccccccccccc}
\hline
\hline
Index & weight & G8 & G9 & K0 & K1 & K2 & K3 & K4 & K5 & K6 & K7 & M0 & Uncertainty & Source \\ 
\hline
Na{\sc I} 5890\AA & 2.0 & 1.424 & 1.971 & 2.519 & 3.066 & 3.613 & 4.160 & 4.708 & 5.255 & 5.802 & 6.349 & 6.896 & 0.112 & This work \\
Ca{\sc I} 6162\AA & 1.0 & 0.850 & 1.029 & 1.207 & 1.386 & 1.564 & 1.743 & 1.921 & 2.099 & 2.278 & 2.456 & 2.635 & 0.033 & This work \\
Ca{\sc I} 6103\AA & 1.0 & 0.323 & 0.358 & 0.393 & 0.428 & 0.463 & 0.498 & 0.533 & 0.567 & 0.602 & 0.637 & 0.672 & 0.018 & This work \\
Na{\sc I} 8183\AA & 1.0 & 0.279 & 0.325 & 0.372 & 0.418 & 0.465 & 0.511 & 0.558 & 0.604 & 0.651 & 0.697 & 0.743 & 0.008 & This work \\
Na{\sc I} 8195\AA & 1.0 & 0.357 & 0.412 & 0.468 & 0.523 & 0.578 & 0.634 & 0.689 & 0.744 & 0.800 & 0.855 & 0.910 & 0.016 & This work \\
Ca{\sc I} 8690\AA & 2.0 & 0.315 & 0.358 & 0.401 & 0.444 & 0.487 & 0.530 & 0.573 & 0.616 & 0.659 & -- & -- & 0.016 & This work \\
Mg{\sc I} 8806\AA & 1.75 & 0.665 & 0.713 & 0.761 & 0.809 & 0.857 & 0.905 & 0.954 & 1.002 & 1.050 & -- & -- & 0.029 & This work \\
Mg{\sc II} 8824\AA & 2.25 & 0.230 & 0.263 & 0.296 & 0.329 & 0.362 & 0.395 & 0.428 & 0.460 & 0.493 & 0.526 & 0.559 & 0.010 & This work \\
K{\sc I} 7665\AA/7699\AA & 1.0 & 2.531 & 2.442 & 2.354 & 2.265 & 2.177 & 2.088 & 2.000 & 1.912 & 1.823 & 1.735 & 1.646 & 0.111 & This work \\
Ca{\sc II} 8663\AA/Ca{\sc I} 8690\AA & 1.75 & 3.639 & 3.447 & 3.256 & 3.065 & 2.873 & 2.682 & 2.490 & 2.299 & 2.108 & 1.916 & 1.725 & 0.189 & This work \\
Mg{\sc I}~8806\AA/Mg{\sc II}~8824\AA & 1.25 & 2.834 & 2.722 & 2.611 & 2.499 & 2.387 & 2.276 & 2.164 & 2.052 & 1.941 & 1.829 & 1.717 & 0.137 & This work \\
\hline
\multirow{2}{*}{TiO 7140\AA} & \multirow{2}{*}{1.5} &\multirow{2}{*}{--} & \multirow{2}{*}{--} & \multirow{2}{*}{--} & \multirow{2}{*}{--} & \multirow{2}{*}{--} & \multirow{2}{*}{--} & \multirow{2}{*}{--} & \multirow{2}{*}{-2} & \multirow{2}{*}{-1.5} & \multirow{2}{*}{-1} & \multirow{2}{*}{0} & \multirow{2}{*}{0.1} & \cite{jeffries2007} \\
& & & & & & & & & & & & & & \cite{oliveira2003} \\
\hline
\end{tabular}}
\tablefoot{Indices listed here are equivalent widths and their ratios, as indicated by the index name. Values were obtained from linear fitting of equivalent widths to spectral types. Listed uncertainties represent the uncertainty of this fit. The final spectral type is a weighted average of corresponding spectral types to each index and their weights. Indices and weights were calibrated on the Class III spectra listed in the Tab. \ref{tab:templ}.}
\end{table*}

\begin{figure*}
\includegraphics[width=\textwidth, trim={0.2 0 1.8cm 0cm}, clip]{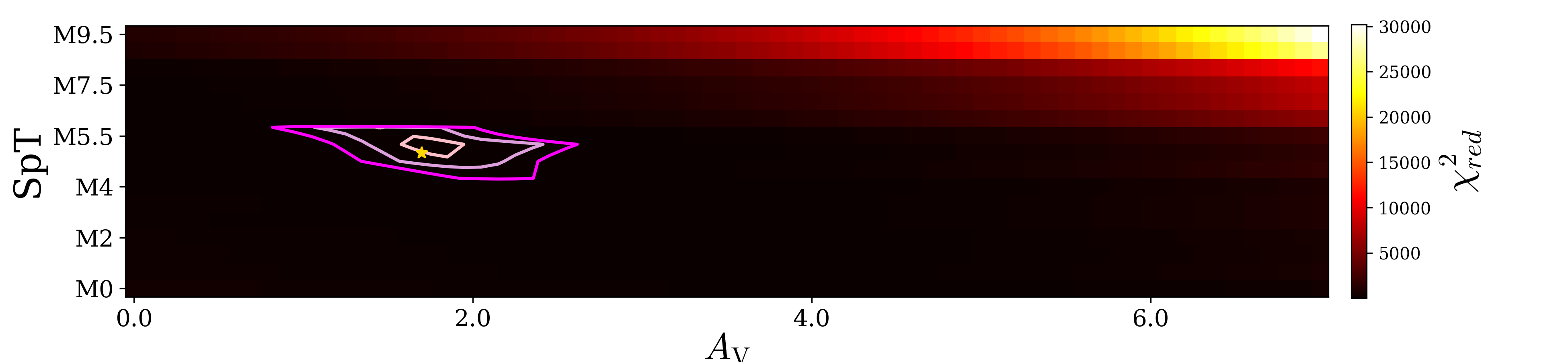}
\includegraphics[width=\textwidth, trim={0.3cm 0 1.8cm .8cm}, clip]{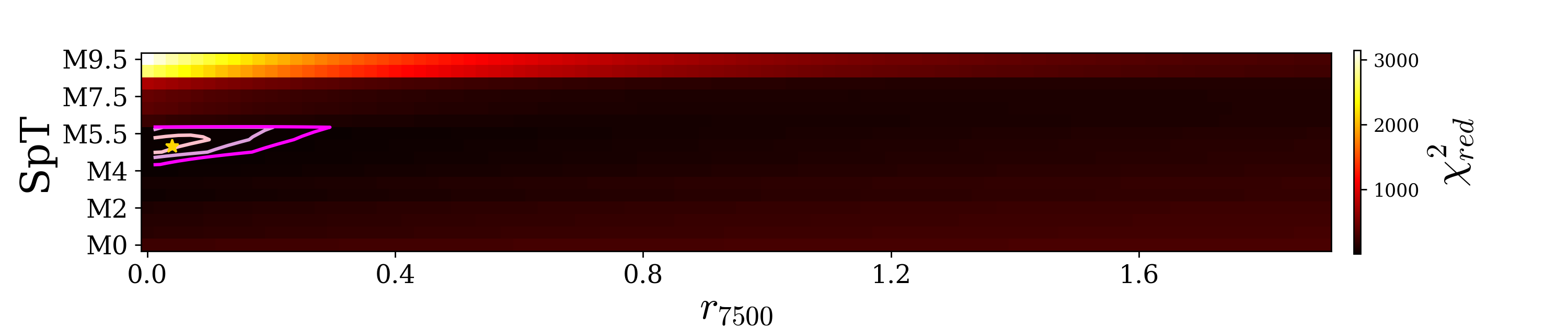}
\caption{The $\chi^2_{\rm red}$ maps corresponding to the fit presented in Fig. \ref{fig:spec-Mstar}. {Top:} $\chi^2_{\rm red}$ as a function of spectral type and visual extinction. {Bottom:} $\chi^2_{\rm red}$ as a function of spectral type and veiling. The yellow star marks the best-fit position in the parameter space. Contours represent 1, 2, and 3$\sigma$ levels.}
\end{figure*}
\addtocounter{figure}{-1}
\begin{figure}
\includegraphics[width=\columnwidth]{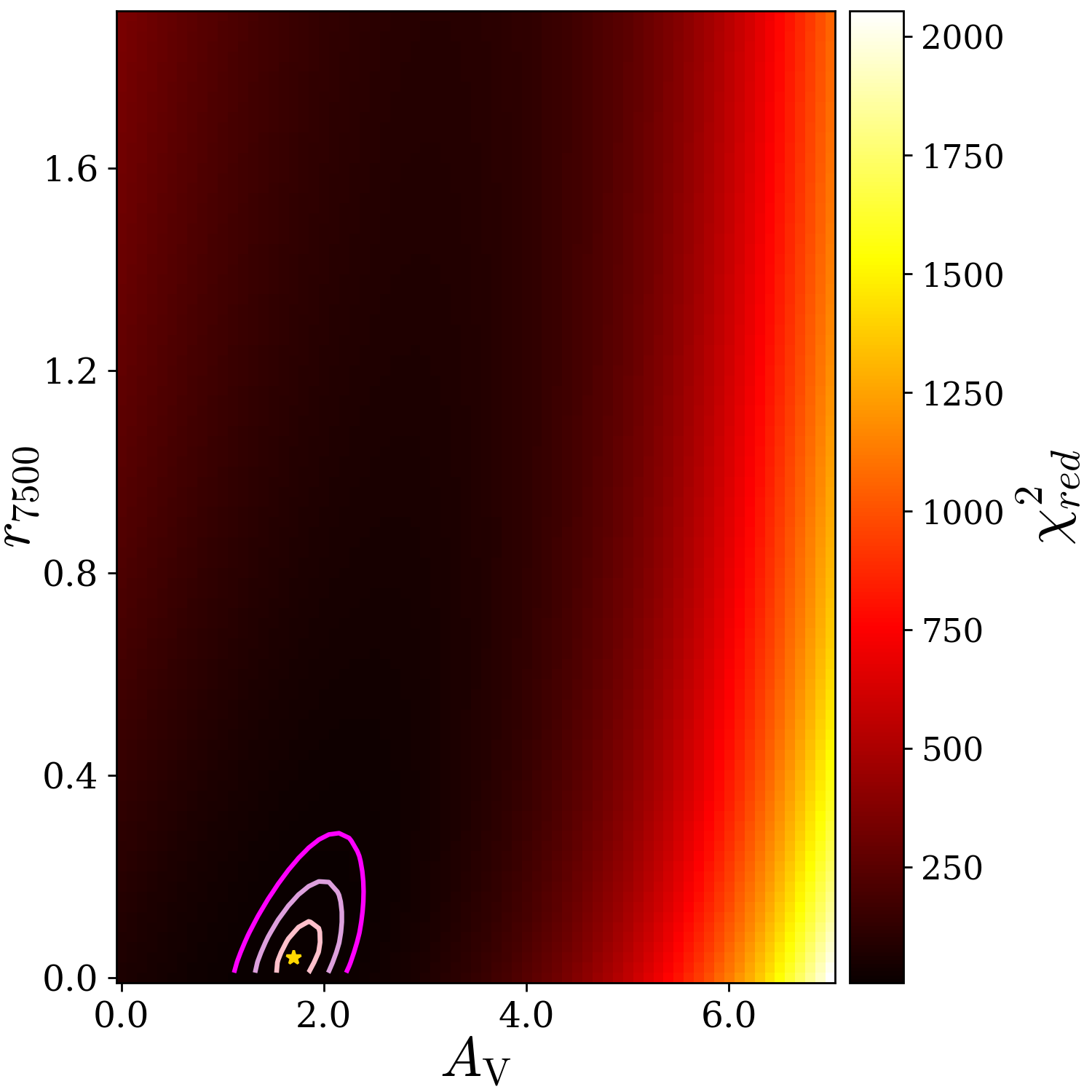}
\caption{{(cont.)} 
The $\chi^2_{\rm red}$ as a function of veiling and extinction. The yellow star marks the best-fit position in the parameter space. Contours represent 1, 2, and 3$\sigma$ levels.}
\label{fig:chi2-Mstar}
\end{figure}

\begin{figure}
\includegraphics[width=0.95\columnwidth]{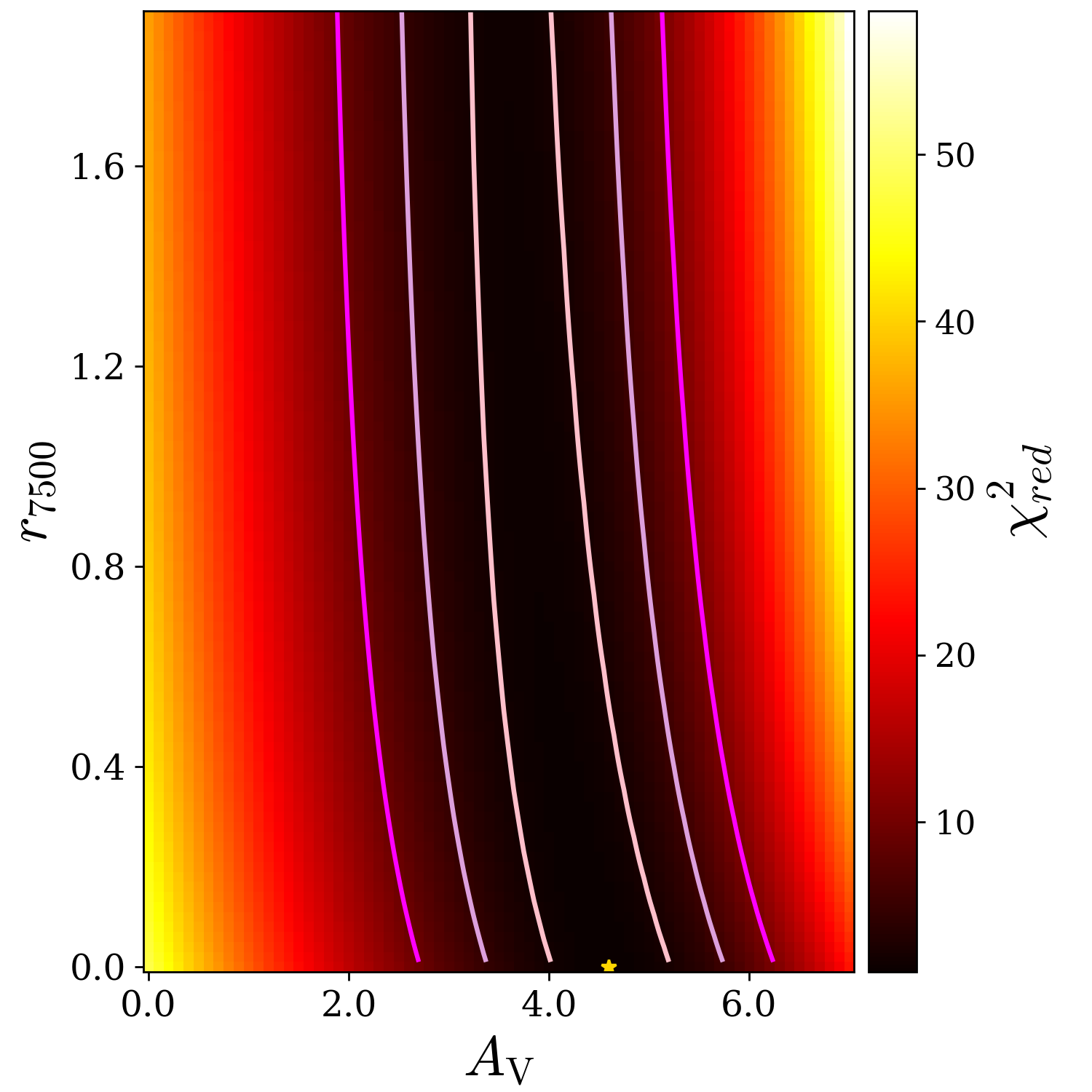}
\caption{The $\chi^2_{red}$ map as a function of veiling and extinction corresponding to the fit presented in Fig. \ref{fig:spec-Kstar}. The yellow star marks the best-fit position in the parameter space. Contours represent 1, 2, and 3$\sigma$ levels. 
}
\label{fig:chi2-Kstar}
\end{figure}

\section{Stellar parameters}
\label{app:tracks}
\indent

\begin{figure}
\includegraphics[width=\columnwidth, trim={0.cm 2.7cm 0cm 2.5cm}, clip]{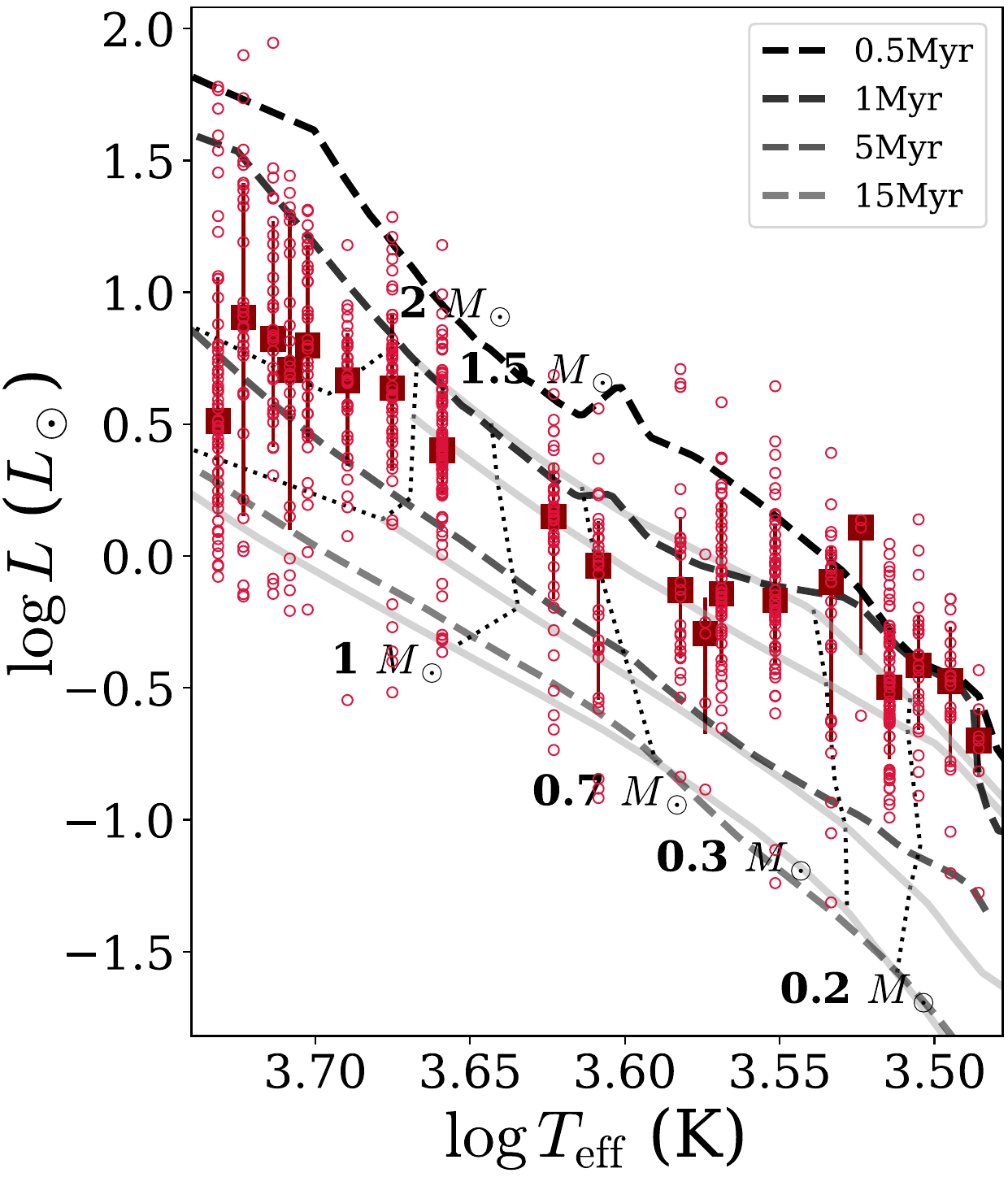}
\caption{HR diagram for low-mass stars of \tr. Empty circles show data points, open squares are median values of the bolometric luminosity for each spectral subclass with errorbars indicating 1-$\sigma$ percentiles. Theoretical tracks from \cite{baraffe2015} are shown as solid grey lines, whereas tracks from \cite{siess2000} are plotted as dashed lines. Dotted lines show tracks for various masses of stars.}
\label{fig:HR-app}
\end{figure}

\begin{figure}
\includegraphics[width=\columnwidth, trim={.2cm 0.1cm .cm .cm}, clip]{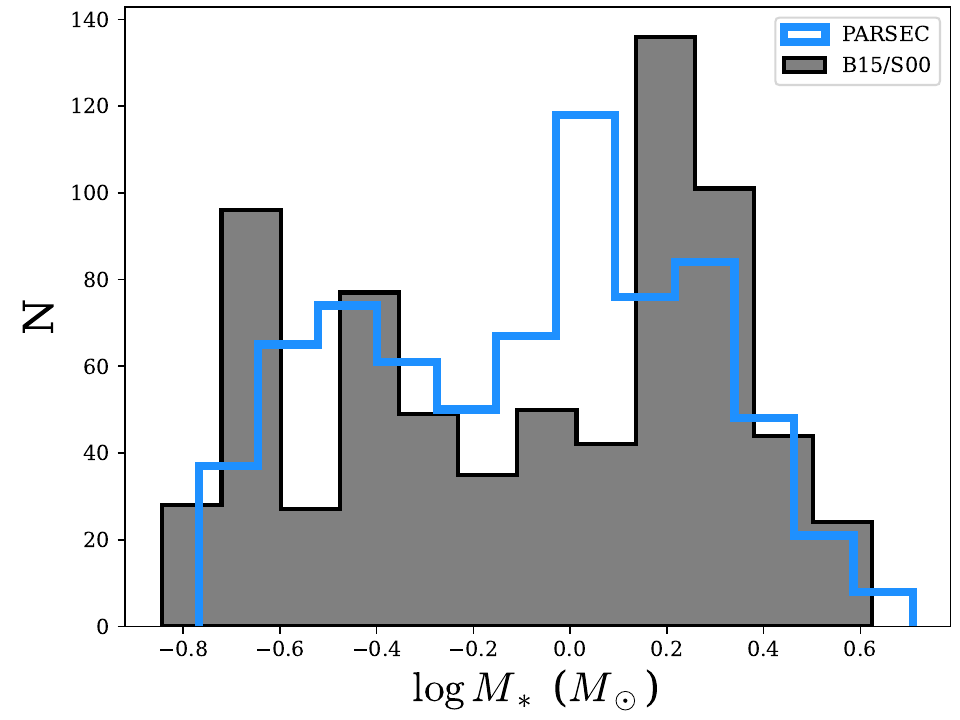}
\caption{Distribution of all stellar masses from the final catalog in logarithmic scale. Black histogram filled with grey shows masses estimated based on evolutionary tracks from \cite{baraffe2015} and \cite{siess2000}, while blue open histogram shows stellar masses based on PARSEC models \citep{bressan2012}.} 
\label{fig:mass-comparison}
\end{figure}

\begin{figure}
\includegraphics[width=\columnwidth, trim={.2cm 0.1cm .cm .cm}, clip]{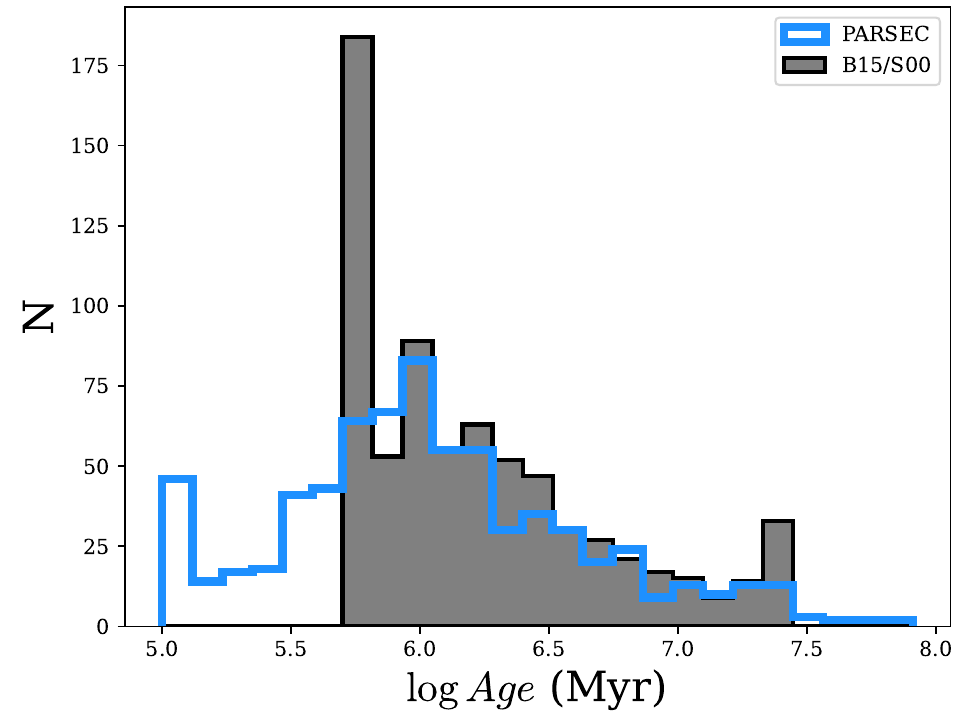}
\caption{Distribution of all stellar ages from the final catalog. Black histogram filled with grey shows ages estimated based on evolutionary tracks from \cite{baraffe2015} and \cite{siess2000}, while blue open histogram shows stellar ages based on PARSEC models \citep{bressan2012}. PARSEC models span ages down to 0.1~Myr, while the other two -- to 0.5~Myr. Both histograms have the same bins fixed to the distribution of PARSEC ages.}
\label{fig:age-comparison}
\end{figure}

\begin{figure}
\includegraphics[width=\columnwidth, trim={.2cm 0.1cm .cm .cm}, clip]{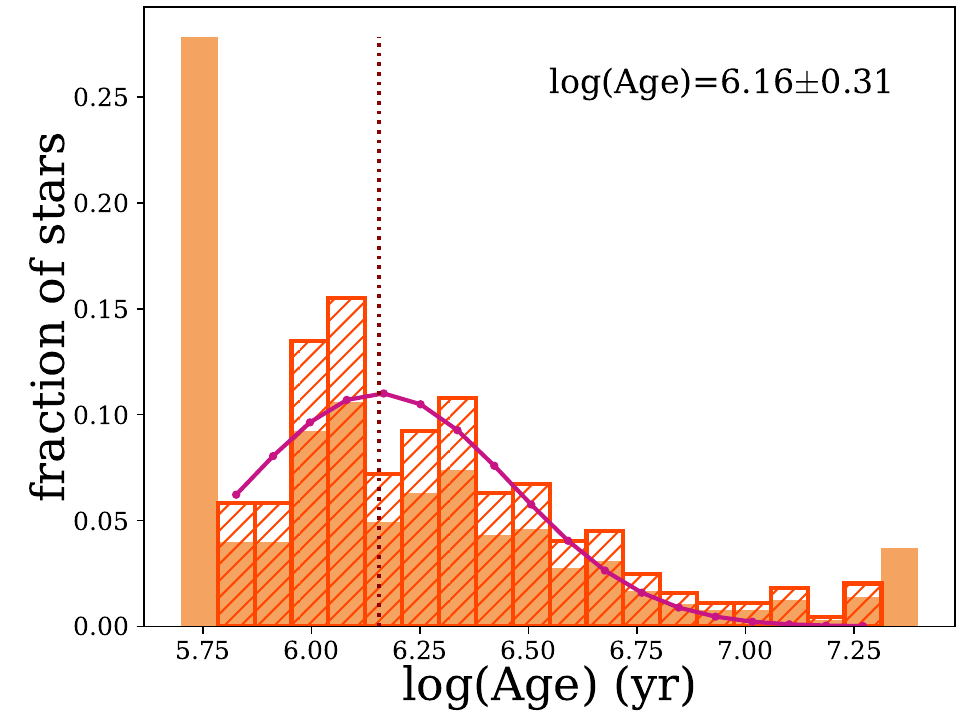}
\caption{Fraction of stellar ages derived from HR diagram for stars with $\log{(T_{\rm eff})}<3.73$, analogous to Fig. \ref{fig:age} but using tracks from \cite{baraffe2015} and \cite{siess2000}. Filled orange histogram shows distribution for the whole sample, while hatched red histogram represents the fraction distribution after removing the extreme bars with respect to the total number of stars within the new age range. The normal fit to the probability density distribution converted into the fraction distribution for the visual purposes is shown as a dark violet curve with mean value of $\log{\rm{(age)}}=6.16\pm0.31$ corresponding to the 1.4$^{+1.5}_{-0.7}$~Myr}. 
\label{fig:age-app}
\end{figure}

{Selection of stellar evolutionary models impact values of derived stellar parameters. In Section \ref{subsec:HR} we constructed HR diagram and employed PARSEC v1.2S tracks \citep{bressan2012} to estimated masses and ages of \tr\ members. Here, we investigate how the choice of tracks impact our results on cluster properties. We use tracks developed for young stars from \cite{siess2000} and \cite{baraffe2015}. The latter are dedicated to low- and very-low mass stars and therefore do not cover star more massive than $\sim$1~$M_\odot$. Thus, we combine them with tracks from \cite{siess2000} and define a border of spectral type K5 between usage of the two models. Neither model explores ages below 0.5~Myr. We restricted the comparison up to 30~Myr, as we do not expect true \tr\ members to be that old. 

We present HR diagram in Figure \ref{fig:HR-app}. Comparison with HR diagram using PARSEC tracks in Figure \ref{fig:HR} shows differences in stellar masses at the lower end. Figure \ref{fig:mass-comparison} compares mass distributions of the two sets of tracks. Distribution of PARSEC masses is shifted with respect to the \cite{baraffe2015} / \cite{siess2000} ones. Differences are the most prominent between PARSEC and \cite{baraffe2015} models up to a factor of $\sim$2. Masses from PARSEC and \cite{siess2000} seem to be consistent with some spread.

The PARSEC models span wider range of stellar ages at the lower end, down to 0.1~Myr, while \cite{baraffe2015} and \cite{siess2000} stop at 0.5~Myr. Therefore, the age distribution from \cite{baraffe2015} / \cite{siess2000} set are affected by the artificial overdensity at the edge of the distribution (Figure \ref{fig:age-comparison}). That feature motivated removal of histogram bars at the borders from the analysis. As we highlighted in Section \ref{subsec:age}, the same method of estimating the cluster age applied to two sets of evolutionary models yields the same within uncertainties cluster age of 1~Myr. We show corresponding distribution in Figure  \ref{fig:age-app}. Individual measurements are affected by the uncorrelated differences between PARSEC and \cite{baraffe2015}, up to a factor of $\sim$3. However, \cite{siess2000} isochrones seem to be offset by a constant factor of 2 towards younger ages.

We have chosen PARSEC models for the analysis because they allow homogeneous treatment of all stars in our sample spamming wide range of masses. We note however, that individual estimates of stellar parameters are uncertain by an unknown value. The short comparison here between two sets of tracks showed that values of both masses and ages can differ by a factor of few depending on a chosen tracks. This significantly exceeds any possible estimates of errors of those parameters done accounting for observational uncertainties and tabulation of spectral templates used here. Therefore, in Tables \ref{tab:cat} and \ref{tab:highbkgcat} we only report estimated values as true uncertainties are impossible to asses.
}

\section{Young stars in Trumpler 14}
\label{app:populations}
\indent

\begin{figure}
\includegraphics[width=\columnwidth, trim={.5cm 0.5cm 0.4cm 0.3cm}, clip]{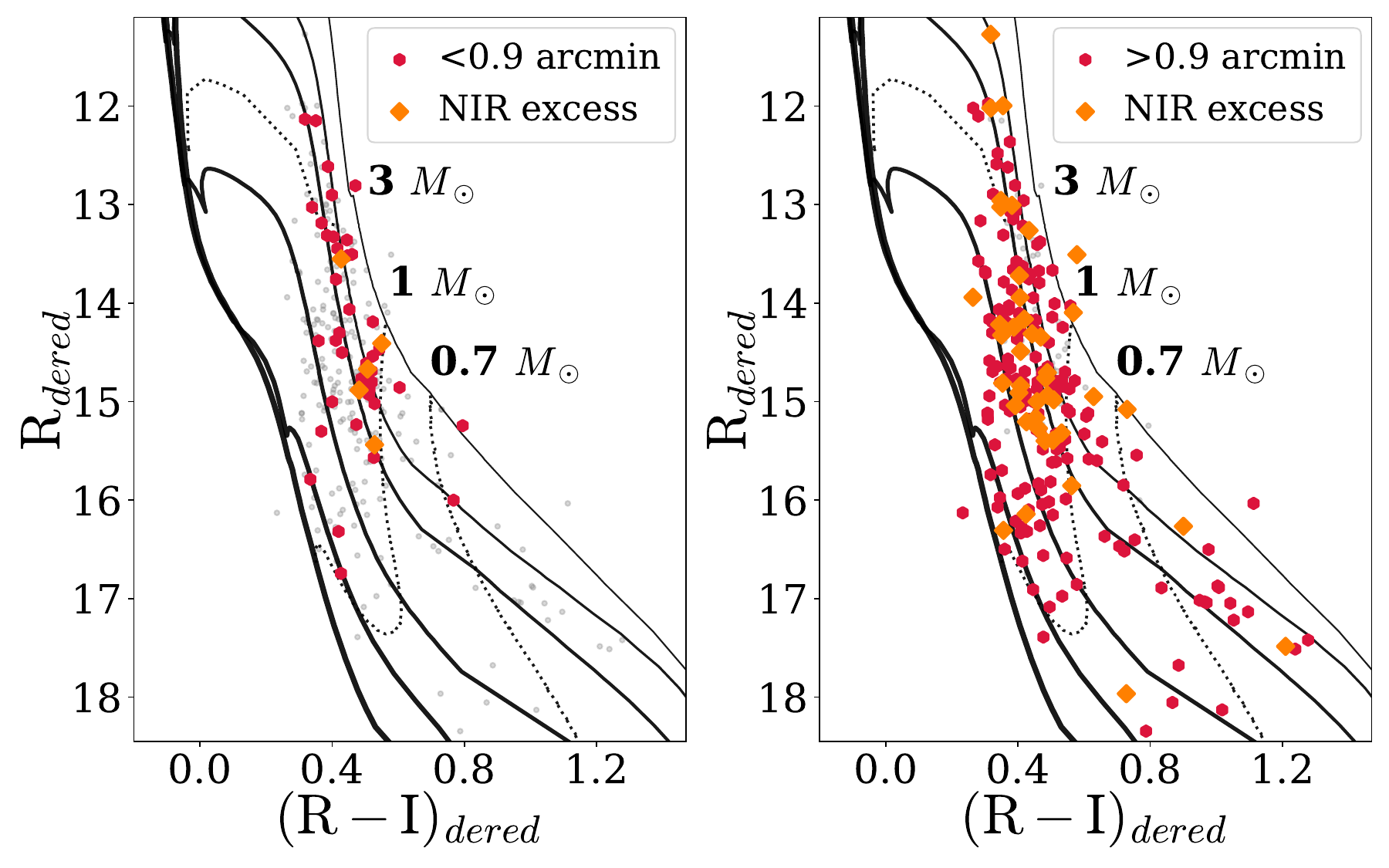}
\includegraphics[width=\columnwidth, trim={.5cm 0.5cm 0.4cm 0.3cm}, clip]{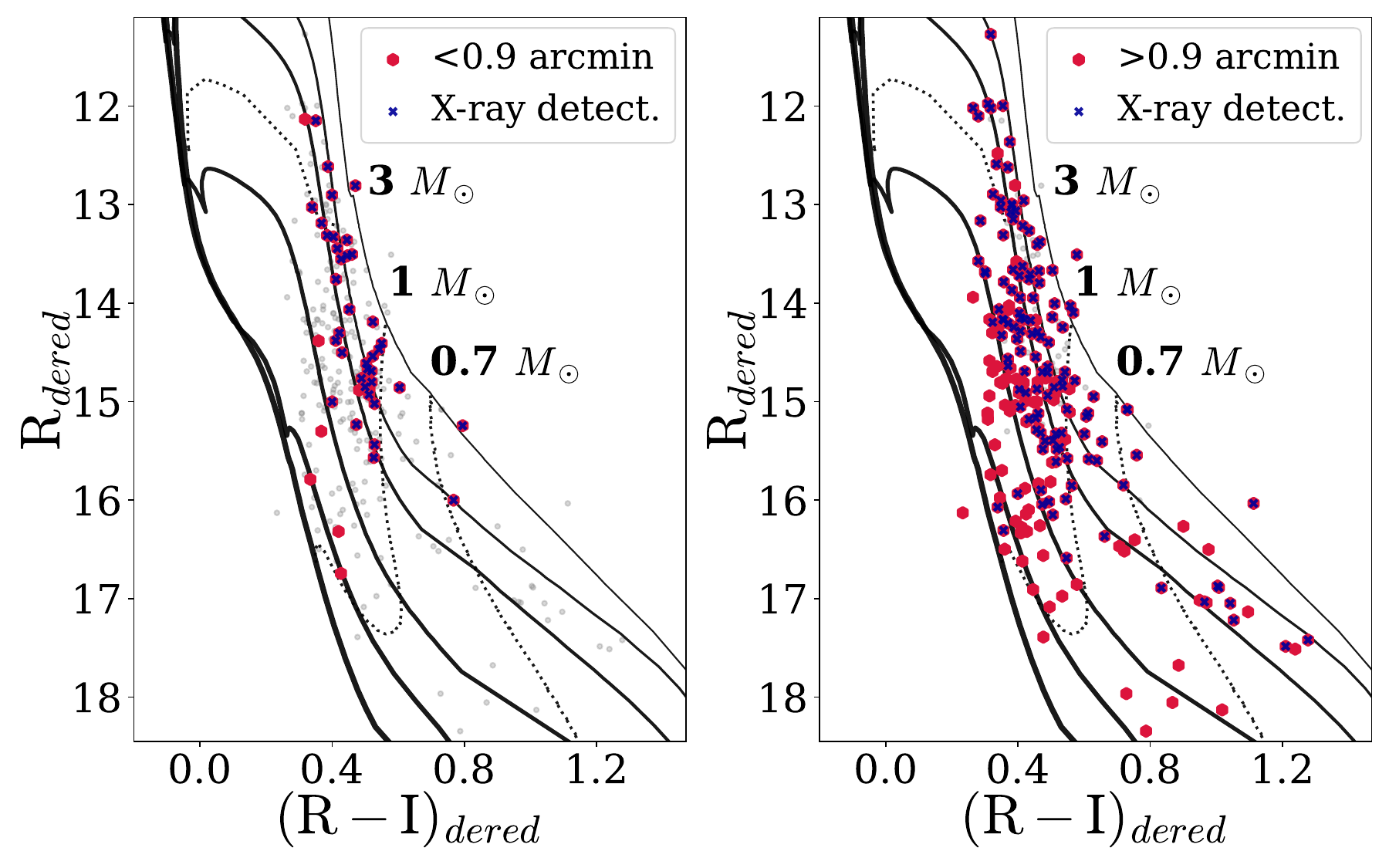}
\caption{Colour-magnitude diagrams for {de-reddened} $R$ and $I$-band magnitudes from MUSE. Red hexagons mark stars within the core of \tr\ ({left}, 0.9\arcmin, \cite{kharchenko2013}) or outside ({right}). The orange diamonds ({top}) indicate the NIR excess stars, while the dark blue crosses mark sources with detected X-ray component \citep[{bottom},][]{townsley2011}.} 
\label{fig:CMD-pop-NIR-X}
\end{figure}

\begin{figure}
\includegraphics[width=\columnwidth, trim={.cm 0 0cm 0cm}, clip]{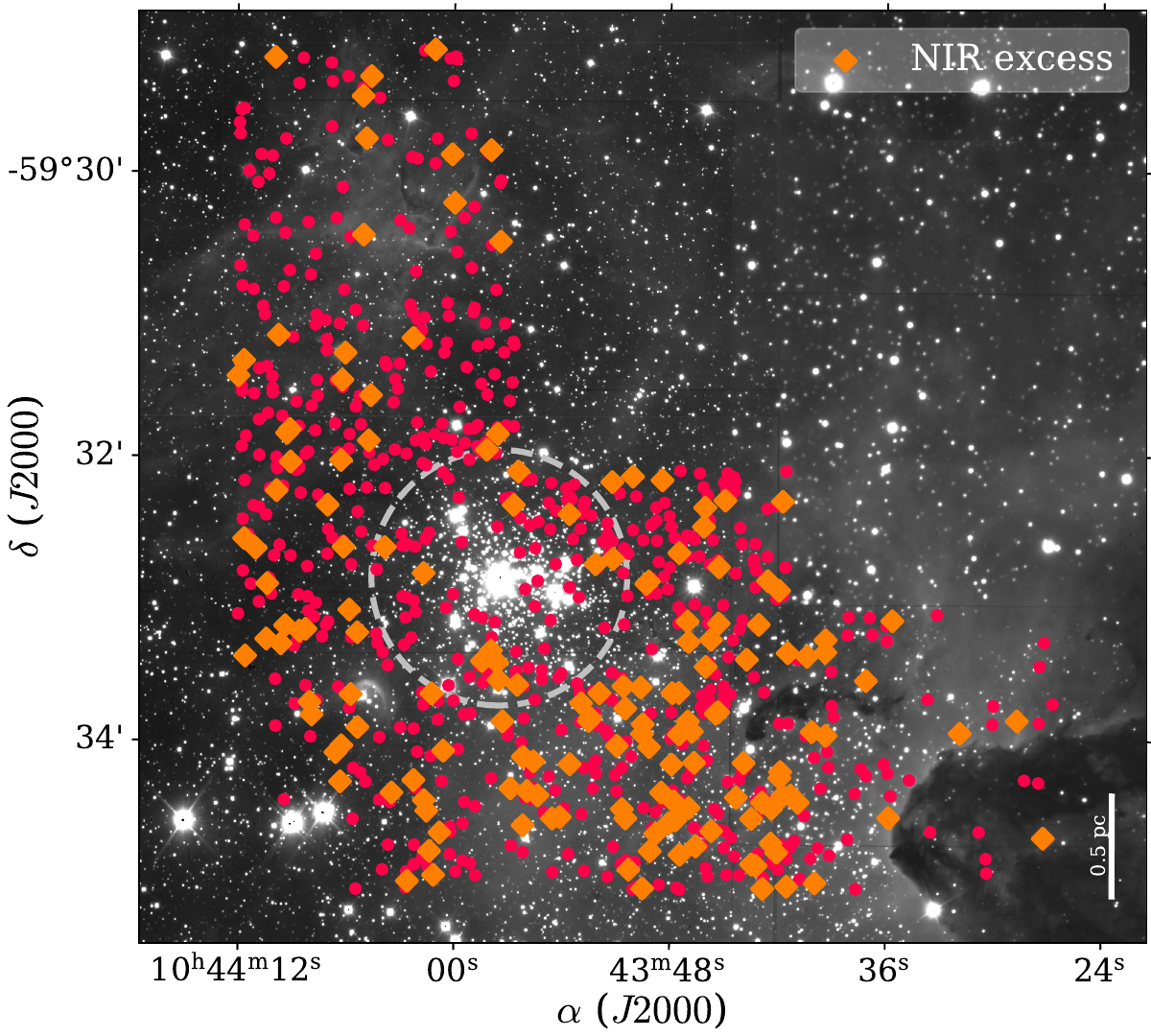}
\includegraphics[width=\columnwidth, trim={.cm 0 0cm 0cm}, clip]{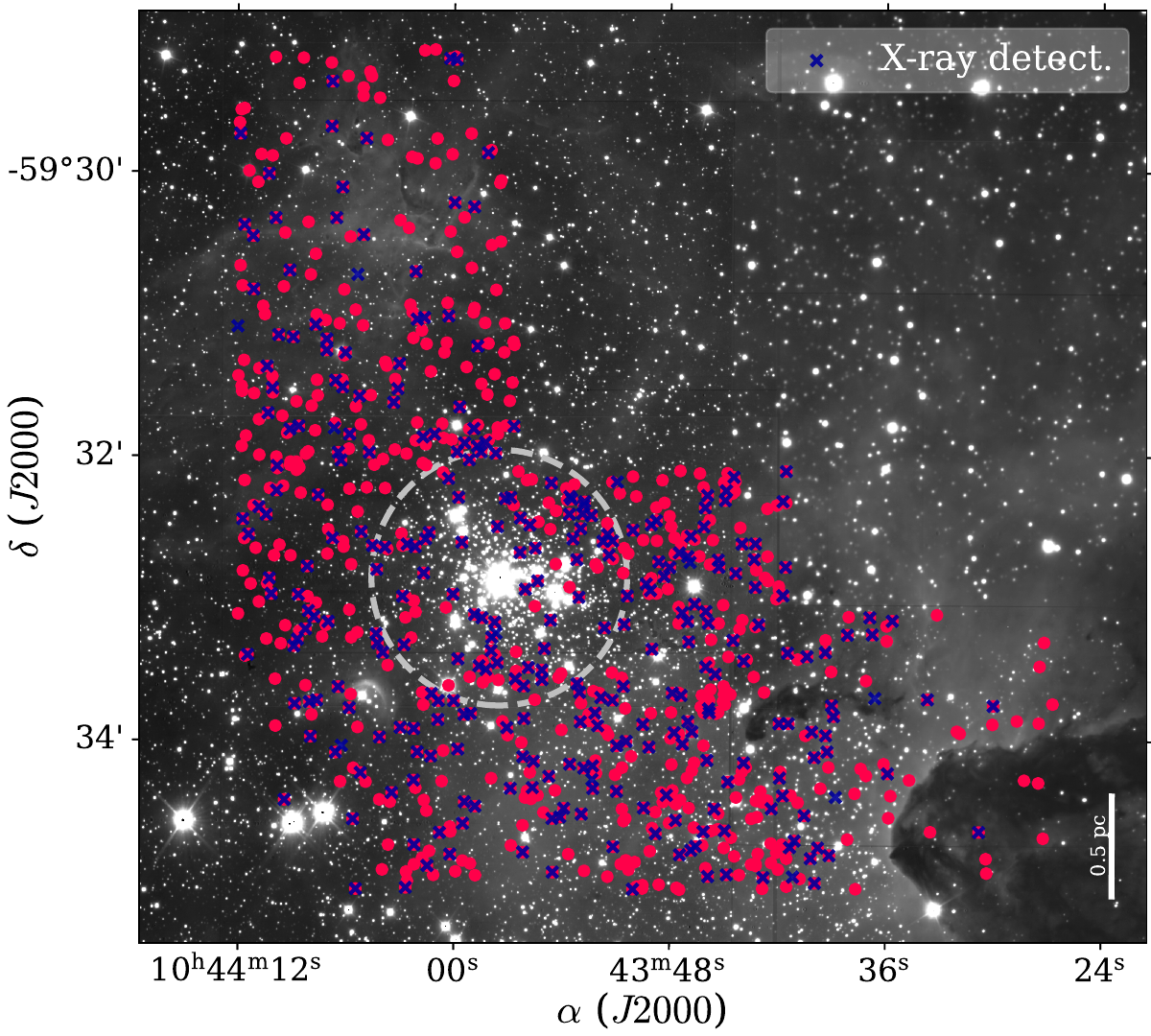}
\caption{Locations of the NIR excess stars (orange diamonds, {top}) and the X-ray detections in \tr\ (dark blue crosses, {bottom}). All stars studied here are marked with red dots, as in Fig. \ref{fig:Tr14detect}. The dashed circle with radius of 0.9\arcmin\ shows the core of the \tr, as defined by \cite{kharchenko2013}. The background image in grey scale is the $H$-band image from HAWK-I \citep{preibisch2011a,preibisch2011b}.} 
\label{fig:sky-NIR-X}
\end{figure}

Here, we complete the discussion in Section \ref{subsec:age} showing the distribution of NIR excess and X-ray sources on the CMD (Fig. \ref{fig:CMD-pop-NIR-X}) and the sky (Fig. \ref{fig:sky-NIR-X}). To define NIR excess stars, we collect the NIR photometry from \cite{preibisch2011a} and \cite{preibisch2014}, and follow the definition of the NIR excess of \cite{zeidler2016}. Whenever photometry from the both catalogues is available, we choose the one with better signal-to-noise ratio. X-ray detections originate in the {\it Chandra} Carina Complex Project \citep[CCCP,][]{townsley2011}. NIR excess is often interpreted as a signpost of the inner circumstellar disk, while strong X-ray emission is expected from low-mass stars. The distributions of both characteristics confirm, that the core of \tr\ consists mainly of young stars ($\sim$1~Myr), while the extended, halo population has more diverse ages, including the very young stars. There is no strong correlation between any of those characteristics and location in the cluster for stars from the extended population.

\end{appendix}

\end{document}